\documentclass[aps,amsmath,amssymb,amsfonts,amsthm,superscriptaddress,preprint,nofootinbib]{revtex4-1}

\usepackage{graphicx}
\usepackage{subfigure}
\usepackage{enumitem}
\usepackage{float}
\usepackage[colorlinks,bookmarks=false,citecolor=blue,linkcolor=blue,urlcolor=blue]{hyperref}
\usepackage{amssymb}
\usepackage{pifont}
\usepackage{bm}
\usepackage{amsmath}
\usepackage{booktabs}
\usepackage{multirow}

\def \k {\bm{k}}

\def \H {\mathcal{H}}

\begin{document}
	
\title{ A Cohomological Framework for Topological Phases from Momentum-Space Crystallographic Groups}

\author{T. R. Liu}
\affiliation{Department of Physics and HK Institute of Quantum Science \& Technology, The University of Hong Kong, Pokfulam Road, Hong Kong, China}

\author{Zheng Zhang}
\affiliation{Department of Physics and HK Institute of Quantum Science \& Technology, The University of Hong Kong, Pokfulam Road, Hong Kong, China}
\affiliation{Department of Physics, School of Science, Lanzhou University of Technology, Lanzhou 730050, China}
	
\author{Y. X. Zhao}
\email[]{yuxinphy@hku.hk}
\affiliation{Department of Physics and HK Institute of Quantum Science \& Technology, The University of Hong Kong, Pokfulam Road, Hong Kong, China}

\begin{abstract}
	Crystallographic groups are conventionally studied in real space to characterize crystal symmetries. Recent work has recognized that when these symmetries are realized projectively, momentum space inherently accommodates nonsymmorphic symmetries, thereby evoking the concept of \textit{momentum-space crystallographic groups} (MCGs). Here, we reveal that the cohomology of MCGs encodes fundamental data of crystalline topological band structures. Specifically, the collection of second cohomology groups, $H^2(\Gamma_F,\mathbb{Z})$, for all MCGs $\Gamma_F$, provides an exhaustive classification of Abelian crystalline topological insulators, serving as an effective approximation to the full crystalline topological classification. Meanwhile, the third cohomology groups $H^3(\Gamma_F,\mathbb{Z})$ across all MCGs exhaustively classify all possible twistings of point-group actions on the Brillouin torus, essential data for twisted equivariant K-theory.
	Furthermore, we establish the isomorphism
	$H^{n+1}(\Gamma_F,\mathbb{Z})\cong H^n\big(\Gamma_F,\operatorname{\mathcal{F}}(\mathbb{R}^d_F,U(1))\big)$ for $ n\ge 1
	$, where $\operatorname{\mathcal{F}}(\mathbb{R}^d_F,U(1))$ denotes the space of continuous $U(1)$-valued functions on the $d$D momentum space $\mathbb{R}^d_F$. The case $n=1$ yields a complete set of topological invariants formulated in purely algebraic terms, which differs fundamentally from the conventional formulation in terms of differential forms. The case $n=2$, analogously, provides a fully algebraic description for all such twistings.
	Thus, the cohomological theory of MCGs serves as a key technical framework for analyzing crystalline topological phases within the general setting of projective symmetry.
\end{abstract}
	
\maketitle

\section{Introduction}
In both electronic and other metamaterials, whether classical or quantum, crystal symmetries are described by crystallographic groups. There are 73 symmorphic space groups out of the total 230 three-dimensional space groups, with the remaining 157 being nonsymmorphic~\cite{bradley2010mathematical,szczepanski2012geometry}. Conventionally, crystallographic groups are only considered in real space, while in momentum space only point groups are taken into account. This is mainly because, for ordinary representations, the $\Gamma$ point, the center of momentum space, is preserved by all point-group symmetries. Therefore, the momentum-space crystallographic groups (MCGs), which are extensions of the point groups by reciprocal-lattice translations, are all symmorphic. This simplification no longer holds for projectively realized crystal symmetries, for which, as revealed by recent developments, it is necessary to consider all 157 nonsymmorphic MCGs and hence all 230 MCGs~\cite{chen2022brillouin,zhang2023nonsymmorphic}. The fractional reciprocal-lattice translations combined with point-group symmetries in momentum space result from the phase factors between real-space translations and point-group symmetries~\cite{mackey1958unitary,mackey1989unitary}.

Nonsymmorphic MCGs have been used to reduce the momentum-space unit cell from the torus to the Klein bottle and, more generally, to all flat compact manifolds, termed platycosms, on which complete topological classifications have been carried out~\cite{chen2022brillouin,Platycosms}. Topological phases protected by nonsymmorphic MCGs have been explored in various condensed-matter and metamaterial systems~\cite{shao2021gauge,xue2022projectively,li2022acoustic,liu2023mobius,meng2023spinful,li2023acoustic,Pu2023acoustic,jiang2023photonic,liu2024topological,Fonseca2024Weyl,Tao2024Higher,zhu2024brillouin,Hu2024higher,Long2024nonabelian,wang2025non,Li2023Klein,huang2025experimental}. Notably, nonsymmorphic MCGs ubiquitously exist in magnetic materials preserving spin-space groups and play a significant role in moiré systems~\cite{xiao2024spin,cualuguaru2025moire}.

Apart from these fascinating applications, we point out in this work that the concept of MCG provides a fresh perspective on the classification of crystalline topological insulators. The MCG framework prompts us to consider band structures over the entire momentum space, on which MCGs act, rather than only on the Brillouin torus. The Brillouin torus is simply the quotient of momentum space by reciprocal translations. An essential advantage is that, if we ignore the MCG action, all vector bundles are topologically trivial over the momentum space $\mathbb{R}_F^d$~\cite{trivial_bundles}. In this sense, the topological classification is entirely determined by the distinct representations of the MCG action on $\mathbb{R}_F^d$.

For Abelian crystalline topological phases, namely, those characterizable by the Abelian Berry connections \cite{XLQi_PRB}, the representations correspond to the first cohomology group $H^1(\Gamma_F, \mathcal{F}(\mathbb{R}_F^d,U(1)))$ consisting of all homomorphisms from $\Gamma_F$ to all continuous $U(1)$-valued functions $\mathcal{F}(\mathbb{R}_F^d,U(1))$ over $\mathbb{R}^d_F$, with the natural $\Gamma_F$-action on $\mathcal{F}(\mathbb{R}_F^d,U(1))$. This will be rigorously justified by the fact that the Borel construction for each point-group action is precisely the classifying space of the corresponding MCG, together with the isomorphism~\cite{Iso_Notes}
\begin{equation}\label{eq:e-Isomorphism}
	H^{n}(\Gamma_F,\mathcal{F}(\mathbb{R}^d_F,U(1)))\cong H^{n+1}(\Gamma_F,\mathbb{Z})
\end{equation}
for $n\ge 1$. Thus, the classification is given by
\begin{equation}\label{eq:cls}
	H^1(\Gamma_F, \mathcal{F}(\mathbb{R}_F^d,U(1)))\cong H^2(\Gamma_F, \mathbb{Z}) .
\end{equation}
The group $H^2(\Gamma_F, \mathbb{Z})$ can be readily computed using GAP~\cite{GAP4}. 

The isomorphism leads to an algebraic formulation of a complete set of topological invariants, radically different from the conventional formulation in terms of differential forms \cite{shiozaki2014topology,alexandradinata2014spin,shiozaki2016topology,shiozaki2022atiyah}. For instance, for all reciprocal lattice translations, $H^2(L_F,\mathbb{Z})\cong \mathbb{Z}^3$ corresponds precisely to the Chern numbers over three 2D sub-tori~\cite{Baile_Nature}. Consequently, the Chern numbers admit an algebraic interpretation as $1$D unitary representations of reciprocal lattice translations over momentum space. Moreover, by incorporating point group symmetries, we present an algebraic formulation of previously unknown topological invariants and provide algebraic formulas for existing ones.

Every crystalline topological phase has an underlying Abelian topological phase, where the Abelian Berry connection is the trace of the non-Abelian Berry connection and symmetry operators are given by the determinants of the full operators \cite{segal1968equivariant}. Thus, Eq.~\eqref{eq:cls} provides an effective approximation for the full classification of crystalline topological phases, which corresponds to the twisted equivariant K-group.

Twistings of point-group actions over the Brillouin torus are essential data for formulating the twisted equivariant K-groups~\cite{atiyah2004twisted,freed2013twisted,gomi2017twists}. 
For each point-group action, the classification of all twistings is given by higher-order cohomology groups of the corresponding MCG, specifically by Eq.~\eqref{eq:e-Isomorphism} with $n=2$. The cohomology groups $H^3(\Gamma_F,\mathbb{Z})$ can be immediately computed by GAP, and the isomorphism provides an algebraic representation of all twistings of point group actions over the Brillouin torus. Previously, only twistings for $2$D symmorphic point-group actions were obtained using the sophisticated Atiyah–Hirzebruch spectral sequence \cite{gomi2017twists}. 

\section{Momentum space crystallographic groups} Let us start with constructing the MCG from a $G$-action $\rho_F$ on $T^d_F$. Here, $G$ is the point group, and each $R\in G$ acts as
\begin{equation}
	\rho_F(R)\bm{k}=R\bm{k}+\bm{\kappa}_R
\end{equation} 
in the general framework of projective crystal symmetries \cite{zhang2023nonsymmorphic}.
Here, $\bm{\kappa}_R$ is a fraction of the reciprocal lattice $L_F$. 
 Then, we can introduce 
\begin{equation}
\bm{\omega}_F(R_2,R_1)=R_2\bm{\kappa}_{R_1}+\bm{\kappa}_{R_2}-\bm{\kappa}_{R_2R_1}\in L_F.
\end{equation}
Consequently, the momentum-space group $\Gamma_F$ is just the twisted semi-direct product,
\begin{equation}\label{eq:Gamma_F}
	\Gamma_F=L_F\rtimes_{(c_F,\bm{\omega}_F)} G.
\end{equation}
Here, $c_F$ denote the arithmetic class of $\Gamma_F$, specifying the $G$-action on $L_F$. All elements can be presented as $\gamma_F=(\bm{l},R)$ with $\bm{l} \in L_F$ and $R\in G$, and the multiplication is represented as
\begin{equation}
	(\bm{l}_2,R_2)(\bm{l}_1,R_1)=(\bm{l}_2+R_2\bm{l}_1+\bm{\omega}_F(R_2,R_1), ~R_2R_1).
\end{equation}
$\gamma_F$ acts on momentum space $\mathbb{R}^d_F$ as
\begin{equation}\label{eq:Trans-gamma}
	\gamma_F\bm{k}=R\bm{k}+\bm{\kappa}_R+\bm{l}.
\end{equation}

For a Hamiltonian $\mathcal{H}(\bm{k})$ in momentum space, its crystallographic symmetries can  represented as
\begin{equation}\label{eq:Symmetry}
	U_{\gamma_F}(\bm{k})\mathcal{H}(\bm{k})U_{\gamma_F}^\dagger(\bm{k})=\mathcal{H}(\gamma_F\bm{k}),
\end{equation} 
where $U_{\gamma_F}(\bm{k})$ is the $\bm{k}$-dependent unitary operator for $\gamma_F$. Usually, $\mathcal{H}(\bm{k})$ and $U_{\gamma_F}(\bm{k})$ are assumed to be invariant under $L_F$, i.e., as functions over the Brillouin torus $T^d_F\cong \mathbb{R}^d_F/L_F$. Here, we consider $\bm{k}$ in the whole momentum space $\mathbb{R}_F^d$ and do not need these periodic conditions.

\begin{figure}
	\includegraphics[width=0.9\textwidth]{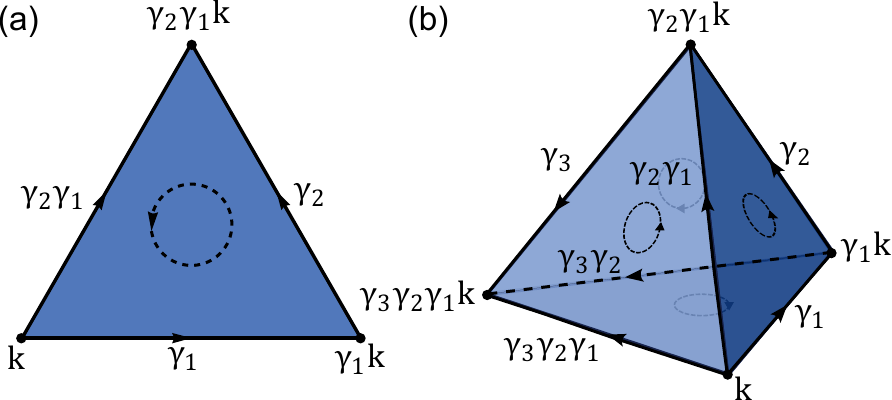}
	\caption{Illustration of the isomorphism Eq.~\eqref{eq:e-Isomorphism} for $n=1$ and $n=2$ in (a) and (b), respectively. $n+1$ ordered group elements form an $(n+1)$-simplex, whose boundary consists of $n+2$ oriented $n$-simplices. Compared with the previous simplexes of group cohomology~\cite{dijkgraaf1990topological,SPT_Wen}, here the vertices are the orbit of $\bm{k}$ under the consecutive action of the $n+1$ group elements.} \label{fig1} 
\end{figure}

\section{Algebraic topological invariants}
For an insulator $\mathcal{H}(\bm{k})$, we can always choose an orthonormal basis $|\psi_\alpha(\bm{k})\rangle$ for the valence bands, which is only required to be continuous in $\mathbb{R}^d_F$, rather than in the Brillouin torus $T_F^d$.
Then, for each $\gamma\in \Gamma_F$, Eq.~\eqref{eq:Symmetry} leads to the transformation,
\begin{equation}
	U_\gamma(\bm{k})|\psi_\alpha(\bm{k})\rangle=\sum_\beta |\psi_{\beta}(\gamma\bm{k})\rangle \mathcal{U}_{\beta\alpha}(\gamma\bm{k})
\end{equation}
with $\mathcal{U}_{\beta\alpha}(\bm{k})$ an unitary matrix for each $\bm{k}$. To simplify the notation, we omit the subscript `$F$' hereafter. We now introduce a phase associated to each $\gamma\in\Gamma_F$ as
\begin{equation}\label{eq:multi_bands}
	e^{2\pi i \phi_\gamma(\gamma\bm{k})}=\mathrm{det}~ \mathcal{U}(\gamma\bm{k})
\end{equation}
where `$\mathrm{det}$' stands for the matrix determinant. Clearly, $\phi_\gamma(\bm{k})$ is continuous in $\mathbb{R}^d_F$. Then, considering two consecutive transformations leads to the $1$-cocycle equation,
\begin{equation}\label{eq:1-cocyle}
	e^{2\pi i  \phi_{\gamma_2}(\gamma_2\gamma_1\bm{k})}e^{2\pi i \phi_{\gamma_1}(\gamma_1\bm{k})}=e^{2\pi i \phi_{\gamma_{2}\gamma_1}(\gamma_2\gamma_1\bm{k})}.
\end{equation}
It is noteworthy that $e^{2\pi i  \phi_{\gamma}(\bm{k})}=e^{2\pi i(\alpha(\gamma^{-1}\bm{k})-\alpha(\bm{k}))}$ for some $e^{2\pi i \alpha(\bm{k})}$ trivially satisfies the above equation. All 1-cocycles, namely solutions of Eq.~\eqref{eq:1-cocyle}, modulo these trivial ones, form the cohomology group $H^1(\Gamma_F,\mathcal{F}(\mathbb{R}^d_F,U(1)))$.
Here, $\mathcal{F}(\mathbb{R}_F^d,U(1))$ denotes the Abelian group of all $U(1)$-valued continuous functions  in momentum space $\mathbb{R}^d_F$, which hosts the natural $\Gamma_F$-action induced from the $\Gamma_F$-action on $\mathbb{R}^d_F$.

Equation~\eqref{eq:1-cocyle} enables us to introduce the integer-valued function,
\begin{equation}\label{eq:integer}
	 N(\gamma_2,\gamma_1)=\phi_{\gamma_1}(\gamma_1\bm{k})-\phi_{\gamma_{2}\gamma_1}(\gamma_2\gamma_1\bm{k})+\phi_{\gamma_2}(\gamma_2\gamma_1\bm{k}),
\end{equation}
as illustrated in Fig.~\ref{fig1}(a).
Here,  $N$ is independent of $\bm{k}$ because of the continuity of $\phi_\gamma$ and the discreteness of integers. Considering three transformations, the associativity of the phase factors lead to the so-called $2$-cocycle equation,
\begin{equation}\label{eq:Z-cocycle}
	N(\gamma_2,\gamma_1)+N(\gamma_3,\gamma_2\gamma_1)=N(\gamma_3,\gamma_2)+N(\gamma_3\gamma_2,\gamma_1).
\end{equation}
with the geometric origin illustrated in Fig.~\ref{fig1}(b).
We can modify $\phi_\gamma(\bm{k})$ to $\phi'_\gamma(\bm{k})=\phi_\gamma(\bm{k})+n(\gamma)$ with $n(\gamma)$ an arbitrary integer. Then, a $2$-cocycle $N$ is accordingly modified to be
\begin{equation}\label{eq:coboundary}
	N'(\gamma_2,\gamma_1)=N(\gamma_2,\gamma_1)+n(\gamma_1)+n(\gamma_2)-n(\gamma_2\gamma_1).
\end{equation}
Thus, $N$ and $N'$ are regarded as equivalent. The equivalence classes of integer-valued $2$-cocycles form the cohomology group $H^2(\Gamma_F,\mathbb{Z})$. 

The above construction of elements in $H^2(\Gamma_F,\mathbb{Z})$ is just the isomorphism Eq.~\eqref{eq:e-Isomorphism} with $n=1$. Thus, we have algebraically formulated a complete set of topological invariants for Abelian crystalline topological insulators. A proof for Eq.~\eqref{eq:e-Isomorphism} is provided in Appendix A. 
 
In practice, we can construct a complete set of cohomological invariants of $H^1(\Gamma_F,\mathcal{F}(\mathbb{R}^d_F,U(1)))$ as a complete set of topological invariants for the classification Eq.~\eqref{eq:cls}. These invariants circumvent a common difficulty for numerical computing: to smooth the valence-band wave functions under certain periodic boundary conditions.   For instance, Kane and Mele's topological invariant requires continuous wavefunctions satisfying the periodic boundary conditions~\cite{kane2005z}. 
Our algebraic formulation is advantageous for numerical computation, since it does not require non-local boundary conditions and only needs a smooth basis over a finite region covering $\mathbb{R}^d_F/\Gamma_F$.

\section{The classifications}
We now elucidate the claimed classifications of Abelian crystalline insulators and point-group action twistings. The starting point is to consider the direct-product space,
\begin{equation}
	EG\times T^d_F.
\end{equation}
Here, $EG$ is any contractible space on which $G$ freely acts~\cite{atiyah1984moment}, and therefore the diagonal $G$-action on $EG\times T^d_F$ is also free. This motivates the orbital space,
\begin{equation}
	EG\times_G T^d_F=(EG\times T^d_F)/G,
\end{equation}
which is called the Borel construction and can be used to formulate the Borel cohomology
\begin{equation}\label{eq:Borel}
	\mathcal{H}^n_G(T^d_F,\mathbb{Z})=\mathcal{H}^n(EG\times_G T^d_F,\mathbb{Z})
\end{equation}
for all $n\ge 0$. Here, `$\mathcal{H}$' stands for topological cohomology.
As shown in Ref.~\cite{atiyah2004twisted}, the classification of equivariant line bundles over $T^d_F$, namely Abelian crystalline topological insulators, is given by 	$\mathcal{H}^2_G(T^d_F,\mathbb{Z})$, and the classification of twistings of the $G$-action over $T^d_F$ is given by $\mathcal{H}^3_G(T^d_F,\mathbb{Z})$.

Significantly, $EG\times_G T^d_F$ can be interpreted as the classifying space of the corresponding MCG $\Gamma_F$ in Eq.~\eqref{eq:Gamma_F}, which we now explain. Consider 
\begin{equation}
	EG\times \mathbb{R}_F^d,
\end{equation}
with $\Gamma_F$ acting on $EG$ through $G$ and on $\mathbb{R}^d_F$ by definition.  Clearly, the diagonal $\Gamma_F$-action is free, and $EG\times \mathbb{R}_F^d$ is contractible. Then, the classifying space $B\Gamma_F$ of $\Gamma_F$ is just the orbital space
\begin{equation}
	B\Gamma_F=(EG\times \mathbb{R}^d_F)/\Gamma_F.
\end{equation}
Factoring out the action of the reciprocal lattice translations $L_F$, we obtain that
\begin{equation}\label{eq:Clf_space}
	B\Gamma_F=EG\times_{G} T_F^d.
\end{equation}
As $\Gamma_F$ is discrete, the topological Borel cohomology Eq.~\eqref{eq:Borel} can be identified with the group cohomology as
\begin{equation}
	\mathcal{H}^n(B\Gamma_F,\mathbb{Z})=H^n(\Gamma_F,\mathbb{Z}).
\end{equation}
Thus, the classifications for Abelian topological insulators and group-action twistings can be readily computed by the available cohomology functions for crystallographic groups in GAP. The results for $2$D and $3$D MCGs have been tabulated in the Supplemental Materials (SM)~\cite{SM}. 

Abelian topological phases are naturally endowed with an Abelian group structure by the direct sum of the valence bands, which corresponds to tensoring the determinant crystalline line bundles. The Abelian group structure is just that of the cohomology groups Eq.~\eqref{eq:cls}.

The torsion component $\mathrm{Tor}\mathrm{Cls}_{(G,\rho_F)}$ is canonically given by $H^1(G,\mathcal{F}(T^d_F,U(1)))$, and the free Abelian quotient group 
\begin{equation}\label{eq:Chern}
\mathrm{Ch}_{(G,\rho_F)}=H^2(\Gamma_F,\mathbb{Z})/\mathrm{Tor}H^2(\Gamma_F,\mathbb{Z})
\end{equation}
corresponds to the Chern numbers. For more detail, see Appendix B.

\section{Classification and representation of twistings} The twistings of point group actions constitute fundamental data in band theory~\cite{freed2013twisted}. All twistings of $G$-action $\rho_F$ over the Brillouin torus are classified by the Borel cohomology group $\mathcal{H}^3_G(T_F^d,\mathbb{Z})=\mathcal{H}^3(EG\times_G T_F^d, \mathbb{Z})$~\cite{atiyah2004twisted}. The Borel cohomology groups were analyzed by sophisticated methods using the spectral sequence, which only solved $13$ symmorphic wallpaper groups in two dimensions~\cite{gomi2017twists}. The concept of MCGs enables us to immediately compute all twistings for all point-group actions via $\mathcal{H}^3_G(T_F^d,\mathbb{Z})\cong H^3(\Gamma_F,\mathbb{Z})$, with results tabulated in the SM.

Moreover, the isomorphism $H^2(\Gamma_F,\mathcal{F}(\mathbb{R}_F^d,U(1)))\cong H^3(\Gamma_F,\mathbb{Z})$ [Eq.~\eqref{eq:e-Isomorphism}] naturally gives rise to a representation of twistings $\Omega$,
\begin{equation}\label{eq:twisting}
	U_{\gamma_2}(\gamma_1\bm{k})U_{\gamma_1}(\bm{k})=\Omega(\gamma_2,\gamma_1|\gamma_2\gamma_1\bm{k})U_{\gamma_2\gamma_1}(\bm{k}).
\end{equation}
Here, $U_{\gamma}(\bm{k})$ are unitary operators for $(\gamma,\bm{k})$, and their associativity leads to the 2-cocycle equation. The construction of an element in $H^3(\Gamma_F,\mathbb{Z})$ from $\Omega$ can be inferred from Fig.~\ref{fig1}(b), analogous to the lower dimensional construction in Eq.~\eqref{eq:integer}. 

In general, the classification of crystalline topological phases corresponds to the twisted equivariant K-group $K^\Omega_G(T_F^d)$, where  $\Omega$ is the twisting of the $G$-action $\rho_F$~\cite{freed2013twisted}. The approximation of $K^\Omega_G(T_F^d)$ by $H^2(\Gamma_F, \mathbb{Z})$ is discussed in Appendix C.

\section{Examples} 
\begin{figure}
    \centering
    \includegraphics[width=0.9\textwidth]{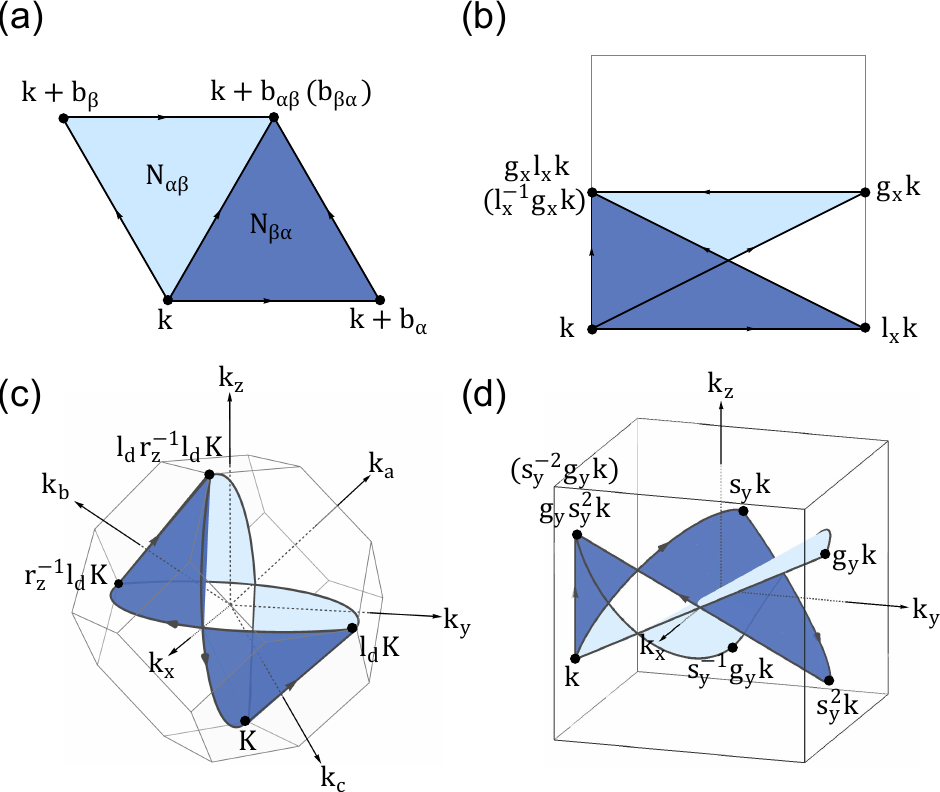}
    \caption{Illustration of $\bm{k}$-orbits associated with the two sides of algebraic relations used to formulate topological invariants. (a), (b) and (d) depict orbits for generic $\bm{k}$, while (c) highlights the orbit for  the high-symmetry momentum $\bm{K}$.The surfaces spanned by the orbits of each side are shaded in dark and light blue, respectively.}
    \label{fig}
\end{figure}
\subsection{$P1$}
This $3$D MCG is just the reciprocal lattice $L_F$. Let $\bm{b}_\alpha$ be three primitive reciprocal lattice vectors with $\alpha=1,2,3$. For a pair $\bm{b}_\alpha$ and $\bm{b}_\beta$, we can first translate $\bm{k}$ by $\bm{b}_\alpha$ and then by $\bm{b}_\beta$ or directly implement the translation by $\bm{b}_{\beta\alpha}=\bm{b}_{\beta}+\bm{b}_\alpha$ [see Fig.~\ref{fig}(a)], which leads to the identity,
\begin{equation}
	e^{2\pi i \phi_{\beta}(\bm{k}+\bm{b}_{\beta\alpha})}e^{2\pi i \phi_{\alpha}(\bm{k}+\bm{b}_\alpha)}=e^{2\pi i \phi_{{\beta\alpha}}(\bm{k}+\bm{b}_{\beta\alpha})}.
\end{equation}
The phase difference of the two sides leads to the integer,
\begin{equation}
	N_{\beta\alpha}=\phi_{\beta}(\bm{k}+\bm{b}_{\beta\alpha})-\phi_{{\beta\alpha}}(\bm{k}+\bm{b}_{\beta\alpha})+\phi_{\alpha}(\bm{k}+\bm{b}_\alpha).
\end{equation}
The two translations can also be implemented in the other order to obtain $N_{\alpha\beta}$ [see Fig.~\ref{fig}(a)].  Therefore, we introduce
\begin{equation}
	C_{\beta\alpha}=N_{\beta\alpha}-N_{\alpha\beta}.
\end{equation}
$C_{\beta\alpha}$ are cohomological invariants, i.e., they are invariant under the transformations Eq.~\eqref{eq:coboundary} and therefore depend only on the cohomology classes in $H^2(L_F,\mathbb{Z})\cong \mathbb{Z}^3$. In fact, they are complete to characterize $H^2(L_F,\mathbb{Z})\cong \mathbb{Z}^3$. In the SM, we show that $C_{\beta\alpha}$ is just the Chern number over the parallelogram spanned by $\bm{b}_\alpha$ and $\bm{b}_\beta$, consistent with the well-known result that Abelian topological phases over $T_F^3$ are completely characterized by the Chern vector $\epsilon^{\alpha\beta\gamma}C_{\beta\gamma}/2$~\cite{SM}.

\subsection{Berbiebach groups} 
The Berbiebach groups $B^\alpha$ can reduce the momentum space to the platycosms $\mathcal{M}^\alpha$ due to their free actions, and for each $\mathcal{M}^\alpha$ the reduced K group $\widetilde{K}(\mathcal{M}^\alpha)$ is isomorphic to $H^2(B^\alpha,\mathbb{Z})$~\cite{Platycosms}. Now, according our theory, $H^2(B^\alpha,\mathbb{Z})\cong H^1(B^\alpha,\mathcal{F}(\mathbb{R_F^d},U(1)))$ leads to a complete set of algebraic topological invariants for the topological classification $\widetilde{K}(\mathcal{M}^\alpha)$.

Here, we demonstrate this by the $2$D Berbiebach group $Pg$, generated by two elements, namely the primitive reciprocal lattice translation $l_x$ along the $x$ direction and the glide reflection $g_x$ inverting $x$, with the actions given by
$
	l_x\bm{k}=(k_x+2\pi,k_y),~g_x\bm{k}=(-k_x,k_y+\pi)
$.
The orbits of $\bm{k}$ under the actions of the two sides are illustrated in Fig.~\ref{fig}(b).
The algebraic $\mathbb{Z}_2$ topological invariant for $H^2(Pg,\mathbb{Z})\cong \mathbb{Z}_2$ can be formulated by considering the relation of generators,
\begin{equation}
g_xl_x=l_x^{-1}g_x.
\end{equation}
The difference between accumulated phases on the two sides,
\begin{equation}\label{eq:Z-valued}
\phi_{g_x}(g_xl_x\bm{k})+\phi_{l_x}(l_x\bm{k})-\phi_{l_x^{-1}}(l_x^{-1}g_x\bm{k})-\phi_{g_x}(g_x\bm{k}),
\end{equation}
is an integer.
Additionally, $l_x^{-1}l_x=1$ implies
$\phi_{l_x^{-1}}(l_x^{-1}\bm{k})+\phi_{l_x}(\bm{k})$ is also an integer. Thus, we can formulate a $\mathbb{Z}_2$ invariant
\begin{equation}
	\nu=\phi_{g_x}(g_xl_x\bm{k})+\phi_{l_x}(l_x\bm{k})+\phi_{l_x}(l_xg_x\bm{k})-\phi_{g_x}(g_x\bm{k})\mod 2.
\end{equation}
Under a coboundary transformation discussed below Eq.~\eqref{eq:Z-cocycle}, $\nu$ changes by an even integer, and therefore $\nu \mod 2$ is manifestly a cohomological invariant for $H^2(Pg,\mathbb{Z})$. In the SM~\cite{SM}, the algebraic topological invariant is shown to be equal to the topological invariant in Ref.~\cite{chen2022brillouin}.

\subsection{$I23$} 
Let us present an algebraic topological invariant for $3$D MCG $I23$. The symmorphic group does not require projective representations and has been previously studied by the Atiyah-Hirzbruch spectral sequence~\cite{shiozaki2022atiyah}. The $\mathbb{Z}_4$ algebraic topological invariant for $H^2(I23, \mathbb{Z}) \cong \mathbb{Z}_3 \times \mathbb{Z}_4$ is out of reach by the sophisticated method there.

The algebraic topological invariant is based on the relation,
\begin{equation}\label{eq:I23-algebraic}
	r_y l_d r_z l_d = l_d^{-1} r_z l_d^{-1} r_y.
\end{equation}
Here, $l_d$ is the diagonal translation, namely $l_d=l_al_bl_c$ with $l_{a,b,c}$ the primitive lattice translations, with the Brillouin zone illustrated in Fig.~\ref{fig}(c). $r_{y}$ ($r_{z}$) is the twofold rotation through the $y$ axis (the $z$ axis). Analogous to the formulation around Eq.~\eqref{eq:Z-valued}, we can translate Eq.~\eqref{eq:I23-algebraic} into a $\mathbb{Z}_4$ invariant, 
\begin{align}
		\nu  = & \phi_{r_y}(r_yl_dr_zl_d\bm{k})+\phi_{l_d}(l_dr_zl_d\bm{k})+\phi_{r_z}(r_zl_d\bm{k})+\phi_{l_d}(l_d\bm{k}) \nonumber \\
		&+\phi_{l_d}(r_zl_d^{-1}r_y\bm{k})-\phi_{r_z}(r_zl_d^{-1}r_y\bm{k}) + \phi_{l_d}(r_y\bm{k})-\phi_{r_y}(r_y\bm{k}) \mod 4,
\end{align}
manifestly a $\mathbb{Z}_4$ cohomology invariant for $H^2(I23,\mathbb{Z})$.

It is noteworthy that the $\mathbb{Z}_2$ subgroup of $\mathbb{Z}_4$ corresponds to the two representations of $(R_y l_d R_z l_d)^2 = 1$ at the high-symmetry momentum $\bm{K} = -\bm{b}_a/2 - 3\bm{b}_b/4-\bm{b}_c/4$. The quotient $\mathbb{Z}_4/\mathbb{Z}_2\cong\mathbb{Z}_2$ is really topological.

In the SM~\cite{SM}, we present topological invariants for $3$D MCGs $P2_1/c$ and $I222$,  both of which contain $\mathbb{Z}_4$ invariants refining the previous $\mathbb{Z}_2$ invariant ~\cite{shiozaki2016topology,shiozaki2022atiyah}. Analogously, the $\mathbb{Z}_4$ invariant for $P2_1/c$  can be constructed from the algebraic relation $g_y s_y^2 = s_y^{-2} g_y$, as illustrated in Fig.~\ref{fig}(d).

\section{Summary and discussions}
In summary, we introduce the concept of MCGs by showing that their cohomology groups provide an exhaustive classification of all Abelian topological phases and all twistings of point-group actions. Furthermore, by establishing an isomorphism to the cohomology groups with functional coefficients, we obtain algebraic formulas for a complete set of topological invariants and representations of all twistings. While $K^\Omega_G(T^d_F)$ assumes the many-band limit, the concept of MCGs leads us to consider $H^{1,\Omega}(\Gamma_F,\mathcal{F}(\mathbb{R}^d_F,U(N)))$ as the classification of $N$-band topological crystalline insulators, where the twisting $\Omega$ is given by Eq.~\eqref{eq:twisting} and $U(N)$ denotes the set of all $N\times N$ unitary matrices.

\bigskip
\noindent \textbf{Acknowledgements}

This work is supported by the GRF of Hong Kong (Nos. 17301224 and 17302525).\\

\section*{Appendix A: The isomorphism}\label{isomorphism}
In this section, we provide a proof to the isomorphism Eq.~\eqref{eq:e-Isomorphism}.
Let us consider the Abelian groups $\mathcal{F}(\mathbb{R}^d_F,A)$ of $A$-valued continuous functions over the momentum space, with $A=\mathbb{Z}$, $\mathbb{R}$ and $U(1)=\mathbb{R}/\mathbb{Z}$. It is clear that $\mathcal{F}(\mathbb{R}^d_F,\mathbb{Z})$ consists of constant functions and therefore $\mathcal{F}(\mathbb{R}^d_F,\mathbb{Z})\cong \mathbb{Z}$.  Since the momentum space $\mathbb{R}^d_F$ is contractible, every $U(1)$-valued function $u(\bm{k})$ can be lifted to an $\mathbb{R}$-valued function $\phi(\bm{k})$ with $u(\bm{k})= e^{2\pi i \phi(\bm{k})}$. Moreover, if $e^{2\pi i \phi(\bm{k})}=1$, the continuous function $\phi(\bm{k})$ is an integral constant. Thus, the three Abelian groups form the short exact sequence,
\begin{equation}
	0\rightarrow \mathbb{Z}\rightarrow \mathcal{F}(\mathbb{R}^d_F,\mathbb{R})\rightarrow \mathcal{F}(\mathbb{R}^d_F,U(1))\rightarrow 1.
\end{equation}
While $\Gamma_F$ trivially acts on $\mathbb{Z}$, its actions on $ \mathcal{F}(\mathbb{R}^d_F,\mathbb{R})$ and $ \mathcal{F}(\mathbb{R}^d_F,U(1))$ are naturally induced from the $\Gamma_F$-action on $\mathbb{R}_F^d$.

Then, the short exact sequence leads to the long exact sequence of cohomology groups, and a segment is given by
\begin{equation}
	H^{n}(\Gamma_F, \mathcal{F}(\mathbb{R}^d_F,\mathbb{R}))\rightarrow H^{n}(\Gamma_F,\mathcal{F}(\mathbb{R}^d_F,U(1)))\rightarrow H^{n+1}(\Gamma_F,\mathbb{Z})\rightarrow H^{n+1}(\Gamma_F,\mathcal{F}(\mathbb{R}^d_F,\mathbb{R}))
\end{equation}
for any integer $n\ge 1$ \cite{Brown1982cohomology,weibel1994introduction}.
Thus, to show the isomorphism Eq.~\eqref{eq:e-Isomorphism}, we only need to show

\begin{equation} \label{eq:vanishing_R}
	H^{n}(\Gamma_F, \mathcal{F}(\mathbb{R}^d_F,\mathbb{R}))=0,
\end{equation}
for $n\ge 1$.

This can be done by considering the Lyndon-Hochschild-Serre spectral sequence \cite{weibel1994introduction}
\begin{equation}
	E_{2}^{p,q}=H^{p}(G, H^{q}(L_F, \mathcal{F}(\mathbb{R}_F^d,\mathbb{R})))\Rightarrow H^{p+q}(\Gamma_F, \mathcal{F}(\mathbb{R}_F^d,\mathbb{R})),
\end{equation}
of the canonical short exact sequence of the momentum-space crystallographic group,
\begin{equation}
	0\rightarrow L_F\rightarrow \Gamma_F\rightarrow G \rightarrow 1.
\end{equation}

The cohomology groups of $L_F$ as the coefficients of the cohomology groups of $G$ are given by
\begin{equation}\label{eq:vanishing_L}
	H^n(L_F,\mathcal{F}(\mathbb{R}_F^d,\mathbb{R}))=0
\end{equation}
for $n\ge 1$. Let us postpone the proof of this result later. The zeroth cohomology group is just the Abelian group of fixed points, and therefore it consists of periodic functions invariant under $L_F$, i.e.,
\begin{equation}
	H^0(L_F,\mathcal{F}(\mathbb{R}_F^d,\mathbb{R}))\cong \mathcal{F}(T_F^d,\mathbb{R}).
\end{equation}
Thus, all sites except the zeroth row on the second page of the spectral sequence host zero. In the zeroth row, the only nonzero element is at the origin with
\begin{equation}
	E_{2}^{0,0}=H^{0}(G, \mathcal{F}(T_F^d,\mathbb{R}))=\mathcal{F}^G(T_F^d,\mathbb{R}),
\end{equation}
while 
\begin{equation}
	E^{p,0}_2= H^{p}(G, \mathcal{F}(T_F^d,\mathbb{R}))=0.
\end{equation}
Above we used the fact that $|G|\cdot [\varphi]=0$ for any $[\varphi]\in H^{p}(G, \mathcal{F}(T_F^d,\mathbb{R}))$. But, clearly this implies $\varphi=0$, as $\varphi$ is valued in $\mathbb{R}$.

The only non-zero site at the second page of the spectral sequence is the $E^{0,0}_{2}=\mathcal{F}^G(T_F^d,\mathbb{R})$, and therefore the stabilization has been reached at the second page. From the zero filtration of $H^{n}(\Gamma_F, \mathcal{F}(\mathbb{R}^d_F,\mathbb{R}))$ for $n\ge 1$, we conclude that Eq.~\eqref{eq:vanishing_R} has been proved.

Now the remaining task is to show Eq.~\eqref{eq:vanishing_L}. It is significant to observe that $L_F$ gives a natural cellular structure for $\mathbb{R}_F^d$. The $d$D cells are just the fundamental domains under $L_F$, and then we can iterate the boundary operation to obtain all lower dimensional cells. We denote $a$D cells as $D_i^{(a)}$ with $i$ labeling all the $a$D cells and $a=d,\cdots,0$. Let the $a$D skeleton $W^{(a)}$ be the union of all cells with dimension equal or lower than $a$.
Then, for each $a>1$, we can form
\begin{equation}\label{eq:Filteration}
	0\rightarrow \mathcal{F}(L_F, \mathcal{F}_0(D_0^{(a)},\mathbb{R})^{\times C^d_a})\rightarrow \mathcal{F}(W^{(a)},\mathbb{R})\rightarrow \mathcal{F}(W^{(a-1)},\mathbb{R})\rightarrow 0.
\end{equation}
It is clear that for each $\mathbb{R}$-valued function on the $(a-1)$D skeleton $W^{(a-1)}$, we can always continuously extend it to be a function over the $a$D skeleton $W^{(a)}$. Inversely, every function over $W^{(a)}$ can be restricted to $W^{(a-1)}$ and the kernel consists of all functions that vanish on $W^{(a-1)}$. Then, $W^{a}-W^{(a-1)}$ is the disconnected union of all internal regions  $\overcirc{D}_i^{(a)}$ of $a$D cells, on which $L_F$ acts freely with $C^d_a$ orbits. Thus, the kernel can be represented as $\mathcal{F}(L_F, \mathcal{F}_0(D_0^{(a)},\mathbb{R})^{\times C^d_a})$, where $\mathcal{F}_0(D_0^{(a)},\mathbb{R})$ denote all functions over a given $a$D cell $D_0^{(a)}$ that vanish on the boundary of the $D_0^{(a)}$. 

Applying Shapiro's Lemma for the trivial subgroup of $L_F$, we know
$
H^n(L_F,\mathcal{F}(L_F, A))=0
$
for $n\ge 1$ for any Abelian group $A$ with trivial $L_F$-action \cite{weibel1994introduction}. Note that $L_F$ now trivially acts on $ \mathcal{F}_0(D_0^{(a)},\mathbb{R})^{\times C^d_a}$, and the action of $L_F$ on the momentum space is now embodied as the natural action of $L_F$ on $L_F$ , which induces the action of $L_F$ on $\mathcal{F}(L_F, \mathcal{F}_0(D_0^{(a)},\mathbb{R})^{\times C^d_a})$. Thus, Shapiro's Lemma implies,
\begin{equation}
	H^n(L_F,\mathcal{F}(L_F, \mathcal{F}_0(D_0^{(a)},\mathbb{R})^{\times C^d_a}))=0
\end{equation}
for $n\ge 1$. Then, from the long exact sequence associated to Eq.~\eqref{eq:Filteration}, we obtain 
\begin{equation}
	H^n(L_F, \mathcal{F}(W^{(a)},\mathbb{R}))\cong H^n(L_F, \mathcal{F}(W^{(a-1)},\mathbb{R})).
\end{equation}
for all $a=1,\cdots,d$. Connecting all such isomorphisms, we see the isomorphism,
\begin{equation}
	H^n(L_F, \mathcal{F}(\mathbb{R}_F^d,\mathbb{R}))\cong H^n(L_F, \mathcal{F}(L_F,\mathbb{R})),
\end{equation}
between the two ends, for $n\ge 1$. Then, we can apply Shapiro's Lemma again for $A=\mathbb{R}$ with $H^n(L_F, \mathcal{F}(L_F,\mathbb{R}))= 0$ for $n\ge 1$ \cite{weibel1994introduction}. Thus, Eq.~\eqref{eq:vanishing_L} has been proved. This completes our proof for the isomorphism Eq.~\eqref{eq:e-Isomorphism}.

\section*{Appendix B: The torsion subgroup of $\mathrm{Cls}(G,\rho_F)$}\label{torsion}
In this section, we specify the torsion subgroup of $\mathrm{Cls}(G,\rho_F)\cong H^1(\Gamma_F, \mathcal{F}(\mathbb{R}^d_F, U(1)))$.  

Let us consider the canonical short exact sequence of a momentum-space crystallographic group $\Gamma_F$,
\begin{equation}
	0\rightarrow L_F\rightarrow \Gamma_F\rightarrow G\rightarrow 1,
\end{equation}
which leads to the exact sequence \cite{Brown1982cohomology,weibel1994introduction},
\begin{equation}
	0\rightarrow H^1(G,\mathcal{F}(T^d_F,U(1)))\rightarrow H^1(\Gamma_F, \mathcal{F}(\mathbb{R}^d_F, U(1)))\rightarrow [H^1(L_F, \mathcal{F}(\mathbb{R}^d_F, U(1)))]^G.
\end{equation}
As proved earlier, the right end has the isomorphism,
\begin{equation}
	H^1(L_F, \mathcal{F}(\mathbb{R}^d_F, U(1)))\cong H^2(L_F,\mathbb{Z}),
\end{equation}
and therefore is a free Abelian group. Together with the fact that $H^1(G,\mathcal{F}(T^d_F,U(1)))$ is a torsion group, we deduce that the torsion subgroup of the classification group has the isomorphism,
\begin{equation}
	\mathrm{Tor} H^1(\Gamma_F, \mathcal{F}(\mathbb{R}^d_F, U(1))) \cong H^1(G,\mathcal{F}(T^d_F,U(1))).
\end{equation}
The physical meaning of $H^1(G,\mathcal{F}(T^d_F,U(1)))$ has been addressed in the main text. 

From the universal coefficient theorem \cite{hatcher2001algebraic}, the torsion subgroup can be identified as
\begin{equation}
	H^1(G,\mathcal{F}(T^d_F,U(1)))\cong \mathrm{Tor} H_1(\Gamma_F,\mathbb{Z}),
\end{equation}
which can be readily produced by GAP \cite{GAP4}.

In the case of a crystalline band trivial as a line bundle, we can choose $|\psi(\bm{k})\rangle$ globally continuous over the Brillouin torus, and  consider all $1$D representations of $G$ over the Brillouin torus. $H^1(G,\mathcal{F}(T_F^d,U(1)))$ is just the Abelian group formed by these single-band representations under tensor product. It is noteworthy that $H^1(G,\mathcal{F}(T_F^d,U(1)))$ contains essential topological information, and cannot be regarded as a mere collection of compatible representations of little co-groups. Especially, even when we fix the representations of $G_{\bm{k}}$ for all  $\bm{k}$, there are still multiple topological configurations. This is clear from our examples.

\section*{Appendix C: Approximation of $K^\Omega_G$-groups}\label{approximation}
In general, we need to consider the twisting $\Omega$ of the $G$-action on the Brillouin torus $T_F^d$ as represented in Eq.~\eqref{eq:twisting}.

Nontrivial twistings lead to band crossings. It is well known that the glide reflection leads to a crossing of two bands and an $N$-fold screw rotation leads to crossings of $N$ bands, due to the nontrivial twistings from real-space fractional translations. 
However, we note that the determinant bundle $\mathrm{Det}E$ of a $\Omega$-twisted $G$-equivariant bundle has a trivial twisting. For instance, for an irreducible representation of the glide reflection,  $\mathcal{G}_x(k_y)^2=e^{ik_y}1_{2\times 2}$, and therefore $(\mathrm{det}\mathcal{G}_x(k_y))^2=e^{2ik_y}$. Then, we can redefine $\mathcal{M}_x(k_y)=e^{-ik_y}\mathrm{det}\mathcal{G}_x(k_y)$ so that $(\mathcal{M}_x(k_y))^2=1$, with the continuity in $T_F^2$ preserved. Thus, for a real-space nonsymmorphic group, a determinant line bundle can be regarded as a single-band representation of the symmorphic group in the same arithmetic crystal class. 

In twisted equivariant K theory,  for a given $\Omega$, the direct sum $E_{1}\oplus E_{2}$ of an $N$-band configuration $E_N$ and an $M$-band configuration $E_M$ is compared with an $(N+M)$-band configuration $E_{3}$ by symmetry-preserving continuous deformations. It is clear that $\mathrm{Det}(E_1\oplus E_2)=\mathrm{Det}E_1\otimes \mathrm{Det}E_2$, since the determinant of the direct sum of two matrices is equal to the product of the determinants of the two matrices.
Thus, there exists a natural group homomorphism from the  twisted equivariant K-group to the group of crystalline single bands, 
\begin{equation}\label{eq:det-homo}
	\mathrm{Det}:~K_{G}^{\Omega}(T_F^d)\rightarrow H^2(\Gamma_F,\mathbb{Z}).
\end{equation}
This is consistent with the definition of the phase factor for each $\gamma\in\Gamma_F$ in Eq.~\eqref{eq:multi_bands}.

While the twisted equivariant K-groups are difficult to compute, the classification of Abelian topological phases can be applied to distill the underlying Abelian topological configuration of any crystalline multi-band structures for all rational projective representations of all real-space crystallographic groups. 

For $d\le 3$, we propose that the topological invariants of Abelian topological phases are sufficient to characterize most crystalline topological phases for the following reasons.  For a crystalline $N$-band structure, its topological class is encoded in the $U(N)$-valued transition functions, but for $n=1,2$ the only nontrivial homotopy group of $U(N)$ is $\pi_1(U(N))$ due to the $U(1)$ center, while $\pi_1(SU(N))=\pi_2(SU(N))=0$.

\onecolumngrid
\renewcommand{\theequation}{S\arabic{equation}}
\setcounter{equation}{0}
\renewcommand{\thefigure}{S\arabic{figure}}
\setcounter{figure}{0}
\setcounter{section}{0} 
\renewcommand{\thesection}{\Roman{section}}
\newpage
\section*{Supplementary Materials}
\section{Basics of cohomology groups}
\subsection{Group cohomology}
	Let us consider a discrete or even a finite group $G$. A $G$-module $A$ is an Abelian group endowed with a $G$-action $T$. The $G$-action $T$ satisfies the following properties. For any $g\in G$, $T_g$ is an automorphism of $A$, such that
	\begin{equation}
		T_g(ab)=T_g(a)T_g(b)
	\end{equation}  
	for any $a,b\in A$. 
	And for any $g_{2},g_1\in G$,
	\begin{equation}
		T_{g_2g_1}=T_{g_2}\circ T_{g_1}.
	\end{equation}
	
	Given such a $G$-module $A$, an $n$-cochain $f  $ is a function from $G^{n}$ to $A$,
	\begin{align}
		f_n : G^{n} \to  A.
	\end{align}
	The set of $n$-cochains $\mathcal{C}^n(G,A)$ forms an Abelian group.
	These cochains are assumed to be normalized, i.e., $f(g_n,\cdots,g_1)= 1$ if $g_i$ = 1 for some $i$.
	 For any two $n$-cochains $f$ and $f'$, their product is given by $(ff')(g_n,\cdots,g_1) = f(g_n,\cdots,g_1)  f'(g_n,\cdots,g_1)$.
	
	The coboundary  homomorphism $\delta: \mathcal{C}^n(G,A) \to \mathcal{C}^{n+1}(G,A)$ is defined as
	\begin{equation}
		\begin{split}
			\delta f (g_{n+1},\cdots,g_1) = T_{g_{n+1}}( f (g_n,\cdots,g_1))  f_n  (g_{n+1},\cdots,g_2)^{(-1)^{n+1}}  \\\prod_{i =1}^{n} f_n (g_{n+1},\cdots,g_{n-i+2} g_{n-i+1},\cdots,  g_1)^{(-1)^i }.
		\end{split}
	\end{equation}
	It is straightforward to check that $\delta^2 f=1$ for any cochain $f$. Here, $1$ is the constant map onto $1\in A$.
	
	An $n$-cochain $c$  is called an n-cocycle if  $\delta c= 1$. The set of cocycles 
	\begin{equation}
		 \mathcal{Z}^n(G,A) = \left\{c\in \mathcal{C}^n(G,A)| \delta c =1\right\}
	\end{equation}
	is a subgroup of $\mathcal{C}^n(G,A)$, which is just the kernel of $\delta$. The coboundary group $\mathcal{B}^n(G,A)$ is the  image of  $\delta$,
	\begin{equation}
		\mathcal{B}^n(G,A) = \{\delta f, f\in \mathcal{C}^{n-1}(G,A) \}.
	\end{equation}
	 Since $\delta^2 f = 1$ for any $f\in \mathcal{C}^n(G,A)$, we  have $\mathcal{B}^n(G,A) \subset  \mathcal{Z}^n(G,A)$, and therefore can form the cohomology group
	 \begin{equation}
	 	H^{n,T}(G,A) = \mathcal{Z}^n(G,A)/\mathcal{B}^n(G,A).
	 \end{equation}
	
In our manuscript, the symbol $G$ denotes a symmetry group, which may be the momentum-space crystallographic group $\Gamma_F$, the translation subgroup $L_F$, or a crystallographic point group. The symbol $A$ represents an Abelian group composed of continuous functions on momentum space (or the Brillouin torus). This structure established $A$ a $G$-module, as $G$ has a natural action on it.

\subsection{Borel cohomology}
Let $G$ be a discrete group. For such $G$, there exists a contractible space $EG$ on which $G$ acts freely. The classifying space $BG$ is defined as the quotient space
\begin{equation}
    BG := EG/G.
\end{equation}
The projection $EG \to BG$ forms a principal $G$-bundle. Any two classifying spaces for a given group $G$ are homotopy equivalent. Moreover, for a discrete group $G$, the classifying space $BG$ can be identified with the Eilenberg-MacLane space $K(G,1)$.

This construction corresponds to the case where $G$ acts on a single point. More generally, for an arbitrary $G$-space $X$, we consider the fibration
\begin{equation}
    X \rightarrow (EG \times X)/G \rightarrow BG.
\end{equation}
According to the Borel construction, the $G$-action on $EG \times X$ is given by  \cite{atiyah1984moment}
\begin{equation}
    g\cdot (e,x) := (e\cdot g^{-1}, g\cdot x),
\end{equation}
for any $g\in G$ and $(e,x)\in EG\times X$.
We denote the orbital space $(EG \times X)/G$ as  $X_G$. 

The Borel cohomology is then defined as the ordinary cohomology of this orbital space,
\begin{equation}
    \H^{*}_G(X,R) := \H^{*}(X_G, R),
\end{equation}
for any coefficient group $R$. 

In general, $X_G$ is a bundle with fiber $X$ over $BG$, as the $G$-action on $EG$ is free. However, if $G$ acts freely on $X$, then $X_G$ can also be regarded as an $EG$-bundle over $X/G$. Since $EG$ is contractible, the equivariant cohomology reduces to the ordinary cohomology $\H^{*}(X/G, R)$.

\section{Classification of equivariant line bundles and twistings }
As established in the main text and Appendix, the complete classification of Abelian crystalline topological insulators is given by the isomorphic cohomology groups,
\begin{equation}
    H^1(\Gamma_F,\mathcal{F}(\mathbb{R}^d_F, U(1))) \cong H^2(\Gamma_F,\mathbb{Z}).
\end{equation}

In twisted equivariant K-theory,  twistings are classified by the equivariant cohomology group $\H_G^3(T^d_F,\mathbb{Z})$ \cite{atiyah2004twisted,freed2013twisted,gomi2017twists}. As shown in the main text, we have the isomorphism,
\begin{equation}
    \H_G^3(T^d_F,\mathbb{Z}) \cong H^3(\Gamma_F, \mathbb{Z}).
\end{equation}

To compute these cohomology groups, we apply the Universal Coefficient Theorem to the homology groups $H_n(\Gamma_F, \mathbb{Z})$ for $n=1,2,3$ \cite{Brown1982cohomology,hatcher2001algebraic}, which can be obtained using the GAP software \cite{GAP4}.

Let $\beta_2$ denote the rank of $H_2(\Gamma_F, \mathbb{Z})$, namely the second Betti number.  The second cohomology group, which classifies topological invariants, decomposes as
\begin{equation}
    H^2(\Gamma_F, \mathbb{Z}) \cong \mathrm{Tor}\big(H_1(\Gamma_F, \mathbb{Z})\big) \oplus \mathbb{Z}^{\beta_2}.
\end{equation}

Similarly, letting $\beta_3$ be the rank of $H_3(\Gamma_F, \mathbb{Z})$, namely the third Betti number, the third cohomology group is given by
\begin{equation}
    H^3(\Gamma_F, \mathbb{Z}) \cong \mathrm{Tor}\big(H_2(\Gamma_F, \mathbb{Z})\big) \oplus \mathbb{Z}^{\beta_3}.
\end{equation}

Based on these formulas, we present the explicit classification results for all $17$ $2$D and $230$ $3$D MCGs in Tabs.~\ref{2D} and~\ref{3D-1}, respectively.

\subsection{Classification table for $2$D momentum-space wallpaper groups}
\begin{table}[h!]
     \caption{Homology and cohomology groups with $\mathbb{Z}$ coefficients for $17$ momentum-space wallpaper groups. For each arithmetic class, we list the corresponding groups $\Gamma_F$, their first three homology groups $H_n(\Gamma_F,\mathbb{Z})$ with $n=1,2,3$ (computed by GAP), and the second and third cohomology groups $H^n(\Gamma_F,\mathbb{Z})$ with $n=2,3$ obtained from the Universal Coefficient Theorem. The torsion subgroups $\mathrm{Tor}H^2(\Gamma_F,\mathbb{Z})$ of $H^2(\Gamma_F,\mathbb{Z})$ are highlighted as a separate column.}
    \label{2D}
    \vspace{8pt}
      \centering
    \scalebox{0.9}{
    \begin{tabular}{|c|c|c|c|c|c|c|c|c|}
        \hline
         Arithmetic class &No.    &  Symbol & $H_1(\Gamma_F,\mathbb{Z})$ & $H_2(\Gamma_F,\mathbb{Z})$ & $H_3(\Gamma_F, \mathbb{Z})$ & $H^2(\Gamma_F,\mathbb{Z})$ & $\mathrm{Tor}H^2(\Gamma_F,\mathbb{Z})$ & $H^3(\Gamma_F, \mathbb{Z})$\\
        \hline
         $1P$ &$1$   &  $P1$ & $\mathbb{Z}^2$ & $\mathbb{Z}$ & $1$  &  $\mathbb{Z}$ & $1$ &  $1$\\
        \hline
        $2P$ & $2$ & $P2$ & $\mathbb{Z}_2^3$ & $\mathbb{Z}$ & $\mathbb{Z}_2^4$ &$\mathbb{Z}_2^3\oplus \mathbb{Z}$ & $\mathbb{Z}_2^3$ & $1$\\
        \hline
        \multirow{2}{*}{$mP$} & $3$ & $Pm$ & $\mathbb{Z}_2^2\oplus\mathbb{Z}$ & $\mathbb{Z}_2^2$ & $\mathbb{Z}_2^2$ & $\mathbb{Z}_2^2$ & $\mathbb{Z}_2^2$ & $\mathbb{Z}_2^2$\\
        \cline{2-9}
        & $4$ & $Pg$ & $\mathbb{Z}_2\oplus\mathbb{Z}$ & $1$ & $1$ & $\mathbb{Z}_2$ & $\mathbb{Z}_2$ & $1$\\
        \hline
        $mC$ & $5$ & $Cm$ & $\mathbb{Z}_2\oplus\mathbb{Z}$ & $\mathbb{Z}_2$ & $\mathbb{Z}_2$ & $\mathbb{Z}_2$ & $\mathbb{Z}_2$ & $\mathbb{Z}_2$\\
        \hline
        \multirow{3}{*}{$mmP$} & $6$ & $Pmm$ & $\mathbb{Z}_2^4$ & $\mathbb{Z}_2^4$ & $\mathbb{Z}_2^8$ &$\mathbb{Z}_2^4$ & $\mathbb{Z}_2^4$ & $\mathbb{Z}_2^4$\\
        \cline{2-9}
        & $7$ & $Pmg$ & $\mathbb{Z}_2^3$ & $\mathbb{Z}_2$ & $\mathbb{Z}_2^3$ & $\mathbb{Z}_2^3$ & $\mathbb{Z}_2^3$ & $\mathbb{Z}_2$\\
        \cline{2-9}
        & $8$ & $Pgg$ & $\mathbb{Z}_2\oplus\mathbb{Z}_4$ & $1$ & $\mathbb{Z}_2^2$ & 
        $\mathbb{Z}_2\oplus\mathbb{Z}_4$ & $\mathbb{Z}_2\oplus\mathbb{Z}_4$ & $1$\\
        \hline
        $mmC$ & $9$ & $Cmm$ & $\mathbb{Z}_2^3$ & $\mathbb{Z}_2^2$ & $\mathbb{Z}_2^5$ & $\mathbb{Z}_2^3$ & $\mathbb{Z}_2^3$ & $\mathbb{Z}_2^2$\\
        \hline
        $4P$ & $10$ & $P4$ & $\mathbb{Z}_2\oplus\mathbb{Z}_4$ & $\mathbb{Z}$ & $\mathbb{Z}_2\oplus\mathbb{Z}_4^2$ & $\mathbb{Z}_2\oplus\mathbb{Z}_4\oplus\mathbb{Z}$ & $\mathbb{Z}_2\oplus\mathbb{Z}_4$ & $1$\\
        \hline
        \multirow{2}{*}{$4mP$} & $11$ & $P4m$ & $\mathbb{Z}_2^3$ & $\mathbb{Z}_2^3$ & $\mathbb{Z}_2^4\oplus\mathbb{Z}_4^2$ & $\mathbb{Z}_2^3$ & $\mathbb{Z}_2^3$ & $\mathbb{Z}_2^3$\\
        \cline{2-9}
        & $12$ & $P4g$ & $\mathbb{Z}_2\oplus\mathbb{Z}_4$ & $\mathbb{Z}_2$ & $\mathbb{Z}_2^2\oplus\mathbb{Z}_4$ & $\mathbb{Z}_2\oplus\mathbb{Z}_4$ & $\mathbb{Z}_2\oplus\mathbb{Z}_4$ & $\mathbb{Z}_2$\\
        \hline
        $3P$ & $13$ & $P3$ & $\mathbb{Z}_3^2$ & $\mathbb{Z}$ & $\mathbb{Z}_3^3$ & $\mathbb{Z}_3^2\oplus\mathbb{Z}$ & $\mathbb{Z}_3^2$ & $1$\\
        \hline
        $3m1P$ & $14$ & $P3m1$ & $\mathbb{Z}_2$ & $\mathbb{Z}_2$ & $\mathbb{Z}_3^2\oplus\mathbb{Z}_6$ & $\mathbb{Z}_2$ & $\mathbb{Z}_2$ & $\mathbb{Z}_2$\\
        \hline
        $31mP$ & $15$ & $P31m$ & $\mathbb{Z}_6$ & $\mathbb{Z}_2$ & $\mathbb{Z}_3\oplus\mathbb{Z}_6$ & $\mathbb{Z}_6$ & $\mathbb{Z}_6$ & $\mathbb{Z}_2$\\
        \hline
        $6P$ & $16$ & $P6$ & $\mathbb{Z}_6$ & $\mathbb{Z}$ & $\mathbb{Z}_6^2$ & $\mathbb{Z}_6\oplus\mathbb{Z}$ & $\mathbb{Z}_6$ & $1$\\
        \hline
        $6mP$ & $17$ & $P6m$ & $\mathbb{Z}_2^2$ & $\mathbb{Z}_2^2$ & $\mathbb{Z}_2^2\oplus\mathbb{Z}_6^2$ & $\mathbb{Z}_2^2$ & $\mathbb{Z}_2^2$ & $\mathbb{Z}_2^2$\\
        \hline
    \end{tabular}}
\end{table}

\newpage
\subsection{Classification table for $3$D momentum-space crystallographic groups}
\begin{table}[H]
    \caption{\small{Homology and cohomology groups with $\mathbb{Z}$ coefficients for $230$ momentum-space crystallographic groups. For each arithmetic class, we list the corresponding groups $\Gamma_F$, their first three homology groups $H_n(\Gamma_F,\mathbb{Z})$ (computed by GAP), and the second and third cohomology groups $H^n(\Gamma_F,\mathbb{Z})$ obtained from the Universal Coefficient Theorem. The torsion subgroups $\mathrm{Tor}H^2(\Gamma_F,\mathbb{Z})$ of $H^2(\Gamma_F,\mathbb{Z})$ are highlighted as a separate column.}}
    \vspace{8pt}
      \centering
      \scalebox{0.9}{
\begin{tabular}{|c|c|c|c|c|c|c|c|c|}
        \hline
         Arithmetic class &No.    &  Symbol & $H_1(\Gamma_F,\mathbb{Z})$ & $H_2(\Gamma_F,\mathbb{Z})$ & $H_3(\Gamma_F, \mathbb{Z})$ & $H^2(\Gamma_F,\mathbb{Z})$ & $\mathrm{Tor}H^2(\Gamma_F,\mathbb{Z})$ & $H^3(\Gamma_F, \mathbb{Z})$\\
        \hline
         $1P$ &$1$   &  $P1$ & $\mathbb{Z}^3$ & $\mathbb{Z}^3$ & $\mathbb{Z}$ &  $\mathbb{Z}^3$ & $1$ & $\mathbb{Z}$\\
        \hline
        $\bar{1}P$ & $2$ & $P\bar{1}$ & $\mathbb{Z}_2^4$ & $\mathbb{Z}^3$ & $\mathbb{Z}_2^8$ & $\mathbb{Z}_2^4\oplus \mathbb{Z}^3$ & $\mathbb{Z}_2^4$ & $1$\\
        \hline
        \multirow{2}{*}{$2P$} & $3$ & $P2$ & $\mathbb{Z}_2^3\oplus\mathbb{Z}$ & $\mathbb{Z}_2^3\oplus\mathbb{Z}$ & $\mathbb{Z}_2^4\oplus\mathbb{Z}$ & $\mathbb{Z}_2^3\oplus \mathbb{Z}$ & $\mathbb{Z}_2^3$ & $\mathbb{Z}_2^3\oplus\mathbb{Z}$\\
        \cline{2-9}
         & $4$ & $P2_1$ & $\mathbb{Z}_2^2\oplus\mathbb{Z}$ & $\mathbb{Z}$ & $\mathbb{Z}$ & $\mathbb{Z}_2^2\oplus \mathbb{Z}$ & $\mathbb{Z}_2^2$ & $\mathbb{Z}$\\
         \hline
        $2C$ & $5$ & $C2$ & $\mathbb{Z}_2^2\oplus \mathbb{Z}$ & $\mathbb{Z}_2\oplus\mathbb{Z}$ & $\mathbb{Z}_2^2\oplus\mathbb{Z}$ & $\mathbb{Z}_2^2\oplus \mathbb{Z}$ & $\mathbb{Z}_2^2$ & $\mathbb{Z}_2\oplus\mathbb{Z}$\\
        \hline
        \multirow{2}{*}{$mP$} & $6$ & $Pm$ & $\mathbb{Z}_2^2\oplus\mathbb{Z}^2$ & $\mathbb{Z}_2^4\oplus\mathbb{Z}$ & $\mathbb{Z}_2^4$ & $\mathbb{Z}_2^2\oplus \mathbb{Z}$ & $\mathbb{Z}_2^2$ & $\mathbb{Z}_2^4$\\
        \cline{2-9}
         & $7$ & $Pc$ & $\mathbb{Z}_2\oplus\mathbb{Z}^2$ & $\mathbb{Z}_2\oplus\mathbb{Z}$ & $1$ & $\mathbb{Z}_2\oplus \mathbb{Z}$ & $\mathbb{Z}_2$ & $\mathbb{Z}_2$\\
\hline
    \end{tabular}
    }
    \label{3D-1}
\end{table}

\newpage
\begin{table}[H]
      \centering
      \scalebox{0.9}{
    \begin{tabular}{|c|c|c|c|c|c|c|c|c|}
        \hline
         Arithmetic class &No.    &  Symbol & $H_1(\Gamma_F,\mathbb{Z})$ & $H_2(\Gamma_F,\mathbb{Z})$ & $H_3(\Gamma_F, \mathbb{Z})$ & $H^2(\Gamma_F,\mathbb{Z})$ & $\mathrm{Tor}H^2(\Gamma_F,\mathbb{Z})$ & $H^3(\Gamma_F, \mathbb{Z})$\\
         \hline
         \multirow{10}{*}{$mm2P$} & $25$ & $Pmm2$ & $\mathbb{Z}_2^4\oplus\mathbb{Z}$ & $\mathbb{Z}_2^8$ & $\mathbb{Z}_2^{12}$ & $\mathbb{Z}_2^4$ & $\mathbb{Z}_2^4$ & $\mathbb{Z}_2^8$\\
         \cline{2-9}
          & $26$ & $Pmc2_1$ & $\mathbb{Z}_2^3\oplus\mathbb{Z}$ & $\mathbb{Z}_2^4$ & $\mathbb{Z}_2^4$ & $\mathbb{Z}_2^3$ & $\mathbb{Z}_2^3$ & $\mathbb{Z}_2^4$\\
          \cline{2-9}
          & $27$ & $Pcc2$ & $\mathbb{Z}_2^3\oplus\mathbb{Z}$ & $\mathbb{Z}_2^4$ & $\mathbb{Z}_2^4$ & $\mathbb{Z}_2^3$ & $\mathbb{Z}_2^3$ & $\mathbb{Z}_2^4$\\
          \cline{2-9}
          & $28$ & $Pma2$ & $\mathbb{Z}_2^3\oplus\mathbb{Z}$ & $\mathbb{Z}_2^4$ & $\mathbb{Z}_2^4$ & $\mathbb{Z}_2^3$ & $\mathbb{Z}_2^3$ & $\mathbb{Z}_2^4$\\
          \cline{2-9}
          & $29$ & $Pca2_1$ & $\mathbb{Z}_2^2\oplus\mathbb{Z}$ & $\mathbb{Z}_2$ & $1$ & $\mathbb{Z}_2^2$ & $\mathbb{Z}_2^2$ & $\mathbb{Z}_2$\\
          \cline{2-9}
          & $30$ & $Pnc2$ & $\mathbb{Z}_2^2\oplus\mathbb{Z}$ & $\mathbb{Z}_2\oplus\mathbb{Z}_4$ & $\mathbb{Z}_2^2$ & $\mathbb{Z}_2^2$ & $\mathbb{Z}_2^2$ & $\mathbb{Z}_2\oplus\mathbb{Z}_4$\\
          \cline{2-9}
          & $31$ & $Pmn2_1$ & $\mathbb{Z}_2^2\oplus\mathbb{Z}$ & $\mathbb{Z}_2^2$ & $\mathbb{Z}_2^2$ & $\mathbb{Z}_2^2$ & $\mathbb{Z}_2^2$ & $\mathbb{Z}_2^2$\\
\hline
    \end{tabular}
    }
    \label{3D-2}
\end{table}

\newpage
\begin{table}[H]
    \centering
    \scalebox{0.9}{
    \begin{tabular}{|c|c|c|c|c|c|c|c|c|}
        \hline
         Arithmetic class &No.    &  Symbol & $H_1(\Gamma_F,\mathbb{Z})$ & $H_2(\Gamma_F,\mathbb{Z})$ & $H_3(\Gamma_F,\mathbb{Z})$ & $H^2(\Gamma_F,\mathbb{Z})$ & $\mathrm{Tor}H^2(\Gamma_F,\mathbb{Z})$ & $H^3(\Gamma_F,\mathbb{Z})$\\
         \hline
        \multirow{6}{*}{$mmmP$} & $57$ & $Pbcm$ & $\mathbb{Z}_2^4$ & $\mathbb{Z}_2^3$ & $\mathbb{Z}_2^5$ & $\mathbb{Z}_2^4$ & $\mathbb{Z}_2^4$ & $\mathbb{Z}_2^3$ \\
         \cline{2-9}
         & $58$ & $Pnnm$ & $\mathbb{Z}_2^2\oplus\mathbb{Z}_4$ & $\mathbb{Z}_2^3$ & $\mathbb{Z}_2^7$ & $\mathbb{Z}_2^2\oplus\mathbb{Z}_4$ & $\mathbb{Z}_2^2\oplus\mathbb{Z}_4$ & $\mathbb{Z}_2^3$\\
         \cline{2-9}
         & $59$ & $Pmmn$ & $\mathbb{Z}_2^4$ & $\mathbb{Z}_2^4$ & $\mathbb{Z}_2^8$ & $\mathbb{Z}_2^4$ & $\mathbb{Z}_2^4$ & $\mathbb{Z}_2^4$ \\
         \cline{2-9}
         & $60$ & $Pbcn$ & $\mathbb{Z}_2^3$ & $\mathbb{Z}_2$ & $\mathbb{Z}_2^3$ & $\mathbb{Z}_2^3$ & $\mathbb{Z}_2^3$ & $\mathbb{Z}_2$\\
         \cline{2-9}
         & $61$ & $Pbca$ & $\mathbb{Z}_2^3$ & $1$ & $\mathbb{Z}_2^2$ & $\mathbb{Z}_2^3$ & $\mathbb{Z}_2^3$ & $1$ \\
         \cline{2-9}
         & $62$ & $Pnma$ & $\mathbb{Z}_2^2\oplus\mathbb{Z}_4$ & $\mathbb{Z}_2^2$ & $\mathbb{Z}_2^4$ & $\mathbb{Z}_2^2\oplus\mathbb{Z}_4$ & $\mathbb{Z}_2^2\oplus\mathbb{Z}_4$ & $\mathbb{Z}_2^2$\\
         \hline
         \multirow{6}{*}{$mmmC$} & $63$ & $Cmcm$ & $\mathbb{Z}_2^4$ & $\mathbb{Z}_2^4$ & $\mathbb{Z}_2^8$ & $\mathbb{Z}_2^4$ & $\mathbb{Z}_2^4$ & $\mathbb{Z}_2^4$\\
\hline    
    \end{tabular}
    }
    \label{3D-3}
\end{table}

\newpage
\begin{table}[H]
    \centering
    \scalebox{0.9}{
    \begin{tabular}{|c|c|c|c|c|c|c|c|c|}
        \hline
         Arithmetic class &No.    &  Symbol & $H_1(\Gamma_F,\mathbb{Z})$ & $H_2(\Gamma_F,\mathbb{Z})$ & $H_3(\Gamma_F,\mathbb{Z})$ & $H^2(\Gamma_F,\mathbb{Z})$ & $\mathrm{Tor}H^2(\Gamma_F,\mathbb{Z})$ & $H^3(\Gamma_F,\mathbb{Z})$\\
         \hline
          \multirow{2}{*}{$4/mI$} & $87$ & $I4/m$ & $\mathbb{Z}_2^2\oplus\mathbb{Z}_4$ & $\mathbb{Z}_2^2\oplus\mathbb{Z}$ & $\mathbb{Z}_2^5\oplus\mathbb{Z}_4^2$ & $\mathbb{Z}_2^2\oplus\mathbb{Z}_4\oplus\mathbb{Z}$ & $\mathbb{Z}_2^2\oplus\mathbb{Z}_4$ & $\mathbb{Z}_2^2$\\
          \cline{2-9}
          & $88$ & $I4_1/a$ & $\mathbb{Z}_2\oplus\mathbb{Z}_4$ & $\mathbb{Z}$ & $\mathbb{Z}_2^3\oplus\mathbb{Z}_4$ & $\mathbb{Z}_2\oplus\mathbb{Z}_4\oplus\mathbb{Z}$ & $\mathbb{Z}_2\oplus\mathbb{Z}_4$ & $1$\\
          \hline
          \multirow{8}{*}{$422P$} & $89$ & $P422$ & $\mathbb{Z}_2^4$ & $\mathbb{Z}_2^4\oplus\mathbb{Z}_4$ & $\mathbb{Z}_2^7\oplus\mathbb{Z}_4^2\oplus\mathbb{Z}$ & $\mathbb{Z}_2^4$ & $\mathbb{Z}_2^4$ & $\mathbb{Z}_2^4\oplus\mathbb{Z}_4\oplus\mathbb{Z}$\\
          \cline{2-9}
          & $90$ & $P42_12$ & $\mathbb{Z}_2^2\oplus\mathbb{Z}_4$ & $\mathbb{Z}_2\oplus\mathbb{Z}_4$ & $\mathbb{Z}_2^3\oplus\mathbb{Z}_4\oplus\mathbb{Z}$ & $\mathbb{Z}_2^2\oplus\mathbb{Z}_4$ & $\mathbb{Z}_2^2\oplus\mathbb{Z}_4$ & $\mathbb{Z}_2\oplus\mathbb{Z}_4\oplus\mathbb{Z}$ \\
          \cline{2-9}
          & $91$ & $P4_122$ & $\mathbb{Z}_2^3$ & $\mathbb{Z}_2^2$ & $\mathbb{Z}_2^3\oplus\mathbb{Z}$ & $\mathbb{Z}_2^3$ & $\mathbb{Z}_2^3$ & $\mathbb{Z}_2^2\oplus\mathbb{Z}$\\
          \cline{2-9}
          & $92$ & $P4_12_12$ & $\mathbb{Z}_2\oplus\mathbb{Z}_4$ & $1$ & $\mathbb{Z}_2\oplus\mathbb{Z}$ & $\mathbb{Z}_2\oplus\mathbb{Z}_4$ & $\mathbb{Z}_2\oplus\mathbb{Z}_4$ & $\mathbb{Z}$ \\
          \cline{2-9}
          & $93$ & $P4_222$ & $\mathbb{Z}_2^4$ & $\mathbb{Z}_2^5$ & $\mathbb{Z}_2^9\oplus\mathbb{Z}$ & $\mathbb{Z}_2^4$ & $\mathbb{Z}_2^4$ & $\mathbb{Z}_2^5\oplus\mathbb{Z}$\\
\hline
    \end{tabular}
    }
    \label{3D-4}
\end{table}

\newpage
\begin{table}[H]
    \centering
    \scalebox{0.9}{
    \begin{tabular}{|c|c|c|c|c|c|c|c|c|}
        \hline
         Arithmetic class &No.    &  Symbol & $H_1(\Gamma_F,\mathbb{Z})$ & $H_2(\Gamma_F,\mathbb{Z})$ & $H_3(\Gamma_F,\mathbb{Z})$ & $H^2(\Gamma_F,\mathbb{Z})$ & $\mathrm{Tor}H^2(\Gamma_F,\mathbb{Z})$ & $H^3(\Gamma_F,\mathbb{Z})$\\
         \hline
        \multirow{2}{*}{$\bar{4}m2I$} & $119$ & $I\bar{4}m2$ & $\mathbb{Z}_2^3$ & $\mathbb{Z}_2^3$ & $\mathbb{Z}_2^5\oplus\mathbb{Z}_4^2$ & $\mathbb{Z}_2^3$ & $\mathbb{Z}_2^3$ & $\mathbb{Z}_2^3$\\
        \cline{2-9}
        & $120$ & $I\bar{4}c2$ & $\mathbb{Z}_2^3$ & $\mathbb{Z}_2^2$ & $\mathbb{Z}_2^2\oplus\mathbb{Z}_4^2$ & $\mathbb{Z}_2^3$ & $\mathbb{Z}_2^3$ & $\mathbb{Z}_2^2$\\
        \hline
        \multirow{2}{*}{$\bar{4}2mI$} & $121$ & $I\bar{4}2m$ & $\mathbb{Z}_2^2\oplus\mathbb{Z}_4$ & $\mathbb{Z}_2^3$ & $\mathbb{Z}_2^4\oplus\mathbb{Z}_4^2$ & $\mathbb{Z}_2^2\oplus\mathbb{Z}_4$ & $\mathbb{Z}_2^2\oplus\mathbb{Z}_4$ & $\mathbb{Z}_2^3$\\
        \cline{2-9}
        & $122$ & $I\bar{4}2d$ & $\mathbb{Z}_2\oplus\mathbb{Z}_4$ & $\mathbb{Z}_2$ & $\mathbb{Z}_2^2\oplus\mathbb{Z}_4$ & $\mathbb{Z}_2\oplus\mathbb{Z}_4$ & $\mathbb{Z}_2\oplus\mathbb{Z}_4$ & $\mathbb{Z}_2$\\
          \hline
        \multirow{16}{*}{$4/mmmP$} & $123$ & $P4/mmm$ & $\mathbb{Z}_2^5$ & $\mathbb{Z}_2^9$ & $\mathbb{Z}_2^{18}\oplus\mathbb{Z}_4^2$ & $\mathbb{Z}_2^5$ & $\mathbb{Z}_2^5$ & $\mathbb{Z}_2^9$\\
        \cline{2-9}
         & $124$ & $P4/mcc$ & $\mathbb{Z}_2^4$ & $\mathbb{Z}_2^5$ & $\mathbb{Z}_2^8\oplus\mathbb{Z}_4^2$ & $\mathbb{Z}_2^4$ & $\mathbb{Z}_2^4$ & $\mathbb{Z}_2^5$\\
         \cline{2-9}
         & $125$ & $P4/nbm$ & $\mathbb{Z}_2^4$ & $\mathbb{Z}_2^4\oplus\mathbb{Z}_4$ & $\mathbb{Z}_2^8\oplus\mathbb{Z}_4^2$ & $\mathbb{Z}_2^4$ & $\mathbb{Z}_2^4$ & $\mathbb{Z}_2^4\oplus\mathbb{Z}_4$ \\
\hline
    \end{tabular}
    }
    \label{3D-5}
\end{table}

\newpage
\begin{table}[H]
 \centering
 \scalebox{0.9}{
    \begin{tabular}{|c|c|c|c|c|c|c|c|c|}
         \hline
         Arithmetic class &No.    &  Symbol & $H_1(\Gamma_F,\mathbb{Z})$ & $H_2(\Gamma_F,\mathbb{Z})$ & $H_3(\Gamma_F, \mathbb{Z})$ & $H^2(\Gamma_F,\mathbb{Z})$ & $\mathrm{Tor}H^2(\Gamma_F,\mathbb{Z})$ & $H^3(\Gamma_F, \mathbb{Z})$\\
         \hline
         \multirow{3}{*}{$312P$} & $149$ & $P312$ & $\mathbb{Z}_2^2$ & $\mathbb{Z}_3\oplus\mathbb{Z}_6$ & $\mathbb{Z}_3\oplus\mathbb{Z}_6^2\oplus\mathbb{Z}$ & $\mathbb{Z}_2^2$ & $\mathbb{Z}_2^2$ & $\mathbb{Z}_3\oplus\mathbb{Z}_6\oplus\mathbb{Z}$\\
         \cline{2-9}
         & $151$ & $P3_112$ & $\mathbb{Z}_2^2$ & $\mathbb{Z}_2$ & $\mathbb{Z}_2^2\oplus\mathbb{Z}$ & $\mathbb{Z}_2^2$ & $\mathbb{Z}_2^2$ & $\mathbb{Z}_2\oplus\mathbb{Z}$\\
         \cline{2-9}
         & $153$ & $P3_212$ & $\mathbb{Z}_2^2$ & $\mathbb{Z}_2$ & $\mathbb{Z}_2^2\oplus\mathbb{Z}$ & $\mathbb{Z}_2^2$ & $\mathbb{Z}_2^2$ & $\mathbb{Z}_2\oplus\mathbb{Z}$\\
         \hline
         \multirow{3}{*}{$321P$} & $150$ & $P321$ & $\mathbb{Z}_2\oplus\mathbb{Z}_6$ & $\mathbb{Z}_6$ & $\mathbb{Z}_6^2\oplus\mathbb{Z}$ & $\mathbb{Z}_2\oplus\mathbb{Z}_6$ & $\mathbb{Z}_2\oplus\mathbb{Z}_6$ & $\mathbb{Z}_6\oplus\mathbb{Z}$\\
         \cline{2-9}
         & $152$ & $P3_121$ & $\mathbb{Z}_2\oplus\mathbb{Z}_6$ & $\mathbb{Z}_2$ & $\mathbb{Z}_2^2\oplus\mathbb{Z}$ & $\mathbb{Z}_2\oplus\mathbb{Z}_6$ & $\mathbb{Z}_2\oplus\mathbb{Z}_6$ & $\mathbb{Z}_2\oplus\mathbb{Z}$\\
         \cline{2-9}
         & $154$ & $P3_221$ & $\mathbb{Z}_2\oplus\mathbb{Z}_6$ & $\mathbb{Z}_2$ & $\mathbb{Z}_2^2\oplus\mathbb{Z}$ & $\mathbb{Z}_2\oplus\mathbb{Z}_6$ & $\mathbb{Z}_2\oplus\mathbb{Z}_6$ & $\mathbb{Z}_2\oplus\mathbb{Z}$\\
         \hline
         $32R$ & $155$ & $R32$ & $\mathbb{Z}_2^2$ & $\mathbb{Z}_2$ & $\mathbb{Z}_2\oplus\mathbb{Z}_6\oplus\mathbb{Z}$ & $\mathbb{Z}_2^2$ & $\mathbb{Z}_2^2$ & $\mathbb{Z}_2\oplus\mathbb{Z}$\\
\hline
    \end{tabular}
    }
    \label{3D-6}
\end{table}

\begin{table}[H]
 \centering
 \scalebox{0.9}{
    \begin{tabular}{|c|c|c|c|c|c|c|c|c|}
         \hline
         Arithmetic class &No.    &  Symbol & $H_1(\Gamma_F,\mathbb{Z})$ & $H_2(\Gamma_F,\mathbb{Z})$ & $H_3(\Gamma_F,\mathbb{Z})$ & $H^2(\Gamma_F,\mathbb{Z})$ & $\mathrm{Tor}H^2(\Gamma_F,\mathbb{Z})$ & $H^3(\Gamma_F,\mathbb{Z})$\\
         \hline
         \multirow{3}{*}{$622P$} & $180$ & $P6_222$ & $\mathbb{Z}_2^3$ & $\mathbb{Z}_2^3$ & $\mathbb{Z}_2^6\oplus\mathbb{Z}$ & $\mathbb{Z}_2^3$ & $\mathbb{Z}_2^3$ & $\mathbb{Z}_2^3\oplus\mathbb{Z}$\\
        \cline{2-9}
         & $181$ & $P6_422$ & $\mathbb{Z}_2^3$ & $\mathbb{Z}_2^3$ & $\mathbb{Z}_2^6\oplus\mathbb{Z}$ & $\mathbb{Z}_2^3$ & $\mathbb{Z}_2^3$ & $\mathbb{Z}_2^3\oplus\mathbb{Z}$\\
        \cline{2-9}
         & $182$ & $P6_322$ & $\mathbb{Z}_2^2$ & $\mathbb{Z}_6$ & $\mathbb{Z}_6^2\oplus\mathbb{Z}$ & $\mathbb{Z}_2^2$ & $\mathbb{Z}_2^2$ & $\mathbb{Z}_6\oplus\mathbb{Z}$\\
         \hline
         \multirow{4}{*}{$6mmP$} & $183$ & $P6mm$ & $\mathbb{Z}_2^2\oplus\mathbb{Z}$ & $\mathbb{Z}_2^4$ & $\mathbb{Z}_2^4\oplus\mathbb{Z}_6^2$ & $\mathbb{Z}_2^2$ & $\mathbb{Z}_2^2$ & $\mathbb{Z}_2^4$\\
         \cline{2-9}
         & $184$ & $P6cc$ & $\mathbb{Z}_2\oplus\mathbb{Z}$ & $\mathbb{Z}_2^2$ & $\mathbb{Z}_6^2$ & $\mathbb{Z}_2$ & $\mathbb{Z}_2$ & $\mathbb{Z}_2^2$\\
         \cline{2-9}
         & $185$ & $P6_3cm$ & $\mathbb{Z}_2\oplus\mathbb{Z}$ & $\mathbb{Z}_2^2$ & $\mathbb{Z}_6^2$ & $\mathbb{Z}_2$ & $\mathbb{Z}_2$ & $\mathbb{Z}_2^2$\\
         \cline{2-9}
         & $186$ & $P6_3mc$ & $\mathbb{Z}_2\oplus\mathbb{Z}$ & $\mathbb{Z}_2^2$ & $\mathbb{Z}_6^2$ & $\mathbb{Z}_2$ & $\mathbb{Z}_2$ & $\mathbb{Z}_2^2$\\
\hline
    \end{tabular}
    }
    \label{3D-7}
\end{table}

\begin{table}[H]
 \centering
 \scalebox{0.9}{
    \begin{tabular}{|c|c|c|c|c|c|c|c|c|}
         \hline
         Arithmetic class &No.    &  Symbol & $H_1(\Gamma_F,\mathbb{Z})$ & $H_2(\Gamma_F,\mathbb{Z})$ & $H_3(\Gamma_F,\mathbb{Z})$ & $H^2(\Gamma_F,\mathbb{Z})$ & $\mathrm{Tor}H^2(\Gamma_F,\mathbb{Z})$ & $H^3(\Gamma_F,\mathbb{Z})$\\
          \hline
         \multirow{2}{*}{$432F$} & $209$ & $F432$ & $\mathbb{Z}_2$ & $\mathbb{Z}_2^3$ & $\mathbb{Z}_2^3\oplus\mathbb{Z}_{12}\oplus\mathbb{Z}$ & $\mathbb{Z}_2$ & $\mathbb{Z}_2$ & $\mathbb{Z}_2^3\oplus\mathbb{Z}$\\
         \cline{2-9}
         & $210$ & $F4_132$ & $\mathbb{Z}_2$ & $\mathbb{Z}_2^2$ & $\mathbb{Z}_2\oplus\mathbb{Z}_6\oplus\mathbb{Z}$ & $\mathbb{Z}_2$ & $\mathbb{Z}_2$ & $\mathbb{Z}_2^2\oplus\mathbb{Z}$\\
         \hline
        \multirow{2}{*}{$432I$} & $211$ & $I432$ & $\mathbb{Z}_2^2$ & $\mathbb{Z}_2^2$ & $\mathbb{Z}_2^3\oplus\mathbb{Z}_{12}\oplus\mathbb{Z}$ & $\mathbb{Z}_2^2$ & $\mathbb{Z}_2^2$ & $\mathbb{Z}_2^2\oplus\mathbb{Z}$\\
         \cline{2-9}
         & $214$ & $I4_132$ & $\mathbb{Z}_2^2$ & $\mathbb{Z}_2$ & $\mathbb{Z}_2^2\oplus\mathbb{Z}_6\oplus\mathbb{Z}$ & $\mathbb{Z}_2^2$ & $\mathbb{Z}_2^2$ & $\mathbb{Z}_2\oplus\mathbb{Z}$\\
         \hline
         \multirow{2}{*}{$\bar{4}3mP$} & $215$ & $P\bar{4}3m$ & $\mathbb{Z}_2^2$ & $\mathbb{Z}_2^3$ & $\mathbb{Z}_2^4\oplus\mathbb{Z}_4\oplus\mathbb{Z}_{12}$ & $\mathbb{Z}_2^2$ & $\mathbb{Z}_2^2$ & $\mathbb{Z}_2^3$\\
         \cline{2-9}
         & $218$ & $P\bar{4}3n$ & $\mathbb{Z}_4$ & $\mathbb{Z}_2^2$ & $\mathbb{Z}_2\oplus\mathbb{Z}_4\oplus\mathbb{Z}_{12}$ & $\mathbb{Z}_4$ & $\mathbb{Z}_4$ & $\mathbb{Z}_2^2$\\
         \hline
         \multirow{2}{*}{$\bar{4}3mF$} & $216$ & $F\bar{4}3m$ & $\mathbb{Z}_2$ & $\mathbb{Z}_2^3$ & $\mathbb{Z}_2^3\oplus\mathbb{Z}_4\oplus\mathbb{Z}_{12}$ & $\mathbb{Z}_2$ & $\mathbb{Z}_2$ & $\mathbb{Z}_2^3$\\
\hline
    \end{tabular}
    }
    \label{3D-8}
\end{table}

\section{Typical examples}
In this section, we provide additional technical details for examples in the main text and introduce more examples. For $P1$, we show that the algebraic invariant $C_{\beta\alpha}$ is equivalent to the Chern number over the sub-torus spanned by $\bm{b}_\alpha$ and $\bm{b}_\beta$. For $Pg$, we show the algebraic invariant is equivalent to the topological $\mathbb{Z}_2$ invariant formulated in Ref.~\cite{chen2022brillouin}, and demonstrate how to numerically compute the algebraic invariant in tight-binding models. The case of $P2_1/c$, briefly mentioned in the main text, and the case of $I222$ are systematically treated here. The algebraic invariants are also formulated as topological invariants in terms of the Berry curvature and connection. The $\mathbb{Z}_4$ algebraic invariant of $I23$ has been formulated in the main text. Here,  we further formulate the $\mathbb{Z}_3$ algebraic invariant. Notably, we show that the $\mathbb{Z}_2$ topological invariants formulated in Refs.~\cite{shiozaki2016topology,shiozaki2022atiyah} correspond to the $\mathbb{Z}_2$ subgroups of the $\mathbb{Z}_4$ components for $I222$ and $I23$.

\subsection{Topological invariants for $P1$}
The momentum-space crystallographic group $P1$ is generated by translations $l_\alpha$ ($\alpha=1,2,3$) and is isomorphic to $L_F$. The classification for Abelian phases is given by
\begin{equation}
    H^2(P1, \mathbb{Z}) \cong \mathbb{Z}^3.
\end{equation}

To derive the topological invariants, consider the phase factors associated with successive translations. First applying the translation by $\bm{b}_\alpha$ and then by $\bm{b}_\beta$ leads to the phase relation
\begin{equation}
e^{2\pi i\phi_{\beta}(\bm{k}+\bm{b}_{\beta\alpha})} e^{2\pi i\phi_{\alpha}(\bm{k}+\bm{b}_{\alpha})}  = e^{2\pi i\phi_{\beta\alpha}(\bm{k}+\bm{b}_{\beta\alpha})}.
\end{equation}
The phase difference yields an integer
\begin{equation}
N_{\beta\alpha} = \phi_{\beta}(\bm{k}+\bm{b}_{\beta\alpha}) - \phi_{\beta\alpha}(\bm{k}+\bm{b}_{\beta\alpha}) + \phi_{\alpha}(\bm{k} + \bm{b}_\alpha).
\end{equation}
Similarly, applying $\bm{b}_\beta$ first and then $\bm{b}_\alpha$ gives
\begin{equation}
N_{\alpha\beta} = \phi_{\alpha}(\bm{k} + \bm{b}_{\alpha\beta}) - \phi_{\alpha\beta}(\bm{k}+\bm{b}_{\alpha\beta}) + \phi_{\beta}(\bm{k}+\bm{b}_\beta).
\end{equation}
Noting that $\bm{b}_{\alpha\beta} = \bm{b}_{\beta\alpha} = \bm{b}_\alpha+\bm{b}_\beta$ and $\phi_{\alpha\beta} = \phi_{\beta\alpha}$, the difference yields
\begin{equation}
C_{\beta\alpha} = N_{\beta\alpha} - N_{\alpha\beta} = \phi_{\beta}(\bm{k}+\bm{b}_{\beta\alpha}) + \phi_{\alpha}(\bm{k} + \bm{b}_\alpha) - \phi_{\alpha}(\bm{k} + \bm{b}_{\alpha\beta}) - \phi_{\beta}(\bm{k}+\bm{b}_\beta).
\end{equation}
$C_{\beta\alpha}$ takes integer values. For different combinations of $\alpha$ and $\beta$, we obtain three independent invariants $C_{\beta\alpha}$, corresponding to the classification $H^2(P1,\mathbb{Z}) \cong \mathbb{Z}^3$.

This topological invariant is precisely the Chern number for the 2D subtorus spanned by $\bm{b}_{\alpha}$ and $\bm{b}_{\beta}$. To see this, we rewrite it as a Berry curvature integral
\begin{equation}
\begin{split}
    C_{\beta\alpha} =& \phi_{\beta}(\bm{k}+\bm{b}_{\beta\alpha}) + \phi_{\alpha}(\bm{k} + \bm{b}_\alpha) - \phi_{\alpha}(\bm{k} + \bm{b}_{\alpha\beta}) - \phi_{\beta}(\bm{k}+\bm{b}_\beta) \\
    =& \left[\phi_{\beta}(\bm{k}+\bm{b}_{\alpha}+\bm{b}_\beta) - \phi_{\beta}(\bm{k}+\bm{b}_\beta)\right] - \left[\phi_{\alpha}(\bm{k}+\bm{b}_{\beta}+\bm{b}_\alpha) - \phi_{\alpha}(\bm{k}+\bm{b}_\alpha)\right] \\
    =& \int_{\bm{k}+\bm{b}_\beta}^{\bm{k}+\bm{b}_{\alpha}+\bm{b}_\beta} dk_\alpha\, \partial_{k_\alpha}\phi_{\beta}(\bm{k}') - \int_{\bm{k}+\bm{b}_\alpha}^{\bm{k}+\bm{b}_{\beta}+\bm{b}_\alpha} dk_\beta\, \partial_{k_\beta}\phi_{\alpha}(\bm{k}') \\
    =& \frac{1}{2\pi}\int_{\bm{k}+\bm{b}_\beta}^{\bm{k}+\bm{b}_{\alpha}+\bm{b}_\beta} dk_\alpha\, \mathcal{A}_\alpha(\bm{k}') - \frac{1}{2\pi}\int_{\bm{k}+\bm{b}_\beta}^{\bm{k}+\bm{b}_{\alpha}+\bm{b}_\beta} dk_\alpha\, \mathcal{A}_\alpha(\bm{k}'-\bm{b}_\beta) \\&- \frac{1}{2\pi}\int_{\bm{k}+\bm{b}_\alpha}^{\bm{k}+\bm{b}_{\beta}+\bm{b}_\alpha} dk_\beta\, \mathcal{A}_\beta(\bm{k}') + \frac{1}{2\pi}\int_{\bm{k}+\bm{b}_\alpha}^{\bm{k}+\bm{b}_{\beta}+\bm{b}_\alpha} dk_\beta\, \mathcal{A}_\beta(\bm{k}'-\bm{b}_\alpha)\\
    =& \frac{1}{2\pi}\int_{\bm{k}}^{\bm{k}+\bm{b}_{\beta}} dk_\beta\, \mathcal{A}_\beta(\bm{k}') + \frac{1}{2\pi}\int_{\bm{k}+\bm{b}_\beta}^{\bm{k}+\bm{b}_{\alpha}+\bm{b}_\beta} dk_\alpha\, \mathcal{A}_\alpha(\bm{k}') \\&- \frac{1}{2\pi}\int_{\bm{k}+\bm{b}_\alpha}^{\bm{k}+\bm{b}_{\beta}+\bm{b}_\alpha} dk_\beta\, \mathcal{A}_\beta(\bm{k}') - \frac{1}{2\pi}\int_{\bm{k}}^{\bm{k}+\bm{b}_{\alpha}} dk_\alpha\, \mathcal{A}_\alpha(\bm{k}')\\
    =& \frac{1}{2\pi} \int_T \mathcal{F}(\bm{k}') \, d^2\bm{k}',
\end{split}
\end{equation}
where $T$ denotes the $2$D torus spanned by $\bm{b}_\alpha$ and $\bm{b}_\beta$, and $\mathcal{F}$ is the Berry curvature.

In deriving this result, we used the identities
\begin{equation}
\begin{split}
    \mathcal{A}_\alpha(\bm{k}) - \mathcal{A}_\alpha(\bm{k}-\bm{b}_\beta) &= 2\pi \partial_{k_\alpha} \phi_{\beta}(\bm{k}), \\
    \mathcal{A}_\beta(\bm{k}) - \mathcal{A}_\beta(\bm{k}-\bm{b}_\alpha) &= 2\pi \partial_{k_\beta} \phi_{\alpha}(\bm{k}),
\end{split}
\end{equation}
which relate the Berry connection $\mathcal{A}_\mu(\bm{k}) = \sum_\nu \langle \psi_\nu(\bm{k})| i\partial_{k_\mu} |\psi_\nu(\bm{k}) \rangle$ to the translation phase factors. We now prove the first identity; the second follows similarly. Under a translation by $\bm{b}_\beta$, the wavefunctions transform as $|\psi_\mu(\bm{k})\rangle = \sum_\nu \mathcal{U}_{\beta,\nu\mu}(\bm{k}+\bm{b}_\beta) |\psi_\nu(\bm{k}+\bm{b}_\beta)\rangle$, where $\mathcal{U}_\beta$ is the unitary representation satisfying $\det \mathcal{U}_\beta = e^{2\pi i \phi_\beta}$. Then,
\begin{equation}
\begin{split}
    \mathcal{A}_\alpha(\bm{k}) &= \sum_\mu \langle \psi_\mu(\bm{k})| i\partial_{k_\alpha} |\psi_\mu(\bm{k}) \rangle \\
    &= \sum_{\mu\nu\rho} \mathcal{U}^\dagger_{\beta,\mu\nu}(\bm{k}+\bm{b}_\beta) \langle \psi_\nu(\bm{k}+\bm{b}_\beta) | i\partial_{k_\alpha} \left( |\psi_\rho(\bm{k}+\bm{b}_\beta)\rangle \mathcal{U}_{\beta,\rho\mu}(\bm{k}+\bm{b}_\beta) \right) \\
    &= \mathcal{A}_\alpha(\bm{k}+\bm{b}_\beta) + i \mathrm{Tr} \left[ \mathcal{U}^\dagger_\beta(\bm{k}+\bm{b}_\beta) \partial_{k_\alpha} \mathcal{U}_\beta(\bm{k}+\bm{b}_\beta) \right].
\end{split}
\end{equation}
Using the identity $\det e^A = e^{\mathrm{Tr} A}$ for any matrix $A$, we have
\begin{equation}
    \mathrm{Tr} \left( \mathcal{U}^\dagger_\beta d\mathcal{U}_\beta \right) = d \log \det \mathcal{U}_\beta = 2\pi i\, d\phi_\beta.
\end{equation}
Thus,
\begin{equation}
    \mathcal{A}_\alpha(\bm{k}) - \mathcal{A}_\alpha(\bm{k}-\bm{b}_\beta) = 2\pi \partial_{k_\alpha} \phi_\beta(\bm{k}).
\end{equation}

\subsection{Topological invariant for $Pg$}
\subsubsection{The expression in terms of phase factors}
In the main text, the phase factor for each $\gamma\in Pg$ is introduced as
\begin{equation}
	U_{\gamma}|\psi(\bm{k})\rangle =e^{2\pi i \phi_\gamma(\gamma\bm{k})}|\psi(\gamma\bm{k})\rangle.
\end{equation}
Accordingly, the integral cocycle is formulated as
\begin{equation}
	N(\gamma_2,\gamma_1)=\phi_{\gamma_1}(\gamma_2^{-1}\bm{k})-\phi_{\gamma_2\gamma_1}(\bm{k})+\phi_{\gamma_2}(\bm{k})\in \mathbb{Z}.
\end{equation}
As mentioned in the main text, the integer $N(\gamma_2,\gamma_1)$ is independent of the choice of $\bm{k}\in \mathbb{R}^d_F$. In the following discussion, we utilize this freedom to simplify several expressions by selecting appropriate momenta.

\begin{equation}
    g_xl_x = l_x^{-1}g_x.
\end{equation}

The topological invariant in the main text involves the following integers,
\begin{eqnarray}
	N(g_x,l_x)&=&\phi_{l_x}(l_x\bm{k})-\phi_{g_xl_x}(g_xl_x\bm{k})+\phi_{g_x}(g_xl_x\bm{k}),\\
	N(l_x^{-1},g_x)&=&\phi_{g_x}(g_x\bm{k})-\phi_{l_x^{-1}g_x}(l_x^{-1}g_x\bm{k})+\phi_{l_x^{-1}}(l_x^{-1}g_x\bm{k}),
\end{eqnarray}
where the reference momenta have been chosen for simplicity. 
Noting that $g_xl_x=l_x^{-1}g_x$, their difference is derived as
\begin{equation}
	\begin{split}
		N(g_x,l_x)-N(l_x^{-1},g_x)&=\phi_{l_x}(l_x\bm{k})+\phi_{g_x}(g_xl_x\bm{k})-\phi_{g_x}(g_x\bm{k})-\phi_{l_x^{-1}}(g_xl_x\bm{k})\\
		&=\phi_{l_x}(g_x^{-1}\bm{k})+\phi_{g_x}(\bm{k})-\phi_{g_x}(l_x\bm{k})-\phi_{l_x^{-1}}(\bm{k}).
	\end{split}
\end{equation}
In the second equality, we replaced $\bm{k}$ in the first line by $l_x^{-1}g_x^{-1}\bm{k}$, and simplified group elements by $g_xl_x=l_x^{-1}g_x$. The integer
\begin{equation}
	N(l_x,l_x^{-1})=\phi_{l_x^{-1}}(\bm{k})+\phi_{l_x}(l_x\bm{k})
\end{equation}
also appears in the topological invariant. Note that for the identity group element, $\phi_{1}=0$. Thus, the topological invariant can be converted as
\begin{equation}
	\begin{split}
		\nu &=N(g_x,l_x)-N(l_x^{-1},g_x)+N(l_x,l_x^{-1})\\
		       &=\phi_{l_x}(g_x^{-1}\bm{k})+\phi_{g_x}(\bm{k})-\phi_{g_x}(l_x\bm{k})+\phi_{l_x}(l_x\bm{k}) \mod 2.
	\end{split}
\end{equation}

\subsubsection{The expression in terms of the Berry connection}
To express the topological invariant in terms of the Berry connection, let us first analyze the constraints imposed by the glide reflection $g_x$ on the Berry connection.

Consider the general case of multiple valence bands. The Berry connection for the determinant line bundle is given by
\begin{equation}
\mathcal{A}_\mu(\bm{k})=\sum_\alpha \langle \psi_\alpha(\bm{k})|i\partial_{k_\mu}|\psi_{\alpha}(\bm{k})\rangle.
\end{equation}
For the $k_x$-component, the constraint from the glide reflection is analyzed as follows.
\begin{equation}
	\begin{split}
			\mathcal{A}_x(\bm{k}) &=\sum_\alpha \langle \psi_\alpha(\bm{k})|U_{g_x}^\dagger U_{g_x}i\partial_{k_x}|\psi_{\alpha}(\bm{k})\rangle\\
			&=\sum_{\alpha\beta\gamma} \mathcal{U}^\dagger_{\alpha\gamma}(g_x\bm{k})\langle \psi_{\gamma}(g_x\bm{k})| i\partial_{k_x}(|\psi_{\beta}(g_x\bm{k})\rangle \mathcal{U}_{\beta\alpha}(g_x\bm{k}))\\
			&=-\mathcal{A}_x(g_x\bm{k})+i\mathrm{Tr} \mathcal{U}^\dagger (g_x\bm{k})\partial_{k_x}\mathcal{U}(g_x\bm{k}).
	\end{split}
\end{equation}
Using the identity for any square matrix $A$,
\begin{equation}
	\det e^{A}=e^{\mathrm{Tr} A},
\end{equation}
we obtain
\begin{equation}
	\det \mathcal{U}^\dagger  d \det \mathcal{U}=\mathrm{Tr}\mathcal{U}^\dagger d\mathcal{U}.
\end{equation}
According to the definition of the phase factor in the main text, we have
\begin{equation}
	\det \mathcal{U}_{g_x}=e^{2\pi i \phi_{g_x}}.
\end{equation}
Thus, the constraint for the Berry connection can be expressed as
\begin{equation}
	\mathcal{A}_x(g_x^{-1}\bm{k}) +\mathcal{A}_x(\bm{k})=2\pi \partial_{k_x}\phi_{g_x}(\bm{k}).
\end{equation}

When formulating the topological invariant for the Klein-bottle insulator, the valence states $|\psi_\alpha(\bm{k})\rangle$ are assumed to be periodic along the $k_x$ direction. Therefore, $e^{2\pi i \phi_{l_x}(\bm{k})}=1$, which implies
\begin{equation}
	\nu= \phi_{g_x}(\bm{k})-\phi_{g_x}(l_x\bm{k}) \mod 2.
\end{equation}
Since the topological invariant is independent of $\bm{k}$, we can choose $\bm{k}=(-\pi,0)$ and proceed as
\begin{equation}
	\begin{split}
			\phi_{g_x}(-\pi, 0)-\phi_{g_x}(\pi, 0) &=-\int_{-\pi}^{\pi} dk_x~\phi_{g_x}(k_x,0)\\
			&= -\frac{1}{2\pi} \int_{-\pi}^{\pi} dk_x ~\mathcal{A}_x(-k_x,-\pi)-\frac{1}{2\pi} \int_{-\pi}^{\pi} dk_x ~\mathcal{A}_x(k_x,0)\\
			&= -\frac{1}{2\pi} \int_{-\pi}^{\pi} dk_x ~\mathcal{A}_x(k_x,-\pi)-\frac{1}{2\pi} \int_{-\pi}^{\pi} dk_x ~\mathcal{A}_x(k_x,0)\\
			&= \frac{1}{2\pi} \int_{-\pi}^{\pi} dk_x ~(\mathcal{A}_x(k_x,0)-\mathcal{A}_x(k_x,-\pi))-\frac{1}{\pi} \int_{-\pi}^{\pi} dk_x ~\mathcal{A}_x(k_x,0)\\
			&= \frac{1}{2\pi} \int_{-\pi}^{\pi} dk_x\int_{-\pi}^0dk_y ~\partial_{k_y}\mathcal{A}_x(k_x,k_y)-\frac{1}{\pi} \int_{-\pi}^{\pi} dk_x ~\mathcal{A}_x(k_x,0).
	\end{split}
\end{equation}
The Berry curvature is defined as
\begin{equation}
	\mathcal{F}(\bm{k})=\partial_{k_x}\mathcal{A}_{y}(\bm{k})-\partial_{k_y}\mathcal{A}_x(\bm{k}).
\end{equation}
Since $\mathcal{A}_{\mu}(\bm{k})$ are periodic in the $k_x$ direction, we can replace $\partial_{k_y}\mathcal{A}_x(k_x,k_y)$ with $-\mathcal{F}(\bm{k})$ in the first integral, noting that $\int_{-\pi}^\pi dk_x~\partial_{k_x}\mathcal{A}_y(\bm{k})=0 $. Thus, the topological invariant becomes
\begin{equation}
	\nu= \frac{1}{2\pi} \int_{-\pi}^{\pi} dk_x\int_{-\pi}^0dk_y~\mathcal{F}(\bm{k})+\frac{1}{\pi} \int_{-\pi}^{\pi} dk_x ~\mathcal{A}_x(k_x,0) \mod 2,
\end{equation}
which is exactly the topological invariant formulated for the Klein-bottle insulator in Ref.~\cite{chen2022brillouin}. Note that for a $\mathbb{Z}_2$-invariant, the overall sign is irrelevant.

\subsubsection{Application to a model for the Klein-bottle insulator}
We consider the four-band Hamiltonian constructed in Ref.~\cite{chen2022brillouin}, which respects momentum-space $Pg$ symmetry. The corresponding symmetry operator $\mathcal{U}_{g_x}$ is given by
\begin{equation}
    \mathcal{U}_{g_x} = \tau_0 \otimes \tau_1~ \mathcal{L}_{G_y/2} \hat{m}_x,
\end{equation}
where $\hat{m}_x$ denotes the mirror operator inverting $k_x$, and $\mathcal{L}_{G_y/2}$ implements a half-translation in reciprocal space along the $G_y$ direction.

The $Pg$ symmetry condition requires
\begin{equation}
    \mathcal{U}_{g_x}(\bm{k})~ \mathcal{H}(\bm{k})~ \mathcal{U}^{\dagger}_{g_x}(\bm{k}) = \mathcal{H}(g_x \bm{k}).
\end{equation}
In Ref.~\cite{chen2022brillouin}, the Hamiltonian $\H(\bm{k})$ is constructed as $\H(\bm{k}) = \H_0(\bm{k}) + \H_1(\bm{k})$.
The first part $\mathcal{H}_0$ reads
\begin{equation}
    \H_0(\bm{k}) =
    \begin{bmatrix}
        \varepsilon & [q^x_1(k_x)]^* & [q^y_+(k_y)]^* & 0 \\
        q^x_1(k_x)  & \varepsilon & 0 & [q^y_-(k_y)]^* \\
        q^y_+(k_y)  & 0 & -\varepsilon & [q^x_2(k_x)]^* \\
        0 & q^y_-(k_y) & q^x_2(k_x) & -\varepsilon \\
    \end{bmatrix},
\end{equation}
where $q^x_a(k_x) = t^x_{a1} + t^x_{a2} e^{ik_x}$ for $a = 1,2$, $q^y_{\pm}(k_y) = t^y_1 \pm t^y_2 e^{ik_y}$, $\pm\varepsilon$ denote the on-site energies.
The second term $\mathcal{H}_1$ is used to break time-reversal symmetry to obtain the most general form
\begin{equation}
    \mathcal{H}_1(\bm{k}) = \lambda \cos k_y~ \tau_1 \otimes \sigma_2 + \lambda \sin k_y~ \tau_2 \otimes \sigma_2.
\end{equation}

\begin{figure}[htbp]
    \centering
    \includegraphics[width=0.4\linewidth]{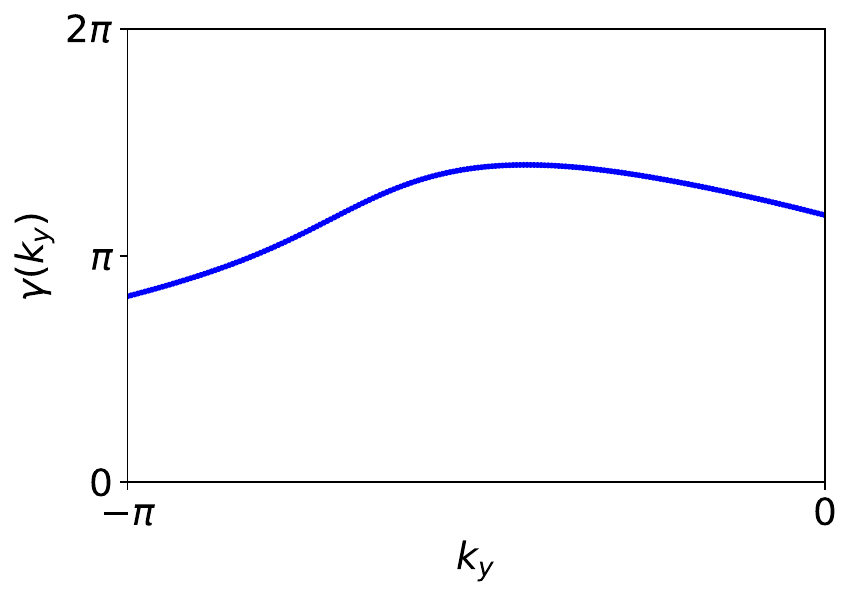}
    \caption{Wilson loop $\gamma(k_y)$ of the valence bands for the given parameters. When $\gamma(k_y)$ crosses $\pi$ an odd number of times on $k_y\in[-\pi,0)$, the $\mathbb{Z}_2$ topological number is nontrivial.}\label{wilson2band}
\end{figure}

Now we compute the topological invariant
\begin{equation}\label{Klein-invariant}
    \nu = \phi_{g_x}(-\pi,0) + \phi_{l_x}(\pi,-\pi) - \phi_{g_x}(\pi,0) + \phi_{l_x}(\pi,0) \mod 2,
\end{equation}
associated with the two valence bands of $\mathcal{H}(\bm{k})$ in the following parameter set
\begin{equation}
    t_{11}^x = t_{22}^x = 1,\quad t_{12}^x = t_{21}^x = 3.5,\quad t_1^y = 2,\quad t_2^y = 1.5,\quad \varepsilon = 1,\quad \lambda = 1.
\end{equation}
For these parameters, the two valence bands exhibit a nontrivial $\mathbb{Z}_2$ topology, as indicated by the Wilson loop in Fig.~\ref{wilson2band}.

To apply Eq.~\eqref{Klein-invariant}, the wavefunction $|\psi(\bm{k})\rangle$ must be continuous over the region $[-\pi, \pi] \times [-\pi, 0]$. We enforce continuity numerically using a gauge-fixing algorithm.
The algorithm first fixes the gauge along the line $k_y = -\pi$ such that $\langle \psi(k_{x,i}, -\pi) | \psi(k_{x,i+1}, -\pi) \rangle$ is real and positive, where $k_{x,i}$ and $k_{x,i+1}$ denote neighboring discrete points along the $k_x$ axis. Then, for each $k_{x,i}$, the gauge is fixed along the $k_y$ direction.

\begin{figure}[htbp]
\centering
\subfigure[]{
\begin{minipage}{0.4\textwidth}\label{phix2band}
\centering
\includegraphics[width=1\linewidth]{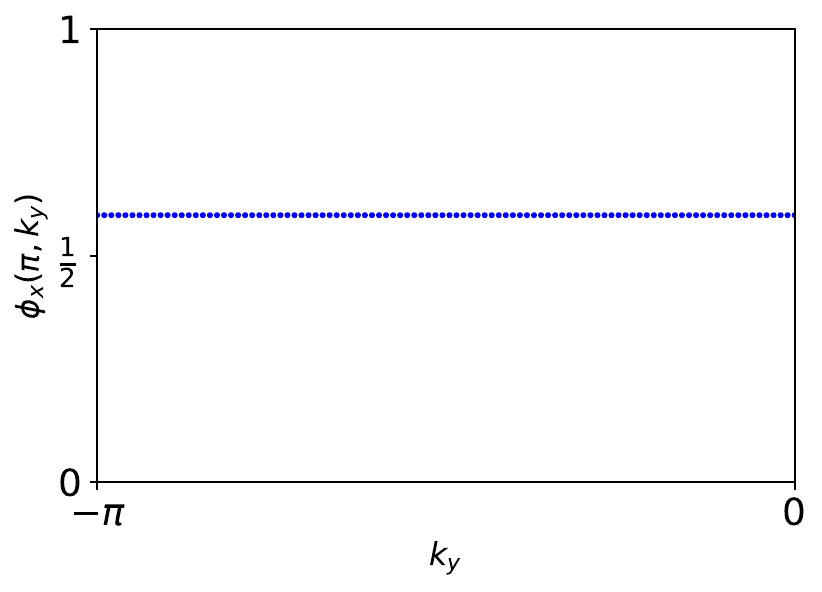} 
\end{minipage}
}
\subfigure[]{
\begin{minipage}{0.4\textwidth}\label{phig2band}
\centering
\includegraphics[width=1\linewidth]{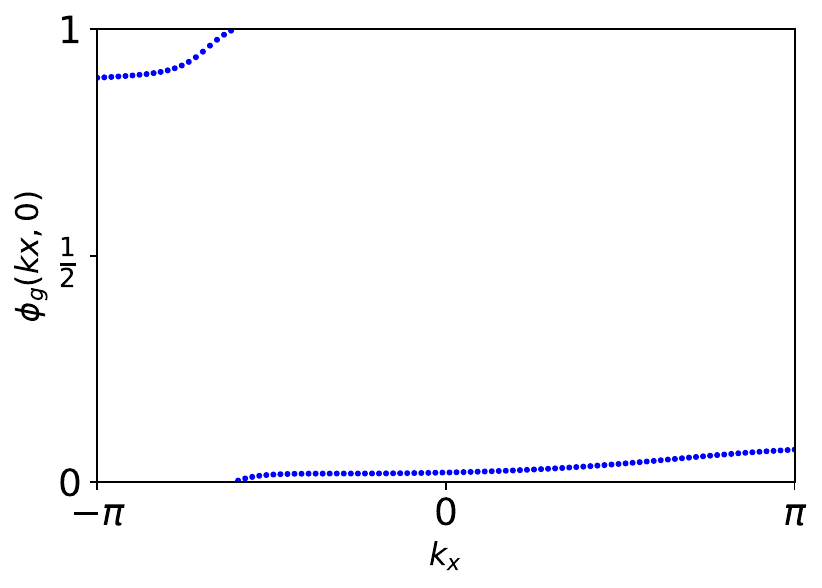} 
\end{minipage}
}
\caption{Continuity of phase functions along symmetry lines: (a) $\phi_{l_x}(k)$ (abbreviated as $\phi_x$) evaluated along $k_x = \pi$; (b) $\phi_{g_x}(k)$ (abbreviated as $\phi_g$) evaluated along $k_y = 0$. A discontinuity appears in panel (b), which can be resolved by shifting the right segment of the curve upward by $2\pi$, thereby rendering $\phi_{g_x}(k)$ continuous. Accordingly, the value $\phi_{g_x}(\pi, 0)$ is taken to exceed $2\pi$ in the main text.}
\end{figure}

After gauge fixing, we evaluate the phase factors in Eq.~\eqref{Klein-invariant}, given by
\begin{equation}\label{phi}
\begin{split}
    e^{2\pi i\phi_{l_x}(\bm{k})} &= \det \mathcal{U}_{l_x}(\bm{k}), \\
    e^{2\pi i\phi_{g_x}(\bm{k})} &= \det \mathcal{U}_{g_x}(\bm{k}).
\end{split}
\end{equation}
Since the phases appear in the exponent, their values are defined modulo $2\pi$. However, this ambiguity does not imply arbitrary choices; according to the definition, $\phi_{l_x}(\bm{k})$ and $\phi_{g_x}(\bm{k})$ must be taken to be continuous across momentum space.
For calculating the topological invariant, we require $\phi_{l_x}(\bm{k})$ to be continuous along $k_x = \pi$, which connects $(\pi, 0)$ and $(\pi, -\pi)$, and $\phi_{g_x}(\bm{k})$ to be continuous along $k_y = 0$, connecting $(-\pi, 0)$ and $(\pi, 0)$. The continuity of $\phi_{l_x}(\bm{k})$ along $k_x = \pi$ is shown in Fig.~\ref{phix2band}, and that of $\phi_{g_x}(\bm{k})$ along $k_y = 0$ is illustrated in Fig.~\ref{phig2band}. Note that in Fig.~\ref{phig2band}, the phase is plotted within the interval $[0, 2\pi)$, resulting in apparent discontinuities. These can be eliminated by shifting the right-hand portion of the curve upward by $2\pi$.

Numerical evaluation yields the following results
\begin{equation} 
\begin{split} 
    \phi_{g_x}(-\pi, 0) &= 0.892887808645513,\\ 
    \phi_{l_x}(\pi, -\pi) &= 0.5895697952404985, \\ 
    \phi_{g_x}(\pi, 0) &= 1.0720273991265099, \\
    \phi_{l_x}(\pi, 0) &= 0.5895697952404985,
\end{split} 
\end{equation} 
which indicates a nontrivial topological invariant of $\nu = 1$, in agreement with the Wilson loop analysis and previous results \cite{chen2022brillouin}.

\subsection{Topological invariants for $P2_1/c$}
The MCG $P2_1/c$ exhibits point group symmetry $C_{2h}$, generated by a screw rotation $s_y$ and a glide reflection $g_y$, defined as
\begin{equation}\label{P21c-generator}
    \begin{split}
        s_y(k_x, k_y, k_z) &= (-k_x, k_y + \pi, -k_z), \\
        g_y(k_x, k_y, k_z) &= (k_x, -k_y, k_z + \pi).
    \end{split}
\end{equation}

The BZ is spanned by the reciprocal lattice vectors
\begin{equation}
    \bm{b}_x = 2\pi(1,0,0), ~\bm{b}_y = 2\pi(0,1,0), ~\bm{b}_z = 2\pi(0,0,1).
\end{equation}

Based on the isomorphism established in the main text, the topological invariants associated with $P2_1/c$ are classified by the group cohomology
\begin{equation}\label{P21c-classify}
    H^2(P2_1/c, \mathbb{Z}) = \mathbb{Z}_2^2 \oplus \mathbb{Z}_4 \oplus \mathbb{Z}.
\end{equation}

\subsubsection{Chern number}
The presence of glide reflection symmetry $g_y$ imposes constraints that force the Chern numbers in the $k_x$–$k_y$ and $k_y$–$k_z$ planes to vanish. However, the Chern number in the $k_x–k_z$ plane remains unconstrained, thereby contributing to the topological classification. This invariant corresponds to the $\mathbb{Z}$ term in Eq.~\eqref{P21c-classify}, and originates from the algebraic relation
\begin{equation}
    [l_x, l_z] = 0.
\end{equation}
Specifically, the difference between the following integers,
\begin{equation}
\begin{split}
    N(l_x, l_z) &= \phi_{l_z}(l_x^{-1}\bm{k}) - \phi_{l_xl_z}(\bm{k}) + \phi_{l_x}(\bm{k}), \\
    N(l_z, l_x) &= \phi_{l_x}(l_z^{-1}\bm{k}) - \phi_{l_zl_x}(\bm{k}) + \phi_{l_z}(\bm{k}),
\end{split}
\end{equation}
defines the topological invariant
\begin{equation}
\begin{split}
    \nu_1 &= N(l_x, l_z) - N(l_z, l_x) \\
    &= \phi_{l_x}(\bm{k}) + \phi_{l_z}(l_x^{-1}\bm{k}) - \phi_{l_x}(l_z^{-1}\bm{k}) - \phi_{l_z}(\bm{k}).
\end{split}
\end{equation}

\begin{figure}[htbp]
    \centering
    \includegraphics[width=0.5\linewidth]{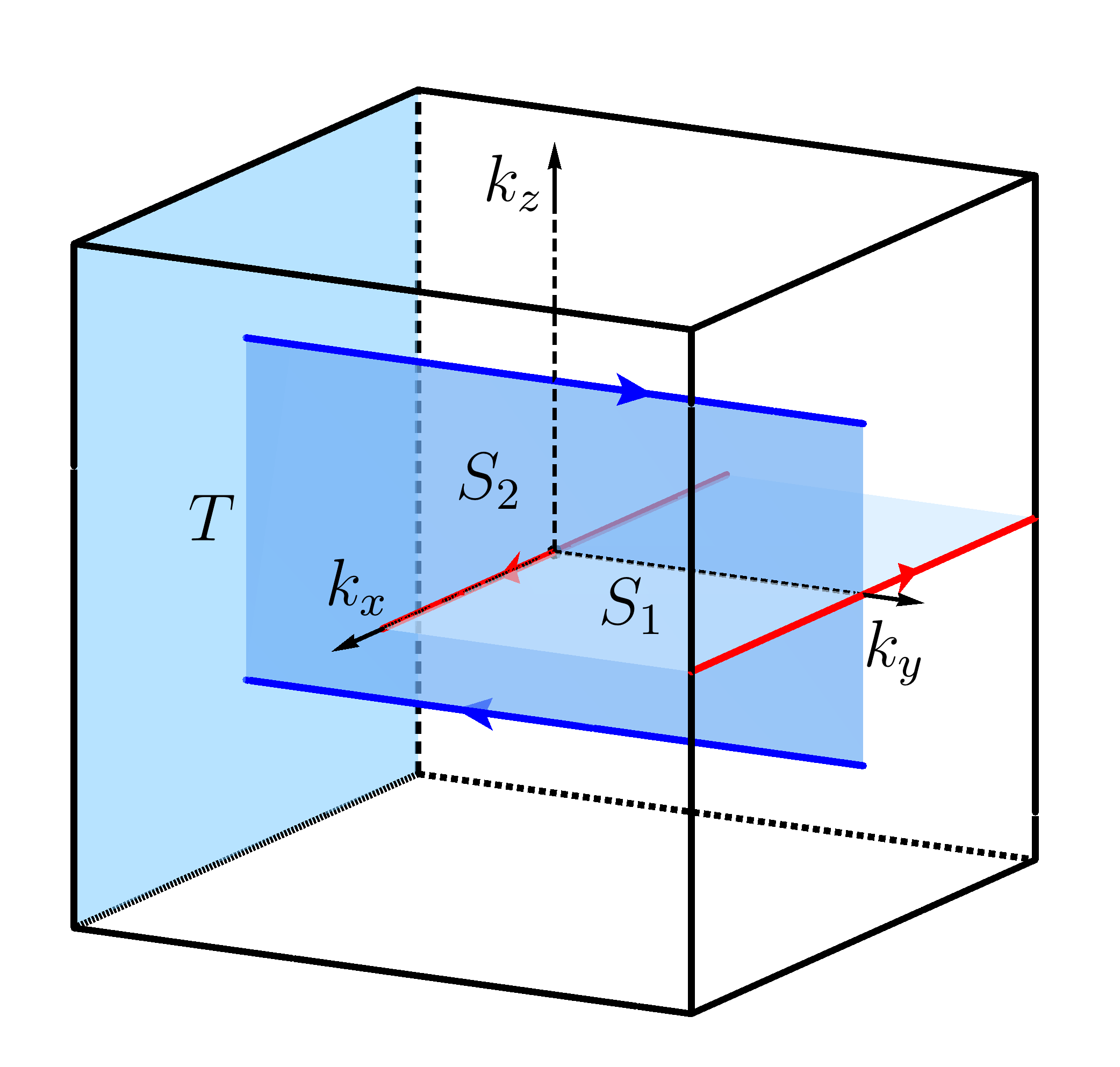}
    \caption{Brillouin zone of momentum-space crystallographic group $P2_1/c$. The region $T = [-\pi, -\pi, -\pi] \times [\pi, -\pi, \pi]$ forms a two-dimensional torus topologically, and the integral of the Berry curvature over $T$ yields the Chern number $\nu_1$. The segment $S_1$ is topologically equivalent to a Klein bottle and contributes to the invariant $\nu_4$. Its boundaries, highlighted in red, are identified under the screw rotation symmetry $s_y$. Similarly, the segment $S_2$ is also a Klein bottle, with boundaries related by either $s_y$ or the glide reflection $g_y$.}
    \label{P21c}
\end{figure}

Equivalently, $\nu_1$ can be expressed as the integral of the Berry curvature over the $k_x$–$k_z$ plane
\begin{equation}
\begin{split}
    \nu_1 &= \phi_{l_x}(\pi,-\pi,\pi) + \phi_{l_z}(-\pi,-\pi,\pi) - \phi_{l_x}(\pi,-\pi,-\pi) - \phi_{l_z}(\pi,-\pi,\pi) \\
    &= -\int_{-\pi}^{\pi} dk_z~ \partial_{k_z}\phi_{l_x}(\pi,-\pi,k_z) - \int_{-\pi}^{\pi} dk_x~ \partial_{k_x}\phi_{l_z}(k_x,-\pi,\pi) \\
    &= -\frac{1}{2\pi} \int_{-\pi}^{\pi} dk_z~ \mathcal{A}_z(\pi,-\pi,k_z) + \frac{1}{2\pi} \int_{-\pi}^{\pi} dk_z~ \mathcal{A}_z(-\pi,-\pi,k_z) \\
    &\quad -\frac{1}{2\pi} \int_{-\pi}^{\pi} dk_x~ \mathcal{A}_x(k_x,-\pi,\pi) + \frac{1}{2\pi} \int_{-\pi}^{\pi} dk_x~ \mathcal{A}_x(k_x,-\pi,-\pi) \\
    &= \frac{1}{2\pi} \int_{-\pi}^{\pi} dk_z \int_{-\pi}^{\pi} dk_x~ \mathcal{F}(k_x,-\pi,k_z) \\
    &= \frac{1}{2\pi} \int_T \mathcal{F}(\bm{k}),
\end{split}
\end{equation}
where $T$ denotes a two-dimensional torus embedded in the BZ, as illustrated in Fig.~\ref{P21c}. In deriving this result, we utilized the following identities
\begin{equation}
\begin{split}
    \mathcal{A}_z(\bm{k}) - \mathcal{A}_z(l_x^{-1}\bm{k}) &= 2\pi~ \partial_{k_z} \phi_{l_x}(\bm{k}), \\
    \mathcal{A}_x(\bm{k}) - \mathcal{A}_x(l_z^{-1}\bm{k}) &= 2\pi~ \partial_{k_x} \phi_{l_z}(\bm{k}).
\end{split}
\end{equation}

\subsubsection{Invariants arising from little co-group representations}
From Eq.~\eqref{P21c-generator}, we observe that
\begin{equation}
    s_yg_y(k_x, k_y, k_z) = (-k_x, -k_y + \pi, -k_z - \pi),
\end{equation}
which leads to the algebraic relation
\begin{equation}
    (s_yg_y)^2 = 1.
\end{equation}
Motivated by this, we consider the integer
\begin{equation}
\begin{split}
    N(s_yg_y, s_yg_y) &= \phi_{s_yg_y}((s_yg_y)^{-1}\bm{k}) - \phi_1(\bm{k}) + \phi_{s_yg_y}(\bm{k}) \\
    &= \phi_{s_yg_y}(\bm{k}) + \phi_{s_yg_y}(s_yg_y\bm{k}),
\end{split}
\end{equation}
which gives rise to the topological invariant
\begin{equation}\label{P21c-rep-1}
    \nu_2 = \phi_{s_yg_y}(\bm{k}) + \phi_{s_yg_y}(s_yg_y\bm{k}) \mod 2.
\end{equation}
Evaluating at the high-symmetry point $\bm{k} = (0, \frac{\pi}{2}, -\frac{\pi}{2})$, we obtain
\begin{equation}
    \nu_2 = 2\phi_{s_yg_y}(0, \frac{\pi}{2}, -\frac{\pi}{2}) \mod 2.
\end{equation}
This indicates that $\nu_2$ captures the parity of the representation of the little co-group $\{1, s_yg_y\}$.

Additionally, there exists another algebraic relation, inequivalent to Eq.~\eqref{P21c-rep-1},
\begin{equation}
    s_yg_y^{-1}(k_x, k_y, k_z) = (-k_x, -k_y + \pi, -k_z + \pi),
\end{equation}
which also satisfies
\begin{equation}
    (s_yg_y^{-1})^2 = 1.
\end{equation}
Following analogous steps, we define the invariant:
\begin{equation}
    \nu_3 = \phi_{s_yg_y^{-1}}(\bm{k}) + \phi_{s_yg_y^{-1}}(s_yg_y^{-1}\bm{k}) \mod 2.
\end{equation}
Evaluating at $\bm{k} = (0, \frac{\pi}{2}, \frac{\pi}{2})$ yields
\begin{equation}
    \nu_3 = 2\phi_{s_yg_y^{-1}}(0, \frac{\pi}{2}, \frac{\pi}{2}) \mod 2,
\end{equation}
which likewise originates from the representation of the little co-group $\{1, s_yg_y^{-1}\}$.

\subsubsection{Invariant arising from screw rotation}
Analogous to the Klein bottle case, where the momentum-space nonsymmorphic operation of glide reflection gives rise to a new topological invariant, the screw rotation symmetry also induces a distinct invariant. Consider the algebraic relation
\begin{equation}
    s_y l_x = l_x^{-1} s_y.
\end{equation}
To explore the associated topology, we introduce the following three integers
\begin{equation}
\begin{split}
    N(s_y, l_x) &= \phi_{l_x}(s_y^{-1}\bm{k}) - \phi_{s_y l_x}(\bm{k}) + \phi_{s_y}(\bm{k}),\\
    N(l_x^{-1}, s_y) &= \phi_{s_y}(l_x \bm{k}) - \phi_{l_x^{-1} s_y}(\bm{k}) + \phi_{l_x^{-1}}(\bm{k}),\\
    N(l_x, l_x^{-1}) &= \phi_{l_x^{-1}}(\bm{k}) + \phi_{l_x}(l_x \bm{k}),
\end{split}
\end{equation}
where, in the final expression, we substitute $\bm{k} \rightarrow l_x \bm{k}$.
Using these terms, we construct the topological invariant
\begin{equation}
\begin{split}
    \nu_4 &= N(s_y, l_x) - N(l_x^{-1}, s_y) + N(l_x, l_x^{-1}) \\
    &= \phi_{s_y}(\bm{k}) + \phi_{l_x}(s_y^{-1}\bm{k}) - \phi_{s_y}(l_x \bm{k}) + \phi_{l_x}(l_x \bm{k}) \mod 2.
\end{split}
\end{equation}

This expression admits a more geometric reformulation. Assuming that the valence bands are periodic along the $k_x$ direction, we have
\begin{equation}
    \nu_4 = \phi_{s_y}(\bm{k}) - \phi_{s_y}(l_x \bm{k}) \mod 2.
\end{equation}

Evaluating at the point $\bm{k} = (-\pi, \pi, 0)$,
\begin{equation}
\begin{split}
    &\phi_{s_y}(-\pi, \pi, 0) - \phi_{s_y}(\pi, \pi, 0) \\=& -\int_{-\pi}^{\pi} dk_x~ \partial_{k_x} \phi_{s_y}(k_x, \pi, 0) \\
    =& -\frac{1}{2\pi} \int_{-\pi}^{\pi} dk_x~ \mathcal{A}_x(k_x, \pi, 0) - \frac{1}{2\pi} \int_{-\pi}^{\pi} dk_x~ \mathcal{A}_x(-k_x, 0, 0) \\
    =& -\frac{1}{2\pi} \int_{-\pi}^{\pi} dk_x~ \mathcal{A}_x(k_x, \pi, 0) - \frac{1}{2\pi} \int_{-\pi}^{\pi} dk_x~ \mathcal{A}_x(k_x, 0, 0) \\
    =& \frac{1}{2\pi} \int_{-\pi}^{\pi} dk_x~ [\mathcal{A}_x(k_x, 0, 0) - \mathcal{A}_x(k_x, \pi, 0)] - \frac{1}{\pi} \int_{-\pi}^{\pi} dk_x~ \mathcal{A}_x(k_x, 0, 0) \\
    =& -\frac{1}{2\pi} \int_{-\pi}^{\pi} dk_x \int_{0}^{\pi} dk_y~ \partial_{k_y} \mathcal{A}_x(k_x, k_y, 0) - \frac{1}{\pi} \int_{-\pi}^{\pi} dk_x~ \mathcal{A}_x(k_x, 0, 0).
\end{split}
\end{equation}
Here, we used the identity
\begin{equation}
    \mathcal{A}_x(\bm{k}) + \mathcal{A}_x(s_y^{-1} \bm{k}) = 2\pi \partial_{k_x} \phi_{s_y}(\bm{k}).
\end{equation}

Using the definition of the Berry curvature
\begin{equation}
    \mathcal{F}_{xy}(\bm{k}) = \partial_{k_x} \mathcal{A}_y(\bm{k}) - \partial_{k_y} \mathcal{A}_x(\bm{k}),
\end{equation}
and with the assumption of $k_x$ periodicity, we can replace $-\partial_{k_y}\mathcal{A}_x(k_x,k_y,0)$ with $\mathcal{F}_{xy}(k_x,k_y,0)$, and arrive at the final expression
\begin{equation}
    \nu_4 = \frac{1}{2\pi} \int_{-\pi}^{\pi} dk_x \int_{0}^{\pi} dk_y~ \mathcal{F}_{xy}(k_x, k_y, 0) - \frac{1}{\pi} \int_{-\pi}^{\pi} dk_x~ \mathcal{A}_x(k_x, 0, 0) \mod 2.
\end{equation}

This procedure is visualized in Fig.~\ref{P21c}, where the segment $S_1$ and its boundaries form a Klein bottle induced by screw rotation $s_y$.

\subsubsection{The reduced $\mathbb{Z}_2$ topological invariant}
Before constructing the final topological invariant, we address a seeming contradiction between the invariants $\nu_2$, $\nu_3$, $\nu_4$ and the classification result in Eq.~\eqref{P21c-classify}. Although our formalism produces three $\mathbb{Z}_2$ invariants, the classification contains only two $\mathbb{Z}_2$ factors. The resolution lies in recognizing that the $\mathbb{Z}_4$ term in the classification encodes both a little co-group representation and a Klein-bottle-type topological invariant.

To illustrate this, we construct the final invariant using the algebraic relation
\begin{equation}
    g_y l_y = l_y^{-1} g_y.
\end{equation}
Following previous methods, we define the $\mathbb{Z}_4$ invariant
\begin{equation}
    \nu_5 = \phi_{g_y}(\bm{k}) + \phi_{l_y}(g_y^{-1} \bm{k}) - \phi_{g_y}(l_y \bm{k}) + \phi_{l_y}(l_y \bm{k}) \mod 4.
\end{equation}
To verify that $\nu_5$ is valued in $\mathbb{Z}_4$ rather than $\mathbb{Z}_2$, consider that
\begin{equation}
    s_y^2 = l_y,
\end{equation}
and under the gauge transformation $\phi_{s_y} \rightarrow \phi_{s_y}+m$ for any integer $m$, the invariant transforms as
\begin{equation}
    \nu_5 \rightarrow \nu_5 + 4m.
\end{equation}

This resolves the earlier contradiction: $\nu_2$, $\nu_3$, and $\nu_5$ are not linearly independent but instead satisfy the constraint
\begin{equation}
    \nu_2 + \nu_3 = \nu_5.
\end{equation}
Therefore, the parity of $\nu_5$ restricts $\nu_2$ and $\nu_3$, reducing the count of independent $\mathbb{Z}_2$ invariants.
To isolate the genuinely $\mathbb{Z}_2$ component within $\nu_5$, we assume an even parity constraint on $\nu_5$. This yields a reduced $\mathbb{Z}_2$ invariant as
\begin{equation}
    \tilde{\nu}_5 = \phi_{g_y}(\bm{k}) + \phi_{l_y}(g_y^{-1} \bm{k}) - \phi_{g_y}(l_y \bm{k}) + \phi_{l_y}(l_y \bm{k}) \mod 2.
\end{equation}

This invariant admits a geometric interpretation. Assuming that the valence bands are periodic along the $k_y$ direction, we define
\begin{equation} 
    \tilde{\nu}_5 = \phi_{g_y}(\bm{k}) - \phi_{g_y}(l_y \bm{k}) \mod 2.
\end{equation} 
Evaluating at $\bm{k} = (0, -\pi, \frac{\pi}{2})$ yields 
\begin{equation} 
    \tilde{\nu}_5 = \phi_{g_y}(0, -\pi, \frac{\pi}{2}) - \phi_{g_y}(0, \pi, \frac{\pi}{2}) \mod 2. 
\end{equation}

We compute the difference as
\begin{equation}
\begin{split}
    &\phi_{g_y}(0, -\pi, \frac{\pi}{2}) - \phi_{g_y}(0, \pi, \frac{\pi}{2}) 
    \\=& -\int_{-\pi}^{\pi} dk_y~ \partial_{k_y} \phi_{g_y}(0, k_y, \frac{\pi}{2}) \\
    =& -\frac{1}{2\pi} \int_{-\pi}^{\pi} dk_y~ \mathcal{A}_y(0, k_y, \frac{\pi}{2}) - \frac{1}{2\pi} \int_{-\pi}^{\pi} dk_y~ \mathcal{A}_y(0, -k_y, -\frac{\pi}{2}) \\
    =& -\frac{1}{2\pi} \int_{-\pi}^{\pi} dk_y~ \mathcal{A}_y(0, k_y, \frac{\pi}{2}) - \frac{1}{2\pi} \int_{-\pi}^{\pi} dk_y~ \mathcal{A}_y(0, k_y, -\frac{\pi}{2}) \\
    =& \frac{1}{2\pi} \int_{-\pi}^{\pi} dk_y~ \left[\mathcal{A}_y(0, k_y, -\frac{\pi}{2}) - \mathcal{A}_y(0, k_y, \frac{\pi}{2})\right] - \frac{1}{\pi} \int_{-\pi}^{\pi} dk_y~ \mathcal{A}_y(0, k_y, -\tfrac{\pi}{2}) \\
    =& -\frac{1}{2\pi} \int_{-\pi}^{\pi} dk_y \int_{-\pi/2}^{\pi/2} dk_z~ \partial_{k_z} \mathcal{A}_y(0, k_y, k_z) - \frac{1}{\pi} \int_{-\pi}^{\pi} dk_y~ \mathcal{A}_y(0, k_y, -\frac{\pi}{2}).
\end{split}
\end{equation}
This derivation uses the identity
\begin{equation}
    \mathcal{A}_y(\bm{k}) + \mathcal{A}_y(g_y^{-1} \bm{k}) = 2\pi \partial_{k_y} \phi_{g_y}(\bm{k}).
\end{equation}

Finally, due to periodicity along the $k_y$ direction, we can replace $-\partial_{k_z}\mathcal{A}_y(0,k_y,k_z)$ with the Berry curvature $\mathcal{F}(0,k_y,k_z)$. The topological invariant is then simplified into
\begin{equation}
    \tilde{\nu}_5 = \frac{1}{2\pi} \int_{-\pi}^{\pi} dk_y \int_{-\pi/2}^{\pi/2} dk_z~ \mathcal{F}_{yz}(0, k_y, k_z) - \frac{1}{\pi} \int_{-\pi}^{\pi} dk_y~ \mathcal{A}_y(0, k_y, -\frac{\pi}{2}) \mod 2.
\end{equation}
This expression is associated with the segment $S_2$ as depicted in Fig.~\ref{P21c}.

\subsection{Topological invariants for $I222$}
The momentum-space crystallographic group $I222$ is characterized by the point group $D_2$, generated by rotations $r_x$ and $r_y$,
\begin{equation}
    r_x(k_x,k_y,k_z) = (k_x,-k_y,-k_z),~r_y(k_x,k_y,k_z) = (-k_x,k_y,-k_z).
\end{equation}
The BZ is spanned by
\begin{equation}\label{I222-basis-1}
    l_a = \pi(-1,1,1), l_b = \pi(1,-1,1), l_c = \pi(1,1,-1).
\end{equation}

The topological classification in this case is given by
\begin{equation}\label{I222-classify}
    H^2(I222,\mathbb{Z}) = \mathbb{Z}_2^2 \oplus \mathbb{Z}_4.
\end{equation}

\subsubsection{Invariants arising from little co-group representations}
In the classification Eq.~\eqref{I222-classify}, the two $\mathbb{Z}_2$ factors arise from the algebraic relations $r_x^2 = r_y^2 = 1$. To illustrate this, consider the relation $r_x^2 = 1$. The corresponding integer
\begin{equation}
    \begin{split}
        N(r_x,r_x) &= \phi_{r_x}(r_x^{-1}\bm{\k}) - \phi_{1}(\bm{k}) + \phi_{r_x}(\bm{k})\\
        &=\phi_{r_x}(\bm{k}) + \phi_{r_x}(r_x\bm{k}),
    \end{split}
\end{equation}
defines the $\mathbb{Z}_2$ cohomological invariant
\begin{equation}
    \nu_1 = \phi_{r_x}(\bm{k}) + \phi_{r_x}(r_x\bm{k}) \mod 2.
\end{equation}
Evaluating at $\bm{k} = (k_x, 0, 0)$, we obtain
\begin{equation}
    \nu_1 = 2\phi_{r_x}(k_x, 0, 0) \mod 2,
\end{equation}
which corresponds to the parity of the representation of the little co-group $\{1, r_x\}$ at the high symmetry line $(k_x,0,0)$.

Analogously, we have the $\mathbb{Z}_2$ invariant
\begin{equation}
    \nu_2 = \phi_{r_y}(\bm{k}) + \phi_{r_y}(r_y\bm{k}) \mod 2,
\end{equation}
which is derived from the algebraic relation $r_y^2 = 1$. This invariant arises from the representation of the little co-group $\{1, r_y\}$ along the high-symmetry line $(0,k_y,0)$.

Apart from these two invariants, there exists another $\mathbb{Z}_2$ invariant from a little co-group representation. To illustrate this, we adopt the momentum-space translation basis
\begin{equation}\label{I222-basis-2}
    l_x = \pi(2,0,0), \quad l_y = \pi(0,2,0), \quad l_d = \pi(1,1,1),
\end{equation}
instead of the one provided in Eq.~\eqref{I222-basis-1}. It is straightforward to verify that this choice is equivalent to Eq.~\eqref{I222-basis-1}.

Observe that
\begin{equation}
    r_xl_y(k_x, k_y, k_z) = (k_x, -k_y - 2\pi, -k_z),
\end{equation}
which leads to the algebraic relation
\begin{equation}
    (r_xl_y)^2 = 1.
\end{equation}
Using the same method as for $r_x^2 = 1$, we construct the topological invariant
\begin{equation}
    \nu_3 = \phi_{r_xl_y}(\bm{k}) + \phi_{r_xl_y}(r_xl_y\bm{k}) \mod 2.
\end{equation}
Evaluating at $\bm{k} = (k_x, -\pi, 0)$, we find that $\nu_3$ corresponds to the representation parity of the little co-group $\{1, r_xl_y\}$ at the high symmetry line $(k_x, -\pi, 0)$.

\subsubsection{The reduced $\mathbb{Z}_2$ topological invariant}
Before proceeding, note that we have constructed three $\mathbb{Z}_2$ invariants, whereas the classification in Eq.~\eqref{I222-classify} contains only two $\mathbb{Z}_2$ factors. This discrepancy is reminiscent of the $P2_1/c$ example, where two $\mathbb{Z}_2$ invariants are linked through a $\mathbb{Z}_4$ invariant. However, the situation here is subtly different. In the previous case, the mixing phenomenon arose from the presence of nonsymmorphic operations. In contrast, $I222$ represents a special instance where such mixing is permitted even within a symmorphic structure. Let us explore the origin of this phenomenon.

The algebraic relation
\begin{equation}\label{I222-Z4-algebraic}
    r_x l_y = l_y^{-1} r_x
\end{equation}
motivates the construction of the following group cohomology integers
\begin{equation}
\begin{split}
    N(r_x, l_y) &= \phi_{l_y}(r_x^{-1}\bm{k}) - \phi_{r_x l_y}(\bm{k}) + \phi_{r_x}(\bm{k}), \\
    N(l_y^{-1}, r_x) &= \phi_{r_x}(l_y\bm{k}) - \phi_{l_y^{-1} r_x}(\bm{k}) + \phi_{l_y^{-1}}(\bm{k}), \\
    N(l_y, l_y^{-1}) &= \phi_{l_y^{-1}}(\bm{k}) + \phi_{l_y}(l_y\bm{k}).
\end{split}
\end{equation}
In the last expression, we evaluate $\bm{k}$ at $l_y\bm{k}$ to simplify the construction of the topological invariant.
Using these relations, we obtain the invariant
\begin{equation}
\begin{split}
    \nu_4 &= N(r_x, l_y) - N(l_y^{-1}, r_x) + N(l_y, l_y^{-1}) \\
        &= \phi_{r_x}(\bm{k}) + \phi_{l_y}(r_x^{-1}\bm{k}) - \phi_{r_x}(l_y\bm{k}) + \phi_{l_y}(l_y\bm{k}) \mod 4.
\end{split}
\end{equation}

This invariant $\nu_4$ is valued in $\mathbb{Z}_4$ rather than $\mathbb{Z}_2$. To demonstrate this, consider the algebraic relation
\begin{equation}\label{I222-ly}
    l_y = r_y l_d r_y^{-1} l_d,
\end{equation}
which implies that under the gauge transformation $\phi_{l_d} \rightarrow \phi_{l_d}+m$, the invariant transforms as
\begin{equation}
    \nu_4 \rightarrow \nu_4 + 4m.
\end{equation}
This confirms that $\nu_4$ changes in increments of 4, establishing its $\mathbb{Z}_4$ character.

The invariant $\nu_4$ incorporates a factor arising from the parity of the little co-group representation. To derive this, we substitute Eq.~\eqref{I222-ly} into Eq.~\eqref{I222-Z4-algebraic}, yielding $(r_z l_d r_y l_d)^2 = 1$. As in previous discussions of little co-groups, this relation gives a $\mathbb{Z}_2$ invariant, corresponding to the one-dimensional representations of the little co-group $\{1, r_z l_d r_y l_d\}$. This $\mathbb{Z}_2$ factor constitutes one component of the full $\mathbb{Z}_4$ invariant.

To isolate the purely topological part of the invariant, we insist that the little co-group representation is trivial and consider
\begin{equation}
    r_x \left(\frac{l_y}{2}\right) r_x^{-1} = \left(\frac{l_y}{2}\right)^{-1},
\end{equation}
which allows us to quotient out the little co-group component and focus on the reduced $\mathbb{Z}_2$ invariant
\begin{equation}
    \tilde{\nu}_4 = \phi_{r_x}(\bm{k}) + \phi_{l_y/2}(r_x \bm{k}) - \phi_{r_x}\left(\frac{l_y}{2} \bm{k}\right) + \phi_{l_y/2}\left(\frac{l_y}{2} \bm{k}\right) \mod 2.
\end{equation}

\begin{figure}[htbp]
    \centering
    \includegraphics[width=0.5\linewidth]{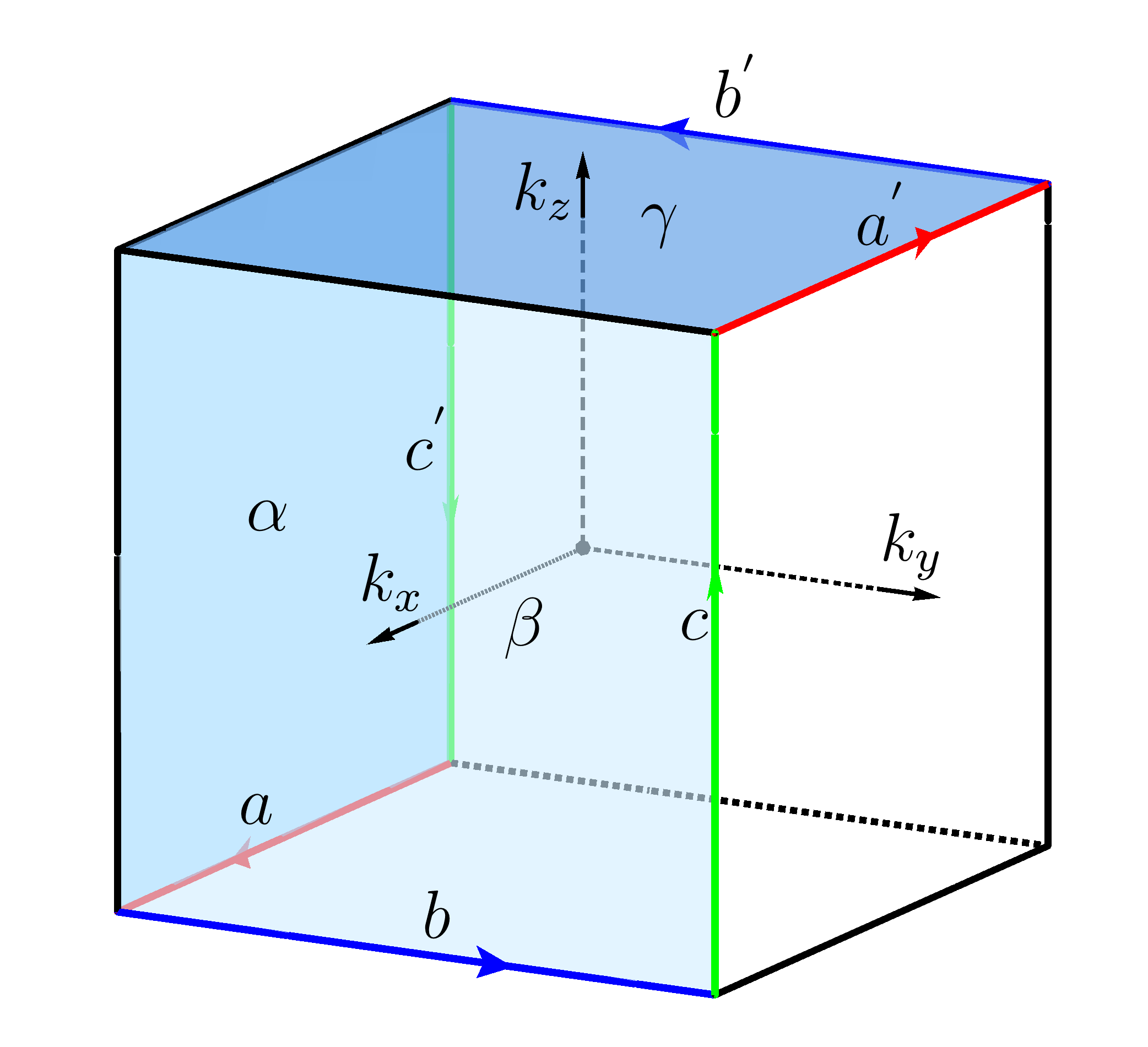}
    \caption{A quarter of the Brillouin zone for momentum-space crystallographic group $I222$. The combined path $a + b + c + a^{\prime} + b^{\prime} + c^{\prime}$ serves as the boundary enclosing the region $\alpha + \beta + \gamma$. Segments $a$ and $a^{\prime}$, $b$ and $b^{\prime}$, and $c$ and $c^{\prime}$ are symmetry-related via rotation operations, as indicated by their matching colors.}
    \label{I222-BZ}
\end{figure}

Assuming the valence bands are periodic along the $k_y$ direction, we evaluate this expression at
\begin{equation}
    \tilde{\nu}_4 = \phi_{r_x} \left( \frac{\pi}{2}, -\frac{\pi}{2}, -\frac{\pi}{2} \right) - \phi_{r_x} \left( \frac{\pi}{2}, \frac{\pi}{2}, -\frac{\pi}{2} \right).
\end{equation}
This phase difference can be reformulated as an integral
\begin{equation}
\begin{split}
    \tilde{\nu}_4 &= - \int_{-\pi/2}^{\pi/2} dk_y~ \partial_{k_y} \phi_{r_x} \left( \frac{\pi}{2}, k_y, -\frac{\pi}{2} \right) \\
    &= -\frac{1}{2\pi} \int_{-\pi/2}^{\pi/2} dk_y~ \mathcal{A}_y \left( \frac{\pi}{2}, k_y, \frac{\pi}{2} \right) 
       -\frac{1}{2\pi} \int_{-\pi/2}^{\pi/2} dk_y~ \mathcal{A}_y \left( \frac{\pi}{2}, -k_y, -\frac{\pi}{2} \right) \\
    &= -\frac{1}{2\pi} \int_{-\pi/2}^{\pi/2} dk_y~ \mathcal{A}_y \left( \frac{\pi}{2}, k_y, \frac{\pi}{2} \right) 
       -\frac{1}{2\pi} \int_{-\pi/2}^{\pi/2} dk_y~ \mathcal{A}_y \left( \frac{\pi}{2}, k_y, -\frac{\pi}{2} \right) \\
    &= \frac{1}{2\pi} \int_{-\pi/2}^{\pi/2} dk_y~ \left[ \mathcal{A}_y \left( \frac{\pi}{2}, k_y, -\frac{\pi}{2} \right) 
     - \mathcal{A}_y \left( \frac{\pi}{2}, k_y, \frac{\pi}{2} \right) \right] 
     - \frac{1}{\pi} \int_{-\pi/2}^{\pi/2} dk_y~ \mathcal{A}_y \left( \frac{\pi}{2}, k_y, -\frac{\pi}{2} \right) \\
    &= - \frac{1}{2\pi} \int_{-\pi/2}^{\pi/2} dk_y \int_{-\pi/2}^{\pi/2} dk_z~ \partial_{k_z} \mathcal{A}_y \left( \frac{\pi}{2}, k_y, k_z \right)
     - \frac{1}{\pi} \int_{-\pi/2}^{\pi/2} dk_y~ \mathcal{A}_y \left( \frac{\pi}{2}, k_y, -\frac{\pi}{2} \right).
\end{split}
\end{equation}
Here, we used the identity
\begin{equation}
    \mathcal{A}_y(\bm{k}) + \mathcal{A}_y(r_x^{-1} \bm{k}) = 2\pi \partial_{k_y} \phi_{r_x}(\bm{k}),
\end{equation}
and invoked periodicity in $k_y$ to replace $\partial_{k_z} \mathcal{A}_y$ with $-\mathcal{F}_{yz}$, giving
\begin{equation}
    \tilde{\nu}_4 = \frac{1}{2\pi}\int_{-\pi/2}^{\pi/2}dk_y\int_{-\pi/2}^{\pi/2}dk_z~\mathcal{F}(\frac{\pi}{2},k_y,k_z)-\frac{1}{\pi}\int_{-\pi/2}^{\pi/2}dk_y~\mathcal{A}_y(\frac{\pi}{2},k_y,-\frac{\pi}{2}).
\end{equation}
Using the notation introduced in Fig.~\ref{I222-BZ}, the invariant $\tilde{\nu}_4$ can be written as 
\begin{equation} 
    \tilde{\nu}_4 = \frac{1}{2\pi} \int_{\beta} \mathcal{F}(\bm{k}) - \frac{1}{\pi} \int_b \mathcal{A}(\bm{k}) \mod 2. 
\end{equation}

Following the same logic but employing alternative symmetry generators, we obtain equivalent expressions
\begin{equation} 
    \tilde{\nu}_4 = \phi_{r_y}(\bm{k}) - \phi_{r_y}\left(\frac{l_z}{2} \bm{k}\right) = \phi_{r_z}(\bm{k}) - \phi_{r_z}\left(\frac{l_x}{2} \bm{k}\right), 
\end{equation} 
which, in integral form, read 
\begin{equation} 
    \tilde{\nu}_4 = \frac{1}{2\pi} \int_{\alpha} \mathcal{F}(\bm{k}) - \frac{1}{\pi} \int_a \mathcal{A}(\bm{k}) = \frac{1}{2\pi} \int_{\gamma} \mathcal{F}(\bm{k}) - \frac{1}{\pi} \int_c \mathcal{A}(\bm{k}) \mod 2. 
\end{equation}

Since $\tilde{\nu}_4$ is a $\mathbb{Z}_2$ quantity, multiplying three such contributions yields no change, allowing for the consolidated expression
\begin{equation} 
    \tilde{\nu}_4 = \frac{1}{2\pi} \int_{\alpha + \beta + \gamma} \mathcal{F}(\bm{k}) - \frac{1}{\pi} \int_{a + b + c} \mathcal{A}(\bm{k}) \mod 2.\end{equation}

Finally, in the $N$-band case, as discussed in the main text, we can take the trace to obtain the Abelian topological invariant
\begin{equation}\label{I222-nonAbelian}
    \tilde{\nu}_4 = \frac{1}{2\pi} \int_{\alpha + \beta + \gamma} \mathrm{tr}~\mathcal{F}(\bm{k})
        - \frac{1}{\pi} \int_{a + b + c} \mathrm{tr}~\mathcal{A}(\bm{k}) \mod 2.
\end{equation}

This reduced $\mathbb{Z}_2$ topological invariant has been previously identified in the literature \cite{shiozaki2016topology,shiozaki2022atiyah}. By examining a quarter of the Brillouin zone, specifically the region $[-\pi/2, \pi/2]^3$ as shown in Fig.~\ref{I222-BZ}, rotational symmetries impose the boundary condition:
\begin{equation} 
    \partial(\alpha + \beta + \gamma) = a + b + c + a' + b' + c'.
\end{equation}
Since the integral of the Berry connection over $a + b + c$ equals that over $a' + b' + c'$, the topological invariant naturally emerges from this symmetry constraint. This observation precisely motivates our adoption of the new basis defined in Eq.~\eqref{I222-nonAbelian}, and our method successfully reconstructs this invariant.

\subsection{Topological invariants for $I23$}
The MCG $I23$ is characterized by the point group $T_h$, generated by the $2$-fold rotations $r_y$ and $r_z$, the $3$-fold rotation $r_3$, and primitive translations $l_a$, $l_b$, and $l_c$. The generators act on momentum space as
\begin{equation}
    r_y(k_x,k_y,k_z) = (-k_x,k_y,-k_z),~r_z(k_x,k_y,k_z) = (-k_x, -k_y, k_z),~r_3(k_x, k_y, k_z) = (k_z, k_x, k_y).
\end{equation}
The BZ is spanned by
\begin{equation}\label{I23-basis-1}
    \bm{b}_a = \pi(-1,1,1),~ \bm{b}_b = \pi(1,-1,1),~ \bm{b}_c = \pi(1,1,-1).
\end{equation}
The topological classification in this case is given by
\begin{equation}\label{I23-classify}
    H^2(I23,\mathbb{Z}) = \mathbb{Z}_3 \oplus \mathbb{Z}_4.
\end{equation}

The $\mathbb{Z}_3$ factor in Eq.~\eqref{I23-classify} originates from the group relation $r_3^3 = 1$. To construct the associated invariant, we introduce the integer
\begin{equation}
    \begin{split}
        N(r_3,r_3,r_3) &= \phi_{r_3}(r_3^2\bm{k}) +\phi_{r_3}(r_3\bm{k}) +\phi_{r_3}(\bm{k}) - \phi_{1}(\bm{k})\\
        &= \phi_{r_3}(r_3^2\bm{k}) +\phi_{r_3}(r_3\bm{k}) +\phi_{r_3}(\bm{k}).
    \end{split}
\end{equation}
This quantity is invariant modulo 3
\begin{equation}
    \nu_1 = \phi_{r_3}(r_3^2\bm{k}) +\phi_{r_3}(r_3\bm{k}) +\phi_{r_3}(\bm{k}) \mod 3.
\end{equation}
Under a coboundary transformation $\phi_{r_3}\rightarrow \phi_{r_3}+n$ for any integer $n$, the invariant transforms as $\nu_{1}\rightarrow \nu_1+3n$, confirming its $\mathbb{Z}_3$ character.

This invariant can be understood in terms of the little co-group. Evaluating $\nu_1$ along the high-symmetry line $\bm{k} = k(1,1,1)$ yields
\begin{equation}
    \nu_1 = 3\phi_{r_3}(\bm{k}) \mod 3.
\end{equation}
This expression corresponds to the three irreducible representations of the little co-group $\{1, r_3, r_3^2\}$ along this diagonal.

Let us turn to the $\mathbb{Z}_4$ factor in the classification. It can be constructed from the algebraic relation
\begin{equation}\label{I23-algebraic-relation1}
    r_yl_dr_zl_d = l_d^{-1}r_zl_d^{-1}r_y.
\end{equation}
Here, $l_d$ is the diagonal translation defined as
\begin{equation}
    l_d:=l_al_bl_c.
\end{equation}

The integer $N_1$ is introduced from the left-hand side as
\begin{equation}
        N_1 =\phi_{l_d}(l_d\bm{k}) + \phi_{r_z}(r_zl_d\bm{k}) + \phi_{l_d}(l_dr_zl_d\bm{k}) + \phi_{r_y}(r_yl_dr_zl_d\bm{k}) - \phi_{r_yl_dr_zl_d}(r_yl_dr_zl_d\bm{k}).
\end{equation}
Similarly, for the right-hand side, we have
\begin{equation}
        N_2 = \phi_{r_y}(r_y\bm{k}) + \phi_{l_d^{-1}}(l_d^{-1}r_y\bm{k}) + \phi_{r_z}(r_zl_d^{-1}r_y\bm{k}) + \phi_{l_d^{-1}}(l_d^{-1}r_zl_d^{-1}r_y\bm{k})-\phi_{l_d^{_1}r_zl_d^{-1}r_y}(l_d^{-1}r_zl_d^{-1}r_y\bm{k}).
\end{equation}
To construct the topological invariant, let us introduce another two integers as
\begin{equation}
\begin{split}
            N(l_d, l_d^{-1}) &= \phi_{l_d}(r_y\bm{k}) + \phi_{l_{d}^{-1}}(l_d^{-1}r_y\bm{k})\\
            N(l_d, l_d^{-1}) &= \phi_{l_d}(r_zl_d^{-1}r_y\bm{k}) + \phi_{l_{d}^{-1}}(l_d^{-1}r_zl_d^{-1}r_y\bm{k}).
\end{split}
\end{equation}
The $\mathbb{Z}_4$ topological invariant is then constructed as
\begin{equation}
\begin{split}
         \nu_2 =& N_1-N_2 + 2N(l_d, l_d^{-1}) \mod 4\\ =& \phi_{l_d}(l_d\bm{k}) + \phi_{l_d}(l_dr_zl_d\bm{k}) + \phi_{l_d}(r_y\bm{k}) + \phi_{l_d}(r_zl_d^{-1}r_y\bm{k})  \\& + \phi_{r_z}(r_zl_d\bm{k}) - \phi_{r_z}(r_zl_d^{-1}r_y\bm{k}) + \phi_{r_y}(r_yl_dr_zl_d\bm{k})- \phi_{r_y}(r_y\bm{k}) \mod 4.
\end{split}
\end{equation}
Under the coboundary transformation $\phi_{l_d}\rightarrow \phi_{l_d}+n$ for any integer $n$, $\nu_2$ transforms as
\begin{equation}
    \nu_2\rightarrow \nu_2+4n.
\end{equation}
Therefore, $\nu_2$ is indeed a $\mathbb{Z}_4$ topological invariant.

Note that Eq.~\eqref{I23-algebraic-relation1} is equivalent to 
\begin{equation}
    (r_yl_dr_zl_d)^2 = 1.
\end{equation}
By repeating the discussion above for the $\mathbb{Z}_3$ invariant, we observe that $\nu_2$ contains a $\mathbb{Z}_2$ component that originates from the little co-group representation of $\{1,r_yl_dr_zl_d\}$. The little co-group representation corresponds precisely to the parity of $\nu_2$. Consequently, a reduced $\mathbb{Z}_2$ invariant remains, satisfying $\mathbb{Z}_4/\mathbb{Z}_2 \cong \mathbb{Z}_2$. In other words, the $\mathbb{Z}_4$ invariant forms a nontrivial extension of two $\mathbb{Z}_2$ invariants. This reduced $\mathbb{Z}_2$ invariant, unsurprisingly, coincides with the one discussed in the $I222$ example and can therefore be expressed as an integral over the $2$D and $1$D subspaces of the BZ.

\bibliographystyle{apsrev4-1}
\bibliography{main}

\begin{thebibliography}{49}%
\makeatletter
\providecommand \@ifxundefined [1]{%
 \@ifx{#1\undefined}
}%
\providecommand \@ifnum [1]{%
 \ifnum #1\expandafter \@firstoftwo
 \else \expandafter \@secondoftwo
 \fi
}%
\providecommand \@ifx [1]{%
 \ifx #1\expandafter \@firstoftwo
 \else \expandafter \@secondoftwo
 \fi
}%
\providecommand \natexlab [1]{#1}%
\providecommand \enquote  [1]{``#1''}%
\providecommand \bibnamefont  [1]{#1}%
\providecommand \bibfnamefont [1]{#1}%
\providecommand \citenamefont [1]{#1}%
\providecommand \href@noop [0]{\@secondoftwo}%
\providecommand \href [0]{\begingroup \@sanitize@url \@href}%
\providecommand \@href[1]{\@@startlink{#1}\@@href}%
\providecommand \@@href[1]{\endgroup#1\@@endlink}%
\providecommand \@sanitize@url [0]{\catcode `\\12\catcode `\$12\catcode `\&12\catcode `\#12\catcode `\^12\catcode `\_12\catcode `\%12\relax}%
\providecommand \@@startlink[1]{}%
\providecommand \@@endlink[0]{}%
\providecommand \url  [0]{\begingroup\@sanitize@url \@url }%
\providecommand \@url [1]{\endgroup\@href {#1}{\urlprefix }}%
\providecommand \urlprefix  [0]{URL }%
\providecommand \Eprint [0]{\href }%
\providecommand \doibase [0]{http://dx.doi.org/}%
\providecommand \selectlanguage [0]{\@gobble}%
\providecommand \bibinfo  [0]{\@secondoftwo}%
\providecommand \bibfield  [0]{\@secondoftwo}%
\providecommand \translation [1]{[#1]}%
\providecommand \BibitemOpen [0]{}%
\providecommand \bibitemStop [0]{}%
\providecommand \bibitemNoStop [0]{.\EOS\space}%
\providecommand \EOS [0]{\spacefactor3000\relax}%
\providecommand \BibitemShut  [1]{\csname bibitem#1\endcsname}%
\let\auto@bib@innerbib\@empty
\bibitem [{\citenamefont {Bradley}\ and\ \citenamefont {Cracknell}(2010)}]{bradley2010mathematical}%
  \BibitemOpen
  \bibfield  {author} {\bibinfo {author} {\bibfnamefont {C.}~\bibnamefont {Bradley}}\ and\ \bibinfo {author} {\bibfnamefont {A.}~\bibnamefont {Cracknell}},\ }\href@noop {} {\emph {\bibinfo {title} {The mathematical theory of symmetry in solids: representation theory for point groups and space groups}}}\ (\bibinfo  {publisher} {Oxford University Press},\ \bibinfo {year} {2010})\BibitemShut {NoStop}%
\bibitem [{\citenamefont {Szczepanski}(2012)}]{szczepanski2012geometry}%
  \BibitemOpen
  \bibfield  {author} {\bibinfo {author} {\bibfnamefont {A.}~\bibnamefont {Szczepanski}},\ }\href@noop {} {\emph {\bibinfo {title} {Geometry of crystallographic groups}}},\ Vol.~\bibinfo {volume} {4}\ (\bibinfo  {publisher} {World scientific},\ \bibinfo {year} {2012})\BibitemShut {NoStop}%
\bibitem [{\citenamefont {Chen}\ \emph {et~al.}(2022)\citenamefont {Chen}, \citenamefont {Yang},\ and\ \citenamefont {Zhao}}]{chen2022brillouin}%
  \BibitemOpen
  \bibfield  {author} {\bibinfo {author} {\bibfnamefont {Z.~Y.}\ \bibnamefont {Chen}}, \bibinfo {author} {\bibfnamefont {S.~A.}\ \bibnamefont {Yang}}, \ and\ \bibinfo {author} {\bibfnamefont {Y.~X.}\ \bibnamefont {Zhao}},\ }\href@noop {} {\bibfield  {journal} {\bibinfo  {journal} {Nat. Commun.}\ }\textbf {\bibinfo {volume} {13}},\ \bibinfo {pages} {2215} (\bibinfo {year} {2022})}\BibitemShut {NoStop}%
\bibitem [{\citenamefont {Zhang}\ \emph {et~al.}(2023)\citenamefont {Zhang}, \citenamefont {Chen}, \citenamefont {Zhang},\ and\ \citenamefont {Zhao}}]{zhang2023nonsymmorphic}%
  \BibitemOpen
  \bibfield  {author} {\bibinfo {author} {\bibfnamefont {C.}~\bibnamefont {Zhang}}, \bibinfo {author} {\bibfnamefont {Z.~Y.}\ \bibnamefont {Chen}}, \bibinfo {author} {\bibfnamefont {Z.}~\bibnamefont {Zhang}}, \ and\ \bibinfo {author} {\bibfnamefont {Y.~X.}\ \bibnamefont {Zhao}},\ }\href@noop {} {\bibfield  {journal} {\bibinfo  {journal} {Phys. Rev. Lett.}\ }\textbf {\bibinfo {volume} {130}},\ \bibinfo {pages} {256601} (\bibinfo {year} {2023})}\BibitemShut {NoStop}%
\bibitem [{\citenamefont {Mackey}(1958)}]{mackey1958unitary}%
  \BibitemOpen
  \bibfield  {author} {\bibinfo {author} {\bibfnamefont {G.~W.}\ \bibnamefont {Mackey}},\ }\href@noop {} {\bibfield  {journal} {\bibinfo  {journal} {Acta. Mathe.}\ }\textbf {\bibinfo {volume} {99}},\ \bibinfo {pages} {265} (\bibinfo {year} {1958})}\BibitemShut {NoStop}%
\bibitem [{\citenamefont {Mackey}(1989)}]{mackey1989unitary}%
  \BibitemOpen
  \bibfield  {author} {\bibinfo {author} {\bibfnamefont {G.}~\bibnamefont {Mackey}},\ }\href@noop {} {\emph {\bibinfo {title} {Unitary group representations in physics, probability, and number theory, Advanced book classics}}}\ (\bibinfo  {publisher} {Addison-Wesley Publishing Company Advanced Book Program, Redwood City, CA},\ \bibinfo {year} {1989})\BibitemShut {NoStop}%
\bibitem [{\citenamefont {Zhang}\ \emph {et~al.}(2025)\citenamefont {Zhang}, \citenamefont {Wang}, \citenamefont {Lyu},\ and\ \citenamefont {Zhao}}]{Platycosms}%
  \BibitemOpen
  \bibfield  {author} {\bibinfo {author} {\bibfnamefont {C.}~\bibnamefont {Zhang}}, \bibinfo {author} {\bibfnamefont {P.}~\bibnamefont {Wang}}, \bibinfo {author} {\bibfnamefont {J.}~\bibnamefont {Lyu}}, \ and\ \bibinfo {author} {\bibfnamefont {Y.~X.}\ \bibnamefont {Zhao}},\ }\href@noop {} {\bibfield  {journal} {\bibinfo  {journal} {Phys. Rev. Lett.}\ }\textbf {\bibinfo {volume} {135}},\ \bibinfo {pages} {136601} (\bibinfo {year} {2025})}\BibitemShut {NoStop}%
\bibitem [{\citenamefont {Shao}\ \emph {et~al.}(2021)\citenamefont {Shao}, \citenamefont {Liu}, \citenamefont {Xiao}, \citenamefont {Yang},\ and\ \citenamefont {Zhao}}]{shao2021gauge}%
  \BibitemOpen
  \bibfield  {author} {\bibinfo {author} {\bibfnamefont {L.~B.}\ \bibnamefont {Shao}}, \bibinfo {author} {\bibfnamefont {Q.}~\bibnamefont {Liu}}, \bibinfo {author} {\bibfnamefont {R.}~\bibnamefont {Xiao}}, \bibinfo {author} {\bibfnamefont {S.~A.}\ \bibnamefont {Yang}}, \ and\ \bibinfo {author} {\bibfnamefont {Y.~X.}\ \bibnamefont {Zhao}},\ }\href@noop {} {\bibfield  {journal} {\bibinfo  {journal} {Phys. Rev. Lett.}\ }\textbf {\bibinfo {volume} {127}},\ \bibinfo {pages} {076401} (\bibinfo {year} {2021})}\BibitemShut {NoStop}%
\bibitem [{\citenamefont {Xue}\ \emph {et~al.}(2022)\citenamefont {Xue}, \citenamefont {Wang}, \citenamefont {Huang}, \citenamefont {Cheng}, \citenamefont {Yu}, \citenamefont {Foo}, \citenamefont {Zhao}, \citenamefont {Yang},\ and\ \citenamefont {Zhang}}]{xue2022projectively}%
  \BibitemOpen
  \bibfield  {author} {\bibinfo {author} {\bibfnamefont {H.}~\bibnamefont {Xue}}, \bibinfo {author} {\bibfnamefont {Z.}~\bibnamefont {Wang}}, \bibinfo {author} {\bibfnamefont {Y.-X.}\ \bibnamefont {Huang}}, \bibinfo {author} {\bibfnamefont {Z.}~\bibnamefont {Cheng}}, \bibinfo {author} {\bibfnamefont {L.}~\bibnamefont {Yu}}, \bibinfo {author} {\bibfnamefont {Y.~X.}\ \bibnamefont {Foo}}, \bibinfo {author} {\bibfnamefont {Y.~X.}\ \bibnamefont {Zhao}}, \bibinfo {author} {\bibfnamefont {S.~A.}\ \bibnamefont {Yang}}, \ and\ \bibinfo {author} {\bibfnamefont {B.}~\bibnamefont {Zhang}},\ }\href@noop {} {\bibfield  {journal} {\bibinfo  {journal} {Phys. Rev. Lett.}\ }\textbf {\bibinfo {volume} {128}},\ \bibinfo {pages} {116802} (\bibinfo {year} {2022})}\BibitemShut {NoStop}%
\bibitem [{\citenamefont {Li}\ \emph {et~al.}(2022)\citenamefont {Li}, \citenamefont {Du}, \citenamefont {Zhang}, \citenamefont {Li}, \citenamefont {Fan}, \citenamefont {Zhang},\ and\ \citenamefont {Qiu}}]{li2022acoustic}%
  \BibitemOpen
  \bibfield  {author} {\bibinfo {author} {\bibfnamefont {T.}~\bibnamefont {Li}}, \bibinfo {author} {\bibfnamefont {J.}~\bibnamefont {Du}}, \bibinfo {author} {\bibfnamefont {Q.}~\bibnamefont {Zhang}}, \bibinfo {author} {\bibfnamefont {Y.}~\bibnamefont {Li}}, \bibinfo {author} {\bibfnamefont {X.}~\bibnamefont {Fan}}, \bibinfo {author} {\bibfnamefont {F.}~\bibnamefont {Zhang}}, \ and\ \bibinfo {author} {\bibfnamefont {C.}~\bibnamefont {Qiu}},\ }\href@noop {} {\bibfield  {journal} {\bibinfo  {journal} {Phys. Rev. Lett.}\ }\textbf {\bibinfo {volume} {128}},\ \bibinfo {pages} {116803} (\bibinfo {year} {2022})}\BibitemShut {NoStop}%
\bibitem [{\citenamefont {Liu}\ \emph {et~al.}(2023)\citenamefont {Liu}, \citenamefont {Wei}, \citenamefont {Wu},\ and\ \citenamefont {Xiao}}]{liu2023mobius}%
  \BibitemOpen
  \bibfield  {author} {\bibinfo {author} {\bibfnamefont {Z.}~\bibnamefont {Liu}}, \bibinfo {author} {\bibfnamefont {G.}~\bibnamefont {Wei}}, \bibinfo {author} {\bibfnamefont {H.}~\bibnamefont {Wu}}, \ and\ \bibinfo {author} {\bibfnamefont {J.-J.}\ \bibnamefont {Xiao}},\ }\href@noop {} {\bibfield  {journal} {\bibinfo  {journal} {Nanophotonics}\ }\textbf {\bibinfo {volume} {12}},\ \bibinfo {pages} {3481} (\bibinfo {year} {2023})}\BibitemShut {NoStop}%
\bibitem [{\citenamefont {Meng}\ \emph {et~al.}(2023)\citenamefont {Meng}, \citenamefont {Lin}, \citenamefont {Shi}, \citenamefont {Wei}, \citenamefont {Yang}, \citenamefont {Yan}, \citenamefont {Zhu}, \citenamefont {Xi}, \citenamefont {Wang}, \citenamefont {Ge}, \citenamefont {Yuan}, \citenamefont {Chen}, \citenamefont {Liu}, \citenamefont {Sun}, \citenamefont {Chen}, \citenamefont {Yang},\ and\ \citenamefont {Gao}}]{meng2023spinful}%
  \BibitemOpen
  \bibfield  {author} {\bibinfo {author} {\bibfnamefont {Y.}~\bibnamefont {Meng}}, \bibinfo {author} {\bibfnamefont {S.}~\bibnamefont {Lin}}, \bibinfo {author} {\bibfnamefont {B.-J.}\ \bibnamefont {Shi}}, \bibinfo {author} {\bibfnamefont {B.}~\bibnamefont {Wei}}, \bibinfo {author} {\bibfnamefont {L.}~\bibnamefont {Yang}}, \bibinfo {author} {\bibfnamefont {B.}~\bibnamefont {Yan}}, \bibinfo {author} {\bibfnamefont {Z.}~\bibnamefont {Zhu}}, \bibinfo {author} {\bibfnamefont {X.}~\bibnamefont {Xi}}, \bibinfo {author} {\bibfnamefont {Y.}~\bibnamefont {Wang}}, \bibinfo {author} {\bibfnamefont {Y.}~\bibnamefont {Ge}}, \bibinfo {author} {\bibfnamefont {S.-Q.}\ \bibnamefont {Yuan}}, \bibinfo {author} {\bibfnamefont {J.}~\bibnamefont {Chen}}, \bibinfo {author} {\bibfnamefont {G.-G.}\ \bibnamefont {Liu}}, \bibinfo {author} {\bibfnamefont {H.-X.}\ \bibnamefont {Sun}}, \bibinfo {author} {\bibfnamefont {H.}~\bibnamefont {Chen}}, \bibinfo {author} {\bibfnamefont {Y.}~\bibnamefont {Yang}}, \ and\ \bibinfo {author}
  {\bibfnamefont {Z.}~\bibnamefont {Gao}},\ }\href@noop {} {\bibfield  {journal} {\bibinfo  {journal} {Phys. Rev. Lett.}\ }\textbf {\bibinfo {volume} {130}},\ \bibinfo {pages} {026101} (\bibinfo {year} {2023})}\BibitemShut {NoStop}%
\bibitem [{\citenamefont {Li}\ \emph {et~al.}(2023{\natexlab{a}})\citenamefont {Li}, \citenamefont {Liu}, \citenamefont {Zhang},\ and\ \citenamefont {Qiu}}]{li2023acoustic}%
  \BibitemOpen
  \bibfield  {author} {\bibinfo {author} {\bibfnamefont {T.}~\bibnamefont {Li}}, \bibinfo {author} {\bibfnamefont {L.}~\bibnamefont {Liu}}, \bibinfo {author} {\bibfnamefont {Q.}~\bibnamefont {Zhang}}, \ and\ \bibinfo {author} {\bibfnamefont {C.}~\bibnamefont {Qiu}},\ }\href@noop {} {\bibfield  {journal} {\bibinfo  {journal} {Commun. Phys.}\ }\textbf {\bibinfo {volume} {6}},\ \bibinfo {pages} {268} (\bibinfo {year} {2023}{\natexlab{a}})}\BibitemShut {NoStop}%
\bibitem [{\citenamefont {Pu}\ \emph {et~al.}(2023)\citenamefont {Pu}, \citenamefont {He}, \citenamefont {Deng}, \citenamefont {Huang}, \citenamefont {Ye}, \citenamefont {Lu}, \citenamefont {Ke},\ and\ \citenamefont {Liu}}]{Pu2023acoustic}%
  \BibitemOpen
  \bibfield  {author} {\bibinfo {author} {\bibfnamefont {Z.}~\bibnamefont {Pu}}, \bibinfo {author} {\bibfnamefont {H.}~\bibnamefont {He}}, \bibinfo {author} {\bibfnamefont {W.}~\bibnamefont {Deng}}, \bibinfo {author} {\bibfnamefont {X.}~\bibnamefont {Huang}}, \bibinfo {author} {\bibfnamefont {L.}~\bibnamefont {Ye}}, \bibinfo {author} {\bibfnamefont {J.}~\bibnamefont {Lu}}, \bibinfo {author} {\bibfnamefont {M.}~\bibnamefont {Ke}}, \ and\ \bibinfo {author} {\bibfnamefont {Z.}~\bibnamefont {Liu}},\ }\href@noop {} {\bibfield  {journal} {\bibinfo  {journal} {Phys. Rev. B}\ }\textbf {\bibinfo {volume} {108}},\ \bibinfo {pages} {L220101} (\bibinfo {year} {2023})}\BibitemShut {NoStop}%
\bibitem [{\citenamefont {Jiang}\ \emph {et~al.}(2023)\citenamefont {Jiang}, \citenamefont {Song}, \citenamefont {Li}, \citenamefont {Lu},\ and\ \citenamefont {Ke}}]{jiang2023photonic}%
  \BibitemOpen
  \bibfield  {author} {\bibinfo {author} {\bibfnamefont {C.}~\bibnamefont {Jiang}}, \bibinfo {author} {\bibfnamefont {Y.}~\bibnamefont {Song}}, \bibinfo {author} {\bibfnamefont {X.}~\bibnamefont {Li}}, \bibinfo {author} {\bibfnamefont {P.}~\bibnamefont {Lu}}, \ and\ \bibinfo {author} {\bibfnamefont {S.}~\bibnamefont {Ke}},\ }\href@noop {} {\bibfield  {journal} {\bibinfo  {journal} {Opt. Lett.}\ }\textbf {\bibinfo {volume} {48}},\ \bibinfo {pages} {2337} (\bibinfo {year} {2023})}\BibitemShut {NoStop}%
\bibitem [{\citenamefont {Liu}\ \emph {et~al.}(2024)\citenamefont {Liu}, \citenamefont {Jiang}, \citenamefont {Wen}, \citenamefont {Song}, \citenamefont {Li}, \citenamefont {Lu},\ and\ \citenamefont {Ke}}]{liu2024topological}%
  \BibitemOpen
  \bibfield  {author} {\bibinfo {author} {\bibfnamefont {Y.}~\bibnamefont {Liu}}, \bibinfo {author} {\bibfnamefont {C.}~\bibnamefont {Jiang}}, \bibinfo {author} {\bibfnamefont {W.}~\bibnamefont {Wen}}, \bibinfo {author} {\bibfnamefont {Y.}~\bibnamefont {Song}}, \bibinfo {author} {\bibfnamefont {X.}~\bibnamefont {Li}}, \bibinfo {author} {\bibfnamefont {P.}~\bibnamefont {Lu}}, \ and\ \bibinfo {author} {\bibfnamefont {S.}~\bibnamefont {Ke}},\ }\href@noop {} {\bibfield  {journal} {\bibinfo  {journal} {Phys. Rev. A}\ }\textbf {\bibinfo {volume} {109}},\ \bibinfo {pages} {013516} (\bibinfo {year} {2024})}\BibitemShut {NoStop}%
\bibitem [{\citenamefont {Fonseca}\ \emph {et~al.}(2024)\citenamefont {Fonseca}, \citenamefont {Vaidya}, \citenamefont {Christensen}, \citenamefont {Rechtsman}, \citenamefont {Hughes},\ and\ \citenamefont {Solja\ifmmode \check{c}\else \v{c}\fi{}i\ifmmode~\acute{c}\else \'{c}\fi{}}}]{Fonseca2024Weyl}%
  \BibitemOpen
  \bibfield  {author} {\bibinfo {author} {\bibfnamefont {A.~G.}\ \bibnamefont {Fonseca}}, \bibinfo {author} {\bibfnamefont {S.}~\bibnamefont {Vaidya}}, \bibinfo {author} {\bibfnamefont {T.}~\bibnamefont {Christensen}}, \bibinfo {author} {\bibfnamefont {M.~C.}\ \bibnamefont {Rechtsman}}, \bibinfo {author} {\bibfnamefont {T.~L.}\ \bibnamefont {Hughes}}, \ and\ \bibinfo {author} {\bibfnamefont {M.}~\bibnamefont {Solja\ifmmode \check{c}\else \v{c}\fi{}i\ifmmode~\acute{c}\else \'{c}\fi{}}},\ }\href@noop {} {\bibfield  {journal} {\bibinfo  {journal} {Phys. Rev. Lett.}\ }\textbf {\bibinfo {volume} {132}},\ \bibinfo {pages} {266601} (\bibinfo {year} {2024})}\BibitemShut {NoStop}%
\bibitem [{\citenamefont {Tao}\ \emph {et~al.}(2024)\citenamefont {Tao}, \citenamefont {Yan}, \citenamefont {Peng}, \citenamefont {Wei}, \citenamefont {Cui}, \citenamefont {Yang}, \citenamefont {Chen},\ and\ \citenamefont {Xu}}]{Tao2024Higher}%
  \BibitemOpen
  \bibfield  {author} {\bibinfo {author} {\bibfnamefont {Y.-L.}\ \bibnamefont {Tao}}, \bibinfo {author} {\bibfnamefont {M.}~\bibnamefont {Yan}}, \bibinfo {author} {\bibfnamefont {M.}~\bibnamefont {Peng}}, \bibinfo {author} {\bibfnamefont {Q.}~\bibnamefont {Wei}}, \bibinfo {author} {\bibfnamefont {Z.}~\bibnamefont {Cui}}, \bibinfo {author} {\bibfnamefont {S.~A.}\ \bibnamefont {Yang}}, \bibinfo {author} {\bibfnamefont {G.}~\bibnamefont {Chen}}, \ and\ \bibinfo {author} {\bibfnamefont {Y.}~\bibnamefont {Xu}},\ }\href@noop {} {\bibfield  {journal} {\bibinfo  {journal} {Phys. Rev. B}\ }\textbf {\bibinfo {volume} {109}},\ \bibinfo {pages} {134107} (\bibinfo {year} {2024})}\BibitemShut {NoStop}%
\bibitem [{\citenamefont {Zhu}\ \emph {et~al.}(2024)\citenamefont {Zhu}, \citenamefont {Yang}, \citenamefont {Wu}, \citenamefont {Meng}, \citenamefont {Xi}, \citenamefont {Yan}, \citenamefont {Chen}, \citenamefont {Lu}, \citenamefont {Huang}, \citenamefont {Deng} \emph {et~al.}}]{zhu2024brillouin}%
  \BibitemOpen
  \bibfield  {author} {\bibinfo {author} {\bibfnamefont {Z.}~\bibnamefont {Zhu}}, \bibinfo {author} {\bibfnamefont {L.}~\bibnamefont {Yang}}, \bibinfo {author} {\bibfnamefont {J.}~\bibnamefont {Wu}}, \bibinfo {author} {\bibfnamefont {Y.}~\bibnamefont {Meng}}, \bibinfo {author} {\bibfnamefont {X.}~\bibnamefont {Xi}}, \bibinfo {author} {\bibfnamefont {B.}~\bibnamefont {Yan}}, \bibinfo {author} {\bibfnamefont {J.}~\bibnamefont {Chen}}, \bibinfo {author} {\bibfnamefont {J.}~\bibnamefont {Lu}}, \bibinfo {author} {\bibfnamefont {X.}~\bibnamefont {Huang}}, \bibinfo {author} {\bibfnamefont {W.}~\bibnamefont {Deng}},  \emph {et~al.},\ }\href@noop {} {\bibfield  {journal} {\bibinfo  {journal} {Sci. Bull.}\ } (\bibinfo {year} {2024})}\BibitemShut {NoStop}%
\bibitem [{\citenamefont {Hu}\ \emph {et~al.}(2024)\citenamefont {Hu}, \citenamefont {Zhuang},\ and\ \citenamefont {Yang}}]{Hu2024higher}%
  \BibitemOpen
  \bibfield  {author} {\bibinfo {author} {\bibfnamefont {J.}~\bibnamefont {Hu}}, \bibinfo {author} {\bibfnamefont {S.}~\bibnamefont {Zhuang}}, \ and\ \bibinfo {author} {\bibfnamefont {Y.}~\bibnamefont {Yang}},\ }\href@noop {} {\bibfield  {journal} {\bibinfo  {journal} {Phys. Rev. Lett.}\ }\textbf {\bibinfo {volume} {132}},\ \bibinfo {pages} {213801} (\bibinfo {year} {2024})}\BibitemShut {NoStop}%
\bibitem [{\citenamefont {Long}\ \emph {et~al.}(2024)\citenamefont {Long}, \citenamefont {Wang}, \citenamefont {Zhang}, \citenamefont {Xue}, \citenamefont {Zhao},\ and\ \citenamefont {Zhang}}]{Long2024nonabelian}%
  \BibitemOpen
  \bibfield  {author} {\bibinfo {author} {\bibfnamefont {Y.}~\bibnamefont {Long}}, \bibinfo {author} {\bibfnamefont {Z.}~\bibnamefont {Wang}}, \bibinfo {author} {\bibfnamefont {C.}~\bibnamefont {Zhang}}, \bibinfo {author} {\bibfnamefont {H.}~\bibnamefont {Xue}}, \bibinfo {author} {\bibfnamefont {Y.~X.}\ \bibnamefont {Zhao}}, \ and\ \bibinfo {author} {\bibfnamefont {B.}~\bibnamefont {Zhang}},\ }\href@noop {} {\bibfield  {journal} {\bibinfo  {journal} {Phys. Rev. Lett.}\ }\textbf {\bibinfo {volume} {132}},\ \bibinfo {pages} {236401} (\bibinfo {year} {2024})}\BibitemShut {NoStop}%
\bibitem [{\citenamefont {Wang}\ \emph {et~al.}(2025)\citenamefont {Wang}, \citenamefont {Fu}, \citenamefont {Ye}, \citenamefont {He}, \citenamefont {Deng}, \citenamefont {Lu}, \citenamefont {Ke},\ and\ \citenamefont {Liu}}]{wang2025non}%
  \BibitemOpen
  \bibfield  {author} {\bibinfo {author} {\bibfnamefont {Q.}~\bibnamefont {Wang}}, \bibinfo {author} {\bibfnamefont {Z.}~\bibnamefont {Fu}}, \bibinfo {author} {\bibfnamefont {L.}~\bibnamefont {Ye}}, \bibinfo {author} {\bibfnamefont {H.}~\bibnamefont {He}}, \bibinfo {author} {\bibfnamefont {W.}~\bibnamefont {Deng}}, \bibinfo {author} {\bibfnamefont {J.}~\bibnamefont {Lu}}, \bibinfo {author} {\bibfnamefont {M.}~\bibnamefont {Ke}}, \ and\ \bibinfo {author} {\bibfnamefont {Z.}~\bibnamefont {Liu}},\ }\href@noop {} {\bibfield  {journal} {\bibinfo  {journal} {Phys. Rev. B}\ }\textbf {\bibinfo {volume} {111}},\ \bibinfo {pages} {L100101} (\bibinfo {year} {2025})}\BibitemShut {NoStop}%
\bibitem [{\citenamefont {Li}\ \emph {et~al.}(2023{\natexlab{b}})\citenamefont {Li}, \citenamefont {Sun}, \citenamefont {Zhang}, \citenamefont {Guo},\ and\ \citenamefont {Trauzettel}}]{Li2023Klein}%
  \BibitemOpen
  \bibfield  {author} {\bibinfo {author} {\bibfnamefont {C.-A.}\ \bibnamefont {Li}}, \bibinfo {author} {\bibfnamefont {J.}~\bibnamefont {Sun}}, \bibinfo {author} {\bibfnamefont {S.-B.}\ \bibnamefont {Zhang}}, \bibinfo {author} {\bibfnamefont {H.}~\bibnamefont {Guo}}, \ and\ \bibinfo {author} {\bibfnamefont {B.}~\bibnamefont {Trauzettel}},\ }\href@noop {} {\bibfield  {journal} {\bibinfo  {journal} {Phys. Rev. B}\ }\textbf {\bibinfo {volume} {108}},\ \bibinfo {pages} {235412} (\bibinfo {year} {2023}{\natexlab{b}})}\BibitemShut {NoStop}%
\bibitem [{\citenamefont {Huang}\ \emph {et~al.}(2025)\citenamefont {Huang}, \citenamefont {Li}, \citenamefont {Jia}, \citenamefont {Hu}, \citenamefont {Li}, \citenamefont {Li}, \citenamefont {Xie}, \citenamefont {Lu}, \citenamefont {Zhan}, \citenamefont {Chen} \emph {et~al.}}]{huang2025experimental}%
  \BibitemOpen
  \bibfield  {author} {\bibinfo {author} {\bibfnamefont {R.}~\bibnamefont {Huang}}, \bibinfo {author} {\bibfnamefont {H.}~\bibnamefont {Li}}, \bibinfo {author} {\bibfnamefont {S.}~\bibnamefont {Jia}}, \bibinfo {author} {\bibfnamefont {J.}~\bibnamefont {Hu}}, \bibinfo {author} {\bibfnamefont {S.}~\bibnamefont {Li}}, \bibinfo {author} {\bibfnamefont {J.}~\bibnamefont {Li}}, \bibinfo {author} {\bibfnamefont {B.}~\bibnamefont {Xie}}, \bibinfo {author} {\bibfnamefont {M.}~\bibnamefont {Lu}}, \bibinfo {author} {\bibfnamefont {P.}~\bibnamefont {Zhan}}, \bibinfo {author} {\bibfnamefont {Y.}~\bibnamefont {Chen}},  \emph {et~al.},\ }\href@noop {} {\bibfield  {journal} {\bibinfo  {journal} {Phys. Rev. Lett.}\ }\textbf {\bibinfo {volume} {135}},\ \bibinfo {pages} {216603} (\bibinfo {year} {2025})}\BibitemShut {NoStop}%
\bibitem [{\citenamefont {Xiao}\ \emph {et~al.}(2024)\citenamefont {Xiao}, \citenamefont {Zhao}, \citenamefont {Li}, \citenamefont {Shindou},\ and\ \citenamefont {Song}}]{xiao2024spin}%
  \BibitemOpen
  \bibfield  {author} {\bibinfo {author} {\bibfnamefont {Z.}~\bibnamefont {Xiao}}, \bibinfo {author} {\bibfnamefont {J.}~\bibnamefont {Zhao}}, \bibinfo {author} {\bibfnamefont {Y.}~\bibnamefont {Li}}, \bibinfo {author} {\bibfnamefont {R.}~\bibnamefont {Shindou}}, \ and\ \bibinfo {author} {\bibfnamefont {Z.-D.}\ \bibnamefont {Song}},\ }\href@noop {} {\bibfield  {journal} {\bibinfo  {journal} {Phys. Rev. X}\ }\textbf {\bibinfo {volume} {14}},\ \bibinfo {pages} {031037} (\bibinfo {year} {2024})}\BibitemShut {NoStop}%
\bibitem [{\citenamefont {C{\u{a}}lug{\u{a}}ru}\ \emph {et~al.}(2025)\citenamefont {C{\u{a}}lug{\u{a}}ru}, \citenamefont {Jiang}, \citenamefont {Hu}, \citenamefont {Pi}, \citenamefont {Yu}, \citenamefont {Vergniory}, \citenamefont {Shan}, \citenamefont {Felser}, \citenamefont {Schoop}, \citenamefont {Efetov} \emph {et~al.}}]{cualuguaru2025moire}%
  \BibitemOpen
  \bibfield  {author} {\bibinfo {author} {\bibfnamefont {D.}~\bibnamefont {C{\u{a}}lug{\u{a}}ru}}, \bibinfo {author} {\bibfnamefont {Y.}~\bibnamefont {Jiang}}, \bibinfo {author} {\bibfnamefont {H.}~\bibnamefont {Hu}}, \bibinfo {author} {\bibfnamefont {H.}~\bibnamefont {Pi}}, \bibinfo {author} {\bibfnamefont {J.}~\bibnamefont {Yu}}, \bibinfo {author} {\bibfnamefont {M.~G.}\ \bibnamefont {Vergniory}}, \bibinfo {author} {\bibfnamefont {J.}~\bibnamefont {Shan}}, \bibinfo {author} {\bibfnamefont {C.}~\bibnamefont {Felser}}, \bibinfo {author} {\bibfnamefont {L.~M.}\ \bibnamefont {Schoop}}, \bibinfo {author} {\bibfnamefont {D.~K.}\ \bibnamefont {Efetov}},  \emph {et~al.},\ }\href@noop {} {\bibfield  {journal} {\bibinfo  {journal} {Nature}\ }\textbf {\bibinfo {volume} {643}},\ \bibinfo {pages} {376} (\bibinfo {year} {2025})}\BibitemShut {NoStop}%
\bibitem [{tri()}]{trivial_bundles}%
  \BibitemOpen
  \href@noop {} {}\bibinfo {note} {Momentum space $R^d_F$ is contractible, and hence homotopic to a point. Consequently, when disregarding the MCG actions, all vector bundles over $R^d_F$ are topologically trivial.}\BibitemShut {Stop}%
\bibitem [{\citenamefont {Qi}\ \emph {et~al.}(2008)\citenamefont {Qi}, \citenamefont {Hughes},\ and\ \citenamefont {Zhang}}]{XLQi_PRB}%
  \BibitemOpen
  \bibfield  {author} {\bibinfo {author} {\bibfnamefont {X.-L.}\ \bibnamefont {Qi}}, \bibinfo {author} {\bibfnamefont {T.~L.}\ \bibnamefont {Hughes}}, \ and\ \bibinfo {author} {\bibfnamefont {S.-C.}\ \bibnamefont {Zhang}},\ }\href@noop {} {\bibfield  {journal} {\bibinfo  {journal} {Phys. Rev. B}\ }\textbf {\bibinfo {volume} {78}},\ \bibinfo {pages} {195424} (\bibinfo {year} {2008})}\BibitemShut {NoStop}%
\bibitem [{Iso()}]{Iso_Notes}%
  \BibitemOpen
  \href@noop {} {}\bibinfo {note} {The isomorphism can be understood as a generalization of the familiar $H^n(G,U(1))\cong H^{n+1}(G,\mathbb{Z})$ for finite groups $G$, which was used in symmetry protected topological phases~\cite{dijkgraaf1990topological,SPT_Wen,SPT_Kapustin,SPT_Spatial}.}\BibitemShut {Stop}%
\bibitem [{GAP(2024)}]{GAP4}%
  \BibitemOpen
  \href@noop {} {\emph {\bibinfo {title} {GAP -- Groups, Algorithms, and Programming, Version 4.13.1}}},\ \bibinfo {organization} {The GAP~Group} (\bibinfo {year} {2024})\BibitemShut {NoStop}%
\bibitem [{\citenamefont {Shiozaki}\ and\ \citenamefont {Sato}(2014)}]{shiozaki2014topology}%
  \BibitemOpen
  \bibfield  {author} {\bibinfo {author} {\bibfnamefont {K.}~\bibnamefont {Shiozaki}}\ and\ \bibinfo {author} {\bibfnamefont {M.}~\bibnamefont {Sato}},\ }\href@noop {} {\bibfield  {journal} {\bibinfo  {journal} {Phys. Rev. B}\ }\textbf {\bibinfo {volume} {90}},\ \bibinfo {pages} {165114} (\bibinfo {year} {2014})}\BibitemShut {NoStop}%
\bibitem [{\citenamefont {Alexandradinata}\ \emph {et~al.}(2014)\citenamefont {Alexandradinata}, \citenamefont {Fang}, \citenamefont {Gilbert},\ and\ \citenamefont {Bernevig}}]{alexandradinata2014spin}%
  \BibitemOpen
  \bibfield  {author} {\bibinfo {author} {\bibfnamefont {A.}~\bibnamefont {Alexandradinata}}, \bibinfo {author} {\bibfnamefont {C.}~\bibnamefont {Fang}}, \bibinfo {author} {\bibfnamefont {M.~J.}\ \bibnamefont {Gilbert}}, \ and\ \bibinfo {author} {\bibfnamefont {B.~A.}\ \bibnamefont {Bernevig}},\ }\href@noop {} {\bibfield  {journal} {\bibinfo  {journal} {Phys. Rev. Lett.}\ }\textbf {\bibinfo {volume} {113}},\ \bibinfo {pages} {116403} (\bibinfo {year} {2014})}\BibitemShut {NoStop}%
\bibitem [{\citenamefont {Shiozaki}\ \emph {et~al.}(2016)\citenamefont {Shiozaki}, \citenamefont {Sato},\ and\ \citenamefont {Gomi}}]{shiozaki2016topology}%
  \BibitemOpen
  \bibfield  {author} {\bibinfo {author} {\bibfnamefont {K.}~\bibnamefont {Shiozaki}}, \bibinfo {author} {\bibfnamefont {M.}~\bibnamefont {Sato}}, \ and\ \bibinfo {author} {\bibfnamefont {K.}~\bibnamefont {Gomi}},\ }\href@noop {} {\bibfield  {journal} {\bibinfo  {journal} {Phys. Rev. B}\ }\textbf {\bibinfo {volume} {93}},\ \bibinfo {pages} {195413} (\bibinfo {year} {2016})}\BibitemShut {NoStop}%
\bibitem [{\citenamefont {Shiozaki}\ \emph {et~al.}(2022)\citenamefont {Shiozaki}, \citenamefont {Sato},\ and\ \citenamefont {Gomi}}]{shiozaki2022atiyah}%
  \BibitemOpen
  \bibfield  {author} {\bibinfo {author} {\bibfnamefont {K.}~\bibnamefont {Shiozaki}}, \bibinfo {author} {\bibfnamefont {M.}~\bibnamefont {Sato}}, \ and\ \bibinfo {author} {\bibfnamefont {K.}~\bibnamefont {Gomi}},\ }\href@noop {} {\bibfield  {journal} {\bibinfo  {journal} {Phys. Rev. B}\ }\textbf {\bibinfo {volume} {106}},\ \bibinfo {pages} {165103} (\bibinfo {year} {2022})}\BibitemShut {NoStop}%
\bibitem [{\citenamefont {Liu}\ \emph {et~al.}(2022)\citenamefont {Liu}, \citenamefont {Gao}, \citenamefont {Wang}, \citenamefont {Xi}, \citenamefont {Hu}, \citenamefont {Wang}, \citenamefont {Liu}, \citenamefont {Lin}, \citenamefont {Deng}, \citenamefont {Yang} \emph {et~al.}}]{Baile_Nature}%
  \BibitemOpen
  \bibfield  {author} {\bibinfo {author} {\bibfnamefont {G.-G.}\ \bibnamefont {Liu}}, \bibinfo {author} {\bibfnamefont {Z.}~\bibnamefont {Gao}}, \bibinfo {author} {\bibfnamefont {Q.}~\bibnamefont {Wang}}, \bibinfo {author} {\bibfnamefont {X.}~\bibnamefont {Xi}}, \bibinfo {author} {\bibfnamefont {Y.-H.}\ \bibnamefont {Hu}}, \bibinfo {author} {\bibfnamefont {M.}~\bibnamefont {Wang}}, \bibinfo {author} {\bibfnamefont {C.}~\bibnamefont {Liu}}, \bibinfo {author} {\bibfnamefont {X.}~\bibnamefont {Lin}}, \bibinfo {author} {\bibfnamefont {L.}~\bibnamefont {Deng}}, \bibinfo {author} {\bibfnamefont {S.~A.}\ \bibnamefont {Yang}},  \emph {et~al.},\ }\href@noop {} {\bibfield  {journal} {\bibinfo  {journal} {Nature}\ }\textbf {\bibinfo {volume} {609}},\ \bibinfo {pages} {925} (\bibinfo {year} {2022})}\BibitemShut {NoStop}%
\bibitem [{\citenamefont {Segal}(1968)}]{segal1968equivariant}%
  \BibitemOpen
  \bibfield  {author} {\bibinfo {author} {\bibfnamefont {G.}~\bibnamefont {Segal}},\ }\href@noop {} {\bibfield  {journal} {\bibinfo  {journal} {Publ. Math. l'IH{\'E}S}\ }\textbf {\bibinfo {volume} {34}},\ \bibinfo {pages} {129} (\bibinfo {year} {1968})}\BibitemShut {NoStop}%
\bibitem [{\citenamefont {Atiyah}\ and\ \citenamefont {Segal}(2004)}]{atiyah2004twisted}%
  \BibitemOpen
  \bibfield  {author} {\bibinfo {author} {\bibfnamefont {M.}~\bibnamefont {Atiyah}}\ and\ \bibinfo {author} {\bibfnamefont {G.}~\bibnamefont {Segal}},\ }\href@noop {} {\bibfield  {journal} {\bibinfo  {journal} {arXiv preprint math/0407054}\ } (\bibinfo {year} {2004})}\BibitemShut {NoStop}%
\bibitem [{\citenamefont {Freed}\ and\ \citenamefont {Moore}(2013)}]{freed2013twisted}%
  \BibitemOpen
  \bibfield  {author} {\bibinfo {author} {\bibfnamefont {D.~S.}\ \bibnamefont {Freed}}\ and\ \bibinfo {author} {\bibfnamefont {G.~W.}\ \bibnamefont {Moore}},\ }\href@noop {} {\bibfield  {journal} {\bibinfo  {journal} {Annales Henri Poincar{\'e}}\ }\textbf {\bibinfo {volume} {14}},\ \bibinfo {pages} {1927} (\bibinfo {year} {2013})}\BibitemShut {NoStop}%
\bibitem [{\citenamefont {Gomi}\ \emph {et~al.}(2017)\citenamefont {Gomi} \emph {et~al.}}]{gomi2017twists}%
  \BibitemOpen
  \bibfield  {author} {\bibinfo {author} {\bibfnamefont {K.}~\bibnamefont {Gomi}} \emph {et~al.},\ }\href@noop {} {\bibfield  {journal} {\bibinfo  {journal} {Symmetry Integr. Geom.: Methods Appl.}\ }\textbf {\bibinfo {volume} {13}},\ \bibinfo {pages} {014} (\bibinfo {year} {2017})}\BibitemShut {NoStop}%
\bibitem [{\citenamefont {Dijkgraaf}\ and\ \citenamefont {Witten}(1990)}]{dijkgraaf1990topological}%
  \BibitemOpen
  \bibfield  {author} {\bibinfo {author} {\bibfnamefont {R.}~\bibnamefont {Dijkgraaf}}\ and\ \bibinfo {author} {\bibfnamefont {E.}~\bibnamefont {Witten}},\ }\href@noop {} {\bibfield  {journal} {\bibinfo  {journal} {Communications in Mathematical Physics}\ }\textbf {\bibinfo {volume} {129}},\ \bibinfo {pages} {393} (\bibinfo {year} {1990})}\BibitemShut {NoStop}%
\bibitem [{\citenamefont {Chen}\ \emph {et~al.}(2013)\citenamefont {Chen}, \citenamefont {Gu}, \citenamefont {Liu},\ and\ \citenamefont {Wen}}]{SPT_Wen}%
  \BibitemOpen
  \bibfield  {author} {\bibinfo {author} {\bibfnamefont {X.}~\bibnamefont {Chen}}, \bibinfo {author} {\bibfnamefont {Z.-C.}\ \bibnamefont {Gu}}, \bibinfo {author} {\bibfnamefont {Z.-X.}\ \bibnamefont {Liu}}, \ and\ \bibinfo {author} {\bibfnamefont {X.-G.}\ \bibnamefont {Wen}},\ }\href@noop {} {\bibfield  {journal} {\bibinfo  {journal} {Phys. Rev. B}\ }\textbf {\bibinfo {volume} {87}},\ \bibinfo {pages} {155114} (\bibinfo {year} {2013})}\BibitemShut {NoStop}%
\bibitem [{\citenamefont {Kane}\ and\ \citenamefont {Mele}(2005)}]{kane2005z}%
  \BibitemOpen
  \bibfield  {author} {\bibinfo {author} {\bibfnamefont {C.~L.}\ \bibnamefont {Kane}}\ and\ \bibinfo {author} {\bibfnamefont {E.~J.}\ \bibnamefont {Mele}},\ }\href@noop {} {\bibfield  {journal} {\bibinfo  {journal} {Phys. Rev. Lett.}\ }\textbf {\bibinfo {volume} {95}},\ \bibinfo {pages} {146802} (\bibinfo {year} {2005})}\BibitemShut {NoStop}%
\bibitem [{\citenamefont {Atiyah}\ and\ \citenamefont {Bott}(1984)}]{atiyah1984moment}%
  \BibitemOpen
  \bibfield  {author} {\bibinfo {author} {\bibfnamefont {M.~F.}\ \bibnamefont {Atiyah}}\ and\ \bibinfo {author} {\bibfnamefont {R.}~\bibnamefont {Bott}},\ }\href@noop {} {\bibfield  {journal} {\bibinfo  {journal} {Topology}\ }\textbf {\bibinfo {volume} {23}},\ \bibinfo {pages} {1} (\bibinfo {year} {1984})}\BibitemShut {NoStop}%
\bibitem [{SM()}]{SM}%
  \BibitemOpen
  \href@noop {} {}\bibinfo {note} {See the Supplemental Materials for basics of group cohomology and Borel cohomology, the classification tables for 2D and 3D MCGs, and topological invariants for MCGs $P1$, $Pg$, $P2_1/c$, $I222$ and $I23$.}\BibitemShut {Stop}%
\bibitem [{\citenamefont {Brown}(1982)}]{Brown1982cohomology}%
  \BibitemOpen
  \bibfield  {author} {\bibinfo {author} {\bibfnamefont {K.~S.}\ \bibnamefont {Brown}},\ }\href@noop {} {\emph {\bibinfo {title} {Cohomology of groups}}}\ (\bibinfo  {publisher} {Springer},\ \bibinfo {year} {1982})\BibitemShut {NoStop}%
\bibitem [{\citenamefont {Weibel}(1994)}]{weibel1994introduction}%
  \BibitemOpen
  \bibfield  {author} {\bibinfo {author} {\bibfnamefont {C.~A.}\ \bibnamefont {Weibel}},\ }\href@noop {} {\emph {\bibinfo {title} {An introduction to homological algebra}}}\ (\bibinfo  {publisher} {Cambridge university press},\ \bibinfo {year} {1994})\BibitemShut {NoStop}%
\bibitem [{\citenamefont {Hatcher}(2001)}]{hatcher2001algebraic}%
  \BibitemOpen
  \bibfield  {author} {\bibinfo {author} {\bibfnamefont {A.}~\bibnamefont {Hatcher}},\ }\href@noop {} {\emph {\bibinfo {title} {Algebraic topology}}}\ (\bibinfo  {publisher} {Cambridge university press},\ \bibinfo {year} {2001})\BibitemShut {NoStop}%
\bibitem [{\citenamefont {Kapustin}\ and\ \citenamefont {Thorngren}(2014)}]{SPT_Kapustin}%
  \BibitemOpen
  \bibfield  {author} {\bibinfo {author} {\bibfnamefont {A.}~\bibnamefont {Kapustin}}\ and\ \bibinfo {author} {\bibfnamefont {R.}~\bibnamefont {Thorngren}},\ }\href@noop {} {\bibfield  {journal} {\bibinfo  {journal} {Phys. Rev. Lett.}\ }\textbf {\bibinfo {volume} {112}},\ \bibinfo {pages} {231602} (\bibinfo {year} {2014})}\BibitemShut {NoStop}%
\bibitem [{SPT()}]{SPT_Spatial}%
  \BibitemOpen
  \href@noop {} {}\bibinfo {note} {Due to the analogy, we propose $H^{d+1}(\Gamma, \mathcal{F}(X,U(1)))$ as a natural group to consider for the classification of bosonic topological phases protected by the spatial group $\Gamma$ acting on the manifold $X$.}\BibitemShut {Stop}%
\end{thebibliography}%

\end{document}